\documentclass[twoside,twocolumn,9pt]{article}
\usepackage{extsizes}
\usepackage[super,sort&compress,comma]{natbib}
\usepackage[version=3]{mhchem}
\usepackage[left=1.5cm, right=1.5cm, top=1.785cm, bottom=2.0cm]{geometry}
\usepackage{balance}
\usepackage{mathptmx}
\usepackage{sectsty}
\usepackage{graphicx}
\usepackage{lastpage}
\usepackage[format=plain,justification=justified,singlelinecheck=false,font={stretch=1.125,small,sf},labelfont=bf,labelsep=space]{caption}
\usepackage{float}
\usepackage{fancyhdr}
\usepackage{fnpos}
\usepackage[english]{babel}
\addto{\captionsenglish}{%
  
}
\usepackage{array}
\usepackage{droidsans}
\usepackage{charter}
\usepackage[T1]{fontenc}
\usepackage[usenames,dvipsnames]{xcolor}
\usepackage{setspace}
\usepackage[compact]{titlesec}
\usepackage{hyperref}

\usepackage{epstopdf}

\definecolor{cream}{RGB}{222,217,201}

\usepackage[separate-uncertainty=true,list-units=repeat,multi-part-units=brackets,range-phrase=--,range-units=single]{siunitx}
\usepackage{color}
\usepackage[normalem]{ulem} 
\usepackage{amssymb}

\newcommand{\avg}[1]{\langle#1\rangle}

\newcommand{\me}{\mathrm{e}}
\newcommand{\Ang}{\si{\angstrom}\;}

\begin{document}

\pagestyle{fancy}
\thispagestyle{plain}
\fancypagestyle{plain}{
\renewcommand{\headrulewidth}{0pt}
}

\makeFNbottom
\makeatletter
\renewcommand\LARGE{\@setfontsize\LARGE{15pt}{17}}
\renewcommand\Large{\@setfontsize\Large{12pt}{14}}
\renewcommand\large{\@setfontsize\large{10pt}{12}}
\renewcommand\footnotesize{\@setfontsize\footnotesize{7pt}{10}}
\makeatother

\renewcommand{\thefootnote}{\fnsymbol{footnote}}
\renewcommand\footnoterule{\vspace*{1pt}%
\color{cream}\hrule width 3.5in height 0.4pt \color{black}\vspace*{5pt}}
\setcounter{secnumdepth}{5}

\makeatletter
\renewcommand\@biblabel[1]{#1}
\renewcommand\@makefntext[1]%
{\noindent\makebox[0pt][r]{\@thefnmark\,}#1}
\makeatother
\renewcommand{\figurename}{\small{Fig.}~}
\sectionfont{\sffamily\Large}
\subsectionfont{\normalsize}
\subsubsectionfont{\bf}
\setstretch{1.125} 
\setlength{\skip\footins}{0.8cm}
\setlength{\footnotesep}{0.25cm}
\setlength{\jot}{10pt}
\titlespacing*{\section}{0pt}{4pt}{4pt}
\titlespacing*{\subsection}{0pt}{15pt}{1pt}

\fancyfoot{}
\fancyfoot[LO,RE]{\vspace{-7.1pt}\includegraphics[height=9pt]{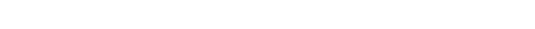}}
\fancyfoot[RO]{\footnotesize{\sffamily{1--\pageref{LastPage} ~\textbar  \hspace{2pt}\thepage}}}
\fancyfoot[LE]{\footnotesize{\sffamily{\thepage~\textbar\hspace{3.45cm} 1--\pageref{LastPage}}}}
\fancyhead{}
\renewcommand{\headrulewidth}{0pt}
\renewcommand{\footrulewidth}{0pt}
\setlength{\arrayrulewidth}{1pt}
\setlength{\columnsep}{6.5mm}
\setlength\bibsep{1pt}

\makeatletter
\newlength{\figrulesep}
\setlength{\figrulesep}{0.5\textfloatsep}

\newcommand{\topfigrule}{\vspace*{-1pt}%
\noindent{\color{cream}\rule[-\figrulesep]{\columnwidth}{1.5pt}} }

\newcommand{\botfigrule}{\vspace*{-2pt}%
\noindent{\color{cream}\rule[\figrulesep]{\columnwidth}{1.5pt}} }

\newcommand{\dblfigrule}{\vspace*{-1pt}%
\noindent{\color{cream}\rule[-\figrulesep]{\textwidth}{1.5pt}} }

\makeatother

\twocolumn[
  \begin{@twocolumnfalse}
{
 \includegraphics[width=18.5cm]{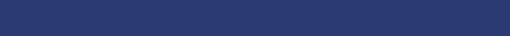}
}\par
\vspace{1em}
\sffamily
\begin{tabular}{m{4.5cm} p{13.5cm} }

  & \noindent\LARGE{
  \textbf{First-passage Fingerprints of Water Diffusion\newline   near Glutamine Surfaces}} \\
\vspace{0.3cm} & \vspace{0.3cm} \\

 & \noindent\large{
    Roman Belousov~$^1$,
    Muhammad Nawaz Qaisrani~$^{1,2}$,
    Ali Hassanali~$^{1\ast}$,
    and \'Edgar Rold\'an~$^{1\ast}$} \\\\ 

& \noindent\normalsize{
    The extent to which biological interfaces affect the  dynamics of water plays a key role in the exchange of matter and chemical interactions that are essential for life. The   density and the mobility of water molecules depend on their proximity to biological
    interfaces and can play an important role in processes such as protein folding
    and aggregation. In this work, we study the dynamics of water near glutamine
    surfaces---a system of interest in studies of neurodegenerative diseases.
     Combining molecular-dynamics simulations and stochastic modelling, we study how
    the mean first-passage time and related statistics of water molecules escaping
    subnanometer-sized regions vary from the interface to the bulk. Our analysis
    reveals a dynamical complexity that reflects underlying chemical and geometrical
    properties of the glutamine surfaces. From the first-passage time statistics of water molecules,
    we infer their space-dependent diffusion coefficient in directions normal to the surfaces.
      Interestingly, our results suggest that the mobility of water varies
    over a longer length scale than the chemical potential associated with the water-protein
    interactions. The synergy of molecular dynamics and first-passage techniques opens
    the possibility for extracting space-dependent diffusion coefficients in more
    complex, inhomogeneous environments that are commonplace in living matter.\looseness-1
  }
\end{tabular}

 \end{@twocolumnfalse} \vspace{0.6cm}

  ]

\renewcommand*\rmdefault{bch}\normalfont\upshape
\rmfamily
\section*{}
\vspace{-1cm}


\footnotetext{\textit{$^{1}$~ICTP - The Abdus Salam International Centre for Theoretical Physics,\\ Strada Costiera 11, 34151, Trieste, Italy.}}
\footnotetext{\textit{$^{2}$~SISSA - International School for Advanced Studies, \\ Via Bonomea 265, 34136 Trieste, Italy.}}
\footnotetext{\textit{$^{\star}$~Corresponding authors: AH (ahassana@ictp.it), and ER (edgar@ictp.it)}. \\ Requests for materials should be addressed to RB (belousov.roman@gmail.com) and MNQ (mqaisran@ictp.it).}





\section{\label{sec:intro} Introduction}
Proteins and DNA molecules execute their functions mostly in aqueous environments
and therefore their interaction with water 
has become a topic of intense
study.~\cite{Fenimore2002,Yang2007,VanHijkoop2007,Best2010,Best2013,Hinczewski2010,VonHansen2010,Wongekkabut2016,Sharma2018}
Numerous experimental and theoretical studies have shown that  the dynamics and thermodynamics
of water are altered when in contact with biomolecules.~\cite{Knapp1993,Muegge1995,Yang2007,VanHijkoop2007,Hinczewski2010,VonHansen2010,Sedlmeier2011,Best2010,Best2013,Wongekkabut2016,
Sharma2018,amann2016x,pearson1979molecular,finer1992solvent,svergun1998protein,merzel2002first,otting1991protein,liepinsh1992nmr,
nucci2011mapping,nucci2011site,brotzakis2016dynamics,russo2004hydration,wood2007coupling,schiro2015translational,russo2005molecular,
bizzarri2002molecular,pizzitutti2007protein,makarov2000residence,marchi2002water,henchman2002structural,li2007hydration,
fogarty2014water,luise2000molecular,rossky1979solvation,heyden2013spatial,garcia2000water,laage2009water,bellissent2016water,rani2015diffusion,
pronk2014dynamic,dellerue2000relaxational,laage2007reorientional}
Typically, the diffusive dynamics of water molecules near a protein surface slows down
by a factor of four to seven relative to the bulk, and also becomes anisotropic.~\cite{Knapp1993,Sedlmeier2011,laagehynes2017} 
The extent of these dynamical perturbations is thought to play an important role
in biological processes.~\cite{PhilipBall2008,Levy2006,laagehynes2017}

A class of biological systems that has recently caught our attention, and forms
the subject of this work, are proteins that have been implicated in numerous neuro-degenerative
diseases---namely, glutamine aggregates.~\cite{Chen2002} Like the amyloids, glutamine
aggregates are stabilized by dense networks of hydrogen bonds and hydrophobic
interactions.~\cite{eisenberg2005,Fitzpatrick5468,roland2016} Besides their role
in biological processes, these systems absorb low-energy photons in the ultraviolet
range of wavelengths and, thus, may have promising applications in bio-nanophotonics.
~\cite{gazitsilk,hassanali1,hassanali2,hassanali3,hassanali4}
Because most of the protein aggregates are formed in aqueous solutions, understanding
the solvent's role in nucleation processes is paramount. Recent experiments point
to the existence of \emph{water pools}, whose properties depend on their proximity to the
protein fibrils.~\cite{wang2017} As the origins of this dependency are still poorly understood,
a quantitative description of the mobility of water close to protein surfaces
remains a challenging task.~\cite{Knapp1993,Liu2004,Hummer2005,Sedlmeier2011,OlivaresRivas2013}

In this work we study the diffusive dynamics of water in contact with
surfaces of glutamine amino acid crystals\cite{hassanali3}. Motivated
by our recent work, which showed a surface-sensitive decrease of the
water mobility near the liquid-crystal interface,~\cite{Qaisrani2019}
here we quantify both the magnitude and the length scales over which
the solvent's diffusivity is altered. Specifically, we perform
molecular-dynamics simulations to extract first-passage time
statistics of water molecules to escape nano-sized regions  near three
different  glutamine crystal structures. Combining measurements of
first-passage times and stochastic modelling, we develop a method to
infer the   space-dependent transverse (i.e. in the direction normal to
the surface) diffusion coefficient of water as a function of the
distance to the interface.

Theory of first-passage times~\cite{redner2001guide,metzler2014first} for stochastic processes provides
a refreshing perspective that has been successful to describe key phenomena in statistical physics~\cite{masoliver1987bistability,hanggi1990reaction,sokolov2003cyclization,condamin2007first,koren2007leapover,mattos2012first,bray2013persistence,pal2017first,hartich2019interlacing},  soft-matter biophysics~\cite{szabo1980first,galburt2007backtracking,roldan2016stochastic,yang2017hydrodynamic,gladrow2019experimental},
astrophysics,~\cite{chandrasekhar1943dynamical,majumdar2005brownian,wergen2011record} finance,~\cite{metzler2014first} and low-temperature electronics.~\cite{singh2019universal}
Simply put, a first-passage time is defined as the  time elapsed until a stochastic
process first reaches a target state, e.g. the first time when a Brownian particle reaches a spatial region.   Examples include: (i) the first-passage
time for one-dimensional (1D) Brownian motion to first cross a threshold located at $L>0$; (ii)  the first-passage time for 1D Brownian motion to first escape through any of the two ends of the interval $[-L,L]$; (iii) the first-passage time for three-dimensional (3D) Brownian motion to escape a cubic cage $[-L,L]\times[-L,L]\times[-L,L]$.

The mobility of water molecules near interfaces, both biological and inorganic, and its relation to first-passage times,
under various thermodynamic conditions has been a subject of recent works.~\cite{berezhkovskii2002single,van2007water,sedlmeier2011water,calero2011first,OlivaresRivas2013,sharma2017hydration} Several approaches, that assume an effective stochastic model of diffusion,
have been introduced in order to estimate the inhomogeneous diffusion coefficient from molecular-dynamics simulations.~\cite{Knapp1993,Liu2004,Hummer2005,Sedlmeier2011,OlivaresRivas2013} In particular, it was shown that conditional mean square displacements~\cite{Hummer2005} and first-passage times subject to various boundary conditions~\cite{sedlmeier2011water,OlivaresRivas2013} provide means to infer the space-dependent diffusion coefficient for  effective models described by Smoluchowski diffusion equations.  However, it remains yet an open problem to analyse the range of validity of Langevin models, to explain water diffusion near proteins, and to
develop robust comparison and inference methods for the space-dependent diffusivities
in soft matter at the nanoscale.

\begin{figure}
\centering
\includegraphics[width=0.65\columnwidth]{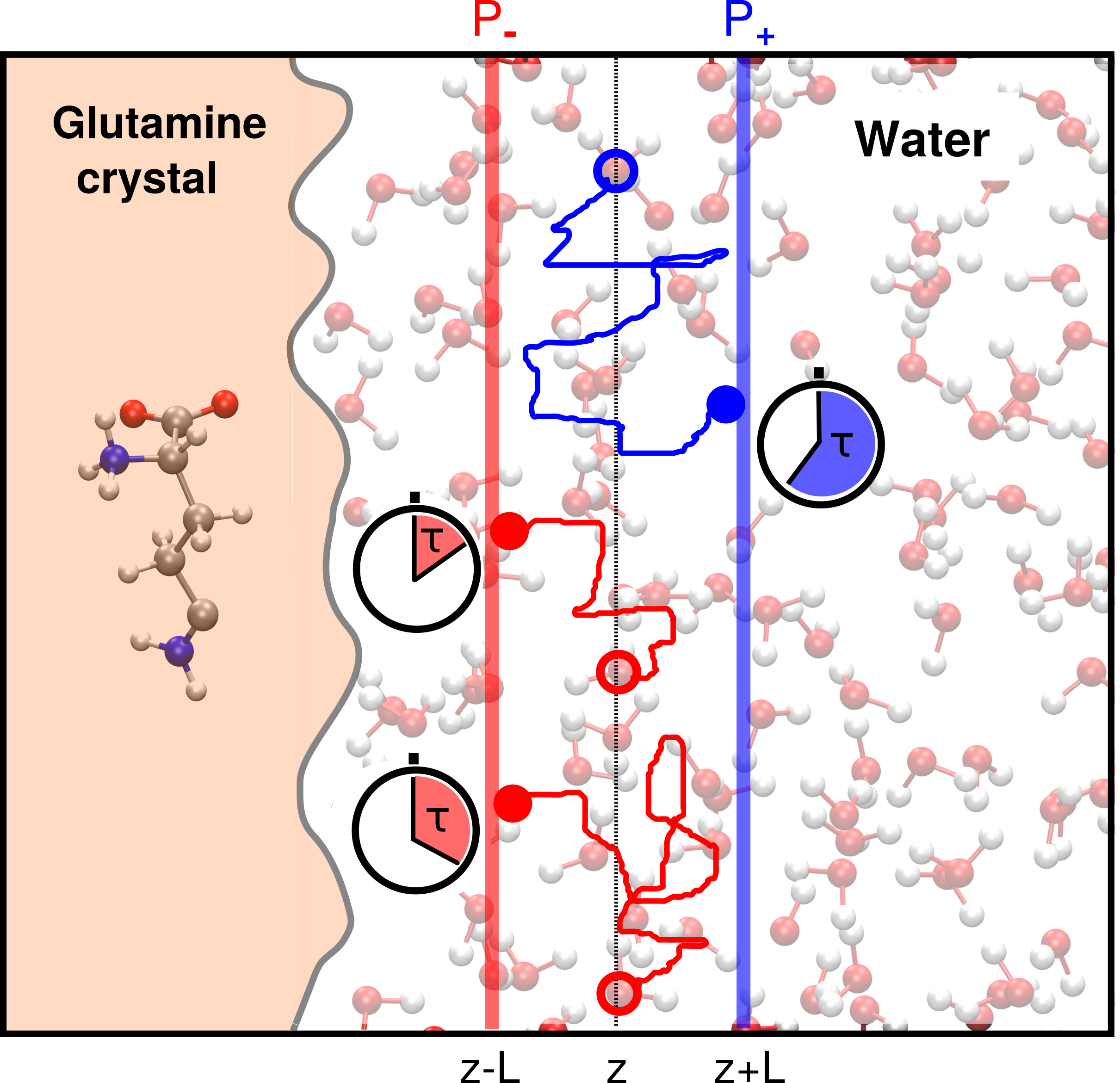}
\caption{Sketch of nanoscale measurements of the water molecules' first-passage times
in one dimension. Near a glutamine crystal (orange shaded area) water oxygen atoms
(open circles) are grouped by their initial positions $z\pm\delta{z}/2$ into slices
of width $\delta{z}>0$ along the $Z$ axis, which is perpendicular to the interface
plane. Molecules in each group diffuse until they reach boundaries of the corridor
$[z -L,z+L]$ at different locations (filled circles) and at a different
first-passage time $\tau$ (clocks). We measure the times $\tau(z)$ and the probabilities
$\mathsf{P}_+(z)$ (blue) and $\mathsf{P}_-(z)$ (red) that the oxygen atoms reach
the absorbing boundary at $z+L$ (blue) or at $z-L$ (red), respectively.}
\label{fig:method}
\end{figure}

Herein we expand the scope of the first-passage time techniques,  to
study water diffusivity in contact with crystalline glutamine. We focus on first-passage
events of water molecules which, initially located in a small spatial window, escape
a subnanometer-sized region. A paradigmatic example is the 1D first-passage
time $\tau(z)$ for a water molecule initially located within a thin shell $[z-\delta z/2,z+\delta z/2]$
of width $\delta z>0$ to cross any of two thresholds located at positions $z-L$ and
$z+L$, with $L>\delta z$ defined as a half width of the first-passage corridor (see Fig.~\ref{fig:method} for an illustration).
The statistics of $\tau(z)$ characterises the kinetics of the ensemble of water molecules
at position $z$ near a surface of the glutamine crystal, which we use to probe the
glutamine-water interaction dynamics.
In particular, we focus on the spatial
dependency of the mean time $\langle\tau(z)\rangle$ and the passage probabilities
to first cross the positive $\mathsf{P}_{+}$ or the negative $\mathsf{P}_{-}$ threshold.
To gain further insights into the three-dimensional
(3D) interactions between the liquid and the protein surface, we extend our approach
by considering first-passage times of water molecules escaping from a three-dimensional
cubic cage.

Our results show that  suitable first-passage statistics reveals "fingerprints" of the underlying
diffusive dynamics and the interactions of  water molecules with glutamine crystals. Interestingly,
parallel to the interface mean first-passage times  exhibit a periodic
pattern that reflects the underlying chemical and geometrical roughness of the protein
surface. Furthermore, the potential arising due to the liquid-crystal interactions
induces anisotropies and asymmetries of the first-passage probabilities, which expose
preferred directions of the motion of water. Using Langevin dynamics simulations
we also investigate the accuracy of the one-dimensional Smoluchowski equation to account for
the inhomogeneous diffusion of water in the direction normal to the three surfaces
of the glutamine crystal. We demonstrate that this model reproduces statistics of
the first-passage events in the water liquid phase above the Gibbs dividing interface.
We use these results to develop an inference method for the space-dependent diffusion
coefficient of water in the direction normal to the surface by fitting
molecular-dynamics mean first-passage times to analytical results derived for the stochastic diffusion model.


The rest of the paper is organised as follows: section~\ref{sec:2} begins analysing
 first-passage time fluctuations in bulk water at atomistic scales.  Section~\ref{sec:3} discusses first-passage statistics of water molecules escaping 3D regions close to
the glutamine crystal---a method to probe fine-scale details of surface heterogeneity.
Section~\ref{sec:4} describes a method to infer space-dependent diffusion coefficient of water near glutamine surfaces, which is applied to the dynamics in the directions perpendicular
to three representative protein structures. In Section~\ref{sec:5} we analyse the
range of validity of state-dependent Markovian diffusion models used in this paper
to describe the first-passage time fluctuations of water molecules.
We close the paper with a discussion and conclusions in Section~\ref{sec:6}.

\section{\label{sec:2} Benchmark: bulk water}

As a benchmark for our approach we first analyse the diffusive dynamics of
bulk liquid water. For this purpose, we  use the open-source package GROMACS~\cite{abraham2015gromacs} to run molecular-dynamics simulations of TIP4P/EW, an empirical potential for water, at density 1 \si{g/cm^3} and at 300 \si{K} (see Appendix~\ref{sec:MD} for computational details). From the molecular dynamics simulations of bulk water, we extract trajectories  containing snapshots of the positions of all the water molecules as a function of time, and then analyse these trajectories as described below.

For the trajectories extracted from molecular dynamics, we evaluate three  1D first-passage times $\tau_X(x)$, $\tau_Y(y)$, and $\tau_Z(z)$ when the oxygen atom of each water molecule travels a distance $L$ from its initial position $q=x,y,\text{ or }z$ along each Cartesian axis $X$, $Y$, and $Z$,  respectively.
Mathematically, the three first-passage times are defined as follows:  $\tau_X(x) = \inf \{ t>0 \;;\; |X(t)-x| \geq L\}$,  $\tau_Y(y) = \inf \{ t>0 \;;\; |Y(t)-y| \geq L\}$ and $\tau_Z(z) = \inf \{ t>0 \;;\; |Z(t)-z)| \geq L\}$, with $x=X(0)$, $y=Y(0)$ and $z=Z(0)$ the initial values of the $(X,Y,Z)$ coordinates of the trajectories (see Fig. 1 for an illustration of the
first-passage time problem along $Z$).
We then combine measurements of $\tau_X(x)$, $\tau_Y(y)$ and $\tau_Z(z)$ in a large
sample of a  single random variable $\tau$ and compute its average $\langle\tau\rangle$,
i.e. the global 1D {\em mean first-passage time}, for different values of $L$.
The mean first-passage time depends on $L$ quadratically $\langle\tau \rangle\sim L^2$
for values of $L$ ranging from $\SI{1}{\angstrom}$\, to $\SI{1}{nm}$. The same scaling law is found
for the mean first-passage time of 1D Brownian motion. In a 1D Brownian motion the
probability density $P(q,t) \equiv P(q,t|q_0,0)$ for a coordinate $q$ of a molecule moving
 with diffusion coefficient $D_\mathrm{bulk}$ evolves according to a Fokker-Planck
equation $\partial_t P(q,t) = D_\mathrm{bulk} \partial^2_q P(q,t)$. In this case
a probability distribution of the mean first-passage time (i.e. time elapsed) for a molecule to escape the interval
$[q_0-L,q_0+L]$ is known and its mean value reads~\cite{redner2001guide}
\begin{equation}
\langle\tau\rangle = \frac{L^2}{2D_\mathrm{bulk}}.
\label{eq:mfptbulk}
\end{equation}
Fitting our measurements to Eq.~\eqref{eq:mfptbulk}, we estimate the diffusion coefficient of bulk water
$D_\mathrm{bulk}=\SI{3.81\pm0.02}{nm^2/ns}$, which is consistent with the value determined
previously  from measurements of mean squared displacements.~\cite{Qaisrani2019}

\begin{figure}
\centering\hspace{-0.5cm}
\includegraphics[width=0.9\columnwidth]{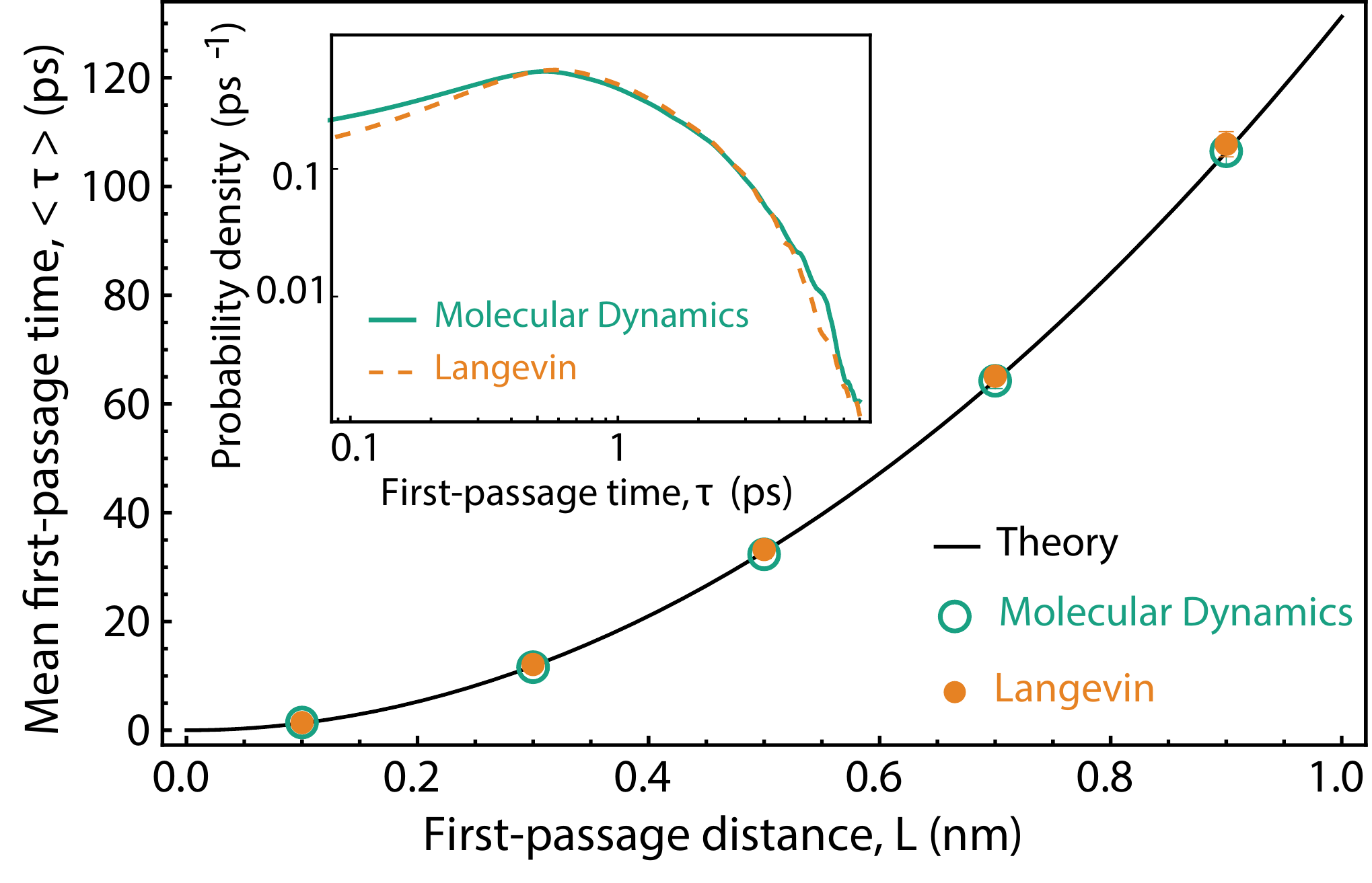}
\caption{Mean first-passage time $\avg{\tau}$ of oxygen atoms in molecules of
bulk water obtained from molecular dynamics (green open circles) and from Langevin
simulations (orange filled circles) as a function of the passage distance $L$. Statistical
averages were taken over fixed-size samples of 12411 events. Error bars of the data
are smaller than the symbol sizes. Fitting the data points of the molecular-dynamics
simulation to a theoretical expression Eq.~\eqref{eq:mfptbulk} (black curve)
yields an estimate of the bulk water self-diffusion coefficient $D_\mathrm{bulk}=\SI{3.81\pm0.02}{nm^2/ns}$. Inset: Probability density of the first-passage time distributions in the molecular
dynamics (green solid line) and in the Langevin dynamics (orange dashed line) simulations
for $L=\SI{1}{\angstrom}$.
\label{fig:bulk}}
\end{figure}


We further examine how accurately the overdamped Langevin
dynamics describes the first-passage time distributions in bulk water using stochastic simulations. To this end
we integrated numerically the stochastic differential equation $\text{d} q/\text{d}t=\sqrt{2D_\mathrm{bulk}}\xi$,
in which $\xi(t)$ is Gaussian white noise with the zero mean $\langle \xi(t)\rangle=0$ and the autocorrelation function $\langle\xi(t)\xi(t')\rangle=\delta(t-t')$. In our Langevin dynamics simulations, we set $D_\mathrm{bulk}=\SI{3.81}{nm^2/ns}$ and use a simulation time step $\Delta t=\SI{1}{fs}$ to harvest a sample of statistically independent first-passage events of the same size
and for the same values of $L$ as in our molecular-dynamics simulations.  The mean first-passage times $\avg{\tau}$ that we obtain from the Langevin dynamics simulations are in excellent agreement with our  results from molecular dynamics, for all the range of $L$ that we explore  ranging  from $0.1$ to $1\,\text{nm}$ (see Fig.~\ref{fig:bulk}). Moreover, even for a distance $L$ as small as \SI{1}{\angstrom},
the probability distributions of $\tau$ obtained from the molecular-dynamics
simulations and Langevin simulations are in excellent agreement (see Fig.~\ref{fig:bulk} inset). This result
lends further credence to the one-dimensional model of diffusion and its ability
to describe the first-passage time statistics down to atomic-sized corridors of  the order of angstroms.

\section{\label{sec:3} 3D first-passage statistics of water molecules near glutamine surfaces}
\begin{figure*}
\centering
 \includegraphics[width=0.85\textwidth]{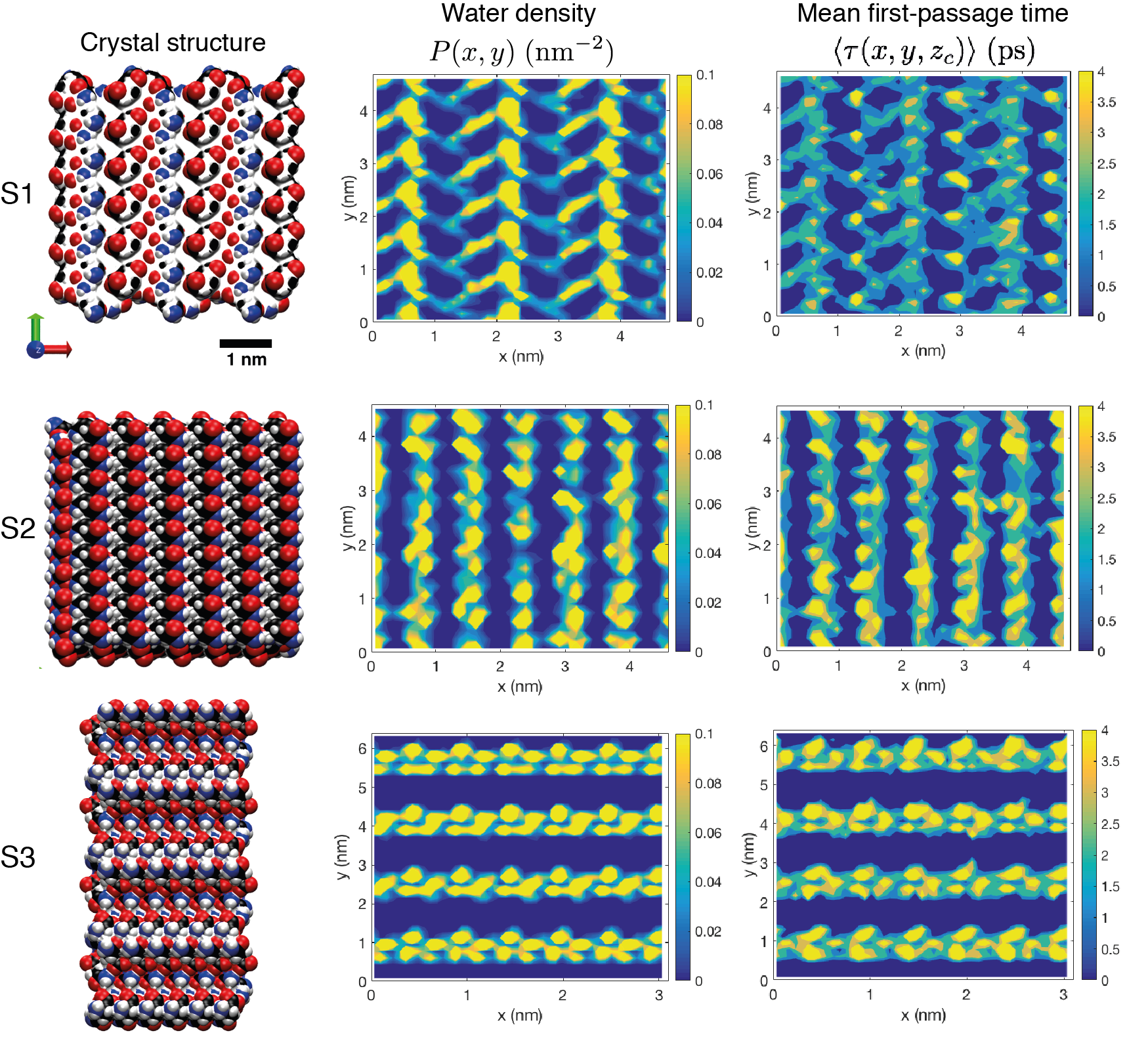}
 \caption{
     First-passage time fingerprints in water molecules diffusing near glutamine crystals.
     Three surfaces of the glutamine crystal slab: crystal structures of the respective
     crystallographic planes perpendicular to the $Z$ axis (left column), density
     histograms $P(x,y)$ of water molecules (centre column) and their mean first-passage
     times $\avg{\tau(x,y)}$ to escape  a cube of edge length $L=\SI{1}{\angstrom}$
     (right column). In the crystal structures, we show hydrogen atoms in white, oxygen
     in red, nitrogen in blue, and carbon in black. The 2D plots represent 3D data
     that were collected in a shell of thickness 1 \si{\angstrom} and projected onto
     the median $XY$ plane centred at the Gibbs dividing interface.
 }
 \label{fig:xy}
\end{figure*}

Having studied the first-passage time dynamics in the bulk,
we move on to discussing the dynamics of water near glutamine. In particular, we analyse in this section  first-passage times of water molecules, which escape three-dimensional regions
near surfaces of crystalline glutamine.
 To this aim, we performed  three equilibrium molecular-dynamics simulations where a slab of crystalline glutamine  (S1, S2, and S3, see  left column of Fig.~\ref{fig:xy}) was exposed to approximately 7000 water molecules. As a reference, we orient the $Z$ axis \emph{perpendicular} to the crystallographic planes of these surfaces (see Fig.~\ref{fig:xy} left column).
The molecular-dynamics simulations of the water-glutamine interfaces, which were previously
reported by some of us, revealed structural and orientational correlations of the liquid within a shell
of $\SI{1}{nm}$ from the crystal surface.~\cite{Qaisrani2019} In order to characterise
both the structure and the dynamics of this shell, we determined the Gibbs dividing
interface---a plane parallel to the surface of the solid at a position $z_\mathrm{GDI}$,
at which the water density is half that of the bulk.
For more details on the simulation
protocol, see Ref.~\cite{Qaisrani2019} and Appendix~\ref{sec:MD}.

Using the molecular-dynamics trajectories, we analyse structural and dynamical features
of a  water shell of 1-\si{\angstrom}  thickness projected onto the Gibbs dividing
interface (GDI), namely: (i) two-dimensional densities of water molecules $P(x,y)$ (Fig.~\ref{fig:xy}, centre column); and (ii)  mean first-passage times $\avg{\tau(x,y)}$ (Fig.~\ref{fig:xy}, right column)
 for the water molecules to escape a cube of edge length $L=\SI{1}{\angstrom}$
centred at $(x,y,z_\mathrm{GDI})$.
Our simulations reveal a clear periodic pattern---at angstrom scales---of the
local density of water and their mean first-passage times (Fig.~\ref{fig:xy}). This is consistent
with the presence of different chemical groups on the glutamine surface, which make
some regions of the surface more accessible than others. In particular,
hydrophilic regions, such as the N- and C-termini and the glutamine side chains,
lead to enhanced local density and increased first-passage times of nearby water molecules. This is
highlighted across the different plots by a the green coloured box allowing the reader
to see the relationship between the chemical groups, the water density and the first
passage times.

\begin{figure}
 \centering
 \includegraphics[width=8cm]{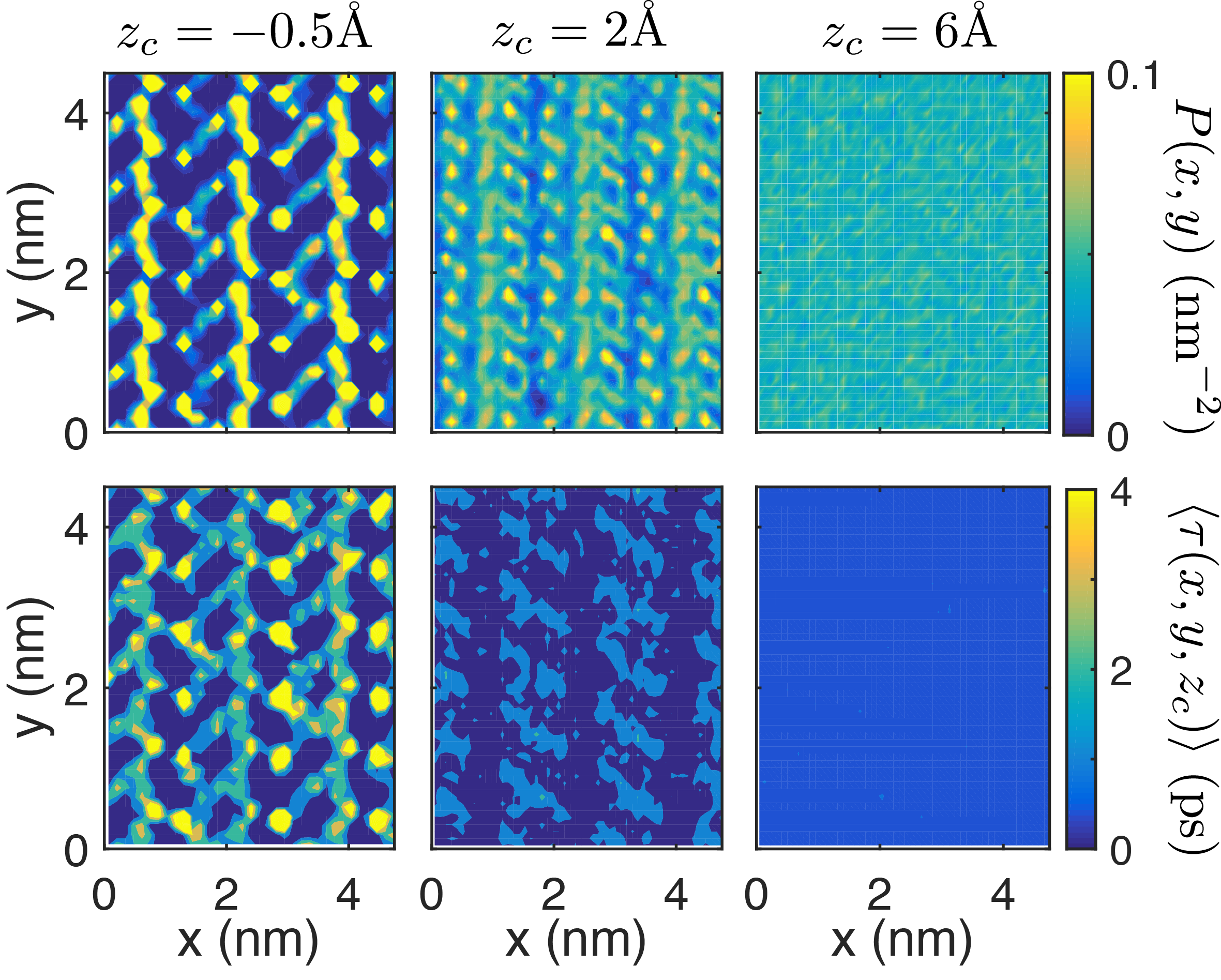}
 \caption{
   Local density $P(x,y)$ (top row) and mean first-passage times $\langle\tau(x,y,z_c)\rangle$ of water molecules to escape
   a cube of edge length \SI{1}{\angstrom} centred at $(x,y,z_c)$ (bottom row)  for various distances
   $z_c$ from the glutamine surface S1: \SI{0.5}{\angstrom} below  (left column), \SI{2}{\angstrom} above (middle column), and \SI{6}{\angstrom} above
   (right column)  the Gibbs dividing interface.
 }
 \label{fig:new2}
\end{figure}

The coupling between the water density and first-passage times,
while interesting, is not so surprising. More intriguing is perhaps the extent to
which the fingerprints of the structure and dynamics can still be observed, as the
projecting plane of the middle and right columns in Fig.~\ref{fig:xy} moves further
away from the crystal surface. At longer distances from the crystal, on the order
of a few angstroms from the Gibbs dividing interface, the periodic pattern of $P(x,y)$
and $\avg{\tau(x,y)}$ fades gradually (Fig.~\ref{fig:new2}). In particular, beyond
$\SI{5}{\angstrom}$ we do not observe any of the patterns in the 2D densities and
the mean first-passage times observed in Fig.~\ref{fig:xy}.

\begin{figure*}
 \centering
 \includegraphics[width=19cm]{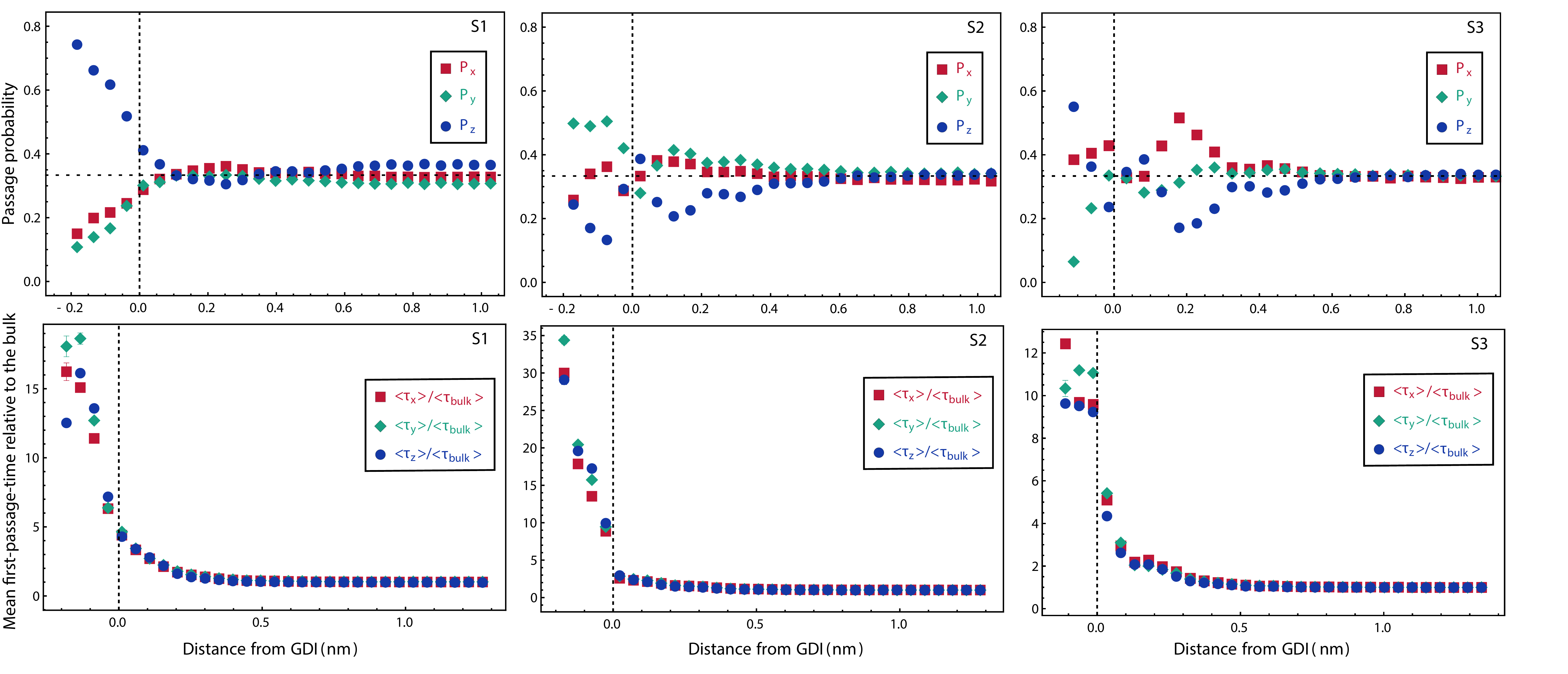}\hfill
 \caption{
Top row: passage probabilities $\mathsf{P}_X(z)$, $\mathsf{P}_Y(z)$ and $\mathsf{P}_Z(z)$
for water molecules to escape a cube of side $L=\SI{0.1}{nm}$ through the faces perpendicular
to $X$, $Y$, and $Z$ axes, respectively, as functions of the distance
from the Gibbs dividing interface with the glutamine surfaces S1 (left), S2 (middle) and S3 (right).
Bottom row: mean first-passage times $\avg{\tau_X(z)}$, $\avg{\tau_Y(z)}$, and $\avg{\tau_Z(z)}$
conditioned on the escape events through the corresponding sides of the cube and normalised
with respect to the bulk-water value $\avg{\tau_\mathrm{bulk}}=\SI{0.623\pm0.007}{ps}$. Both
the probabilities and the mean first-passage times are obtained by averaging
over all events detected in cubes, whose geometric centres are located at $(x,y,z_c)$
with $z_c$ fixed at a given distance from the Gibbs dividing interface.}
 \label{fig:new}
\end{figure*}

In order to establish a  measure on the range at which the
3D passage statistics provide a fingerprint of the interaction between water and glutamine surface,
we examined two first-passage  parameters as a function of separation
distance between the first-passage cube and the glutamine surfaces:
(i) the passage probabilities $\mathsf{P}_x$,  $\mathsf{P}_y$ and  $\mathsf{P}_z$ for water molecules to first exit a \SI{1}{\angstrom^3} volume cube through any of its faces perpendicular to the $X$, $Y$ or $Z$ axes; and (ii)  the corresponding
\emph{conditional} mean first-passage times $\langle \tau_X\rangle,\langle \tau_Y\rangle$
and $\langle \tau_Z\rangle$ associated with these events. We report in  Fig.~\ref{fig:new} the value of
these  parameters  as a function of distance from the surfaces S1, S2 and S3.

The structure of the
3D passage probabilities is quite sensitive to the distance from the protein surface:
for  the three glutamine surfaces, the passage probabilities are asymmetric below
the Gibbs dividing interface, and they all saturate to the bulk value of~1/3
at a distance of $\sim\SI{5}{\angstrom}$.
While for surface S1 all the three components ($\mathsf{P}_X$, $\mathsf{P}_Y$ and $\mathsf{P}_Z$)
are quite similar at about \SI{0.2}{nm}, in the case of S2 and especially S3, the passage
probabilities exhibit a rather pronounced symmetry breaking.
On the other hand, the conditional mean first-passage times  appear to be characterised by a different
phenomenology. Below the Gibbs dividing interface, the mean  first-passage times are
increased up to one order of magnitude with respect to their bulk value consistent with our earlier studies,
where we found an increase of the water molecules' residence times
near the surface.~\cite{Qaisrani2019} Above the interface, our simulations reveal a symmetry $\langle\tau_X\rangle = \langle\tau_Y\rangle=\langle\tau_Z\rangle$
between the conditional
mean first-passage times in different directions.
This symmetry extends also to distances below \SI{5}{\angstrom},
at which the passage probabilities are \emph{not} symmetric. It is clear
that the
first-passage statistics lose their x and y dependence as we move further
away from the surface and hence this motivated us to study the dynamics of
water along $Z$ direction. This forms the subject of the following section.

\section{\label{sec:4} 1D first-passage of water near glutamine: method to infer space-dependent diffusivity}
\begin{figure*}
 \centering
 \includegraphics[width=7cm]{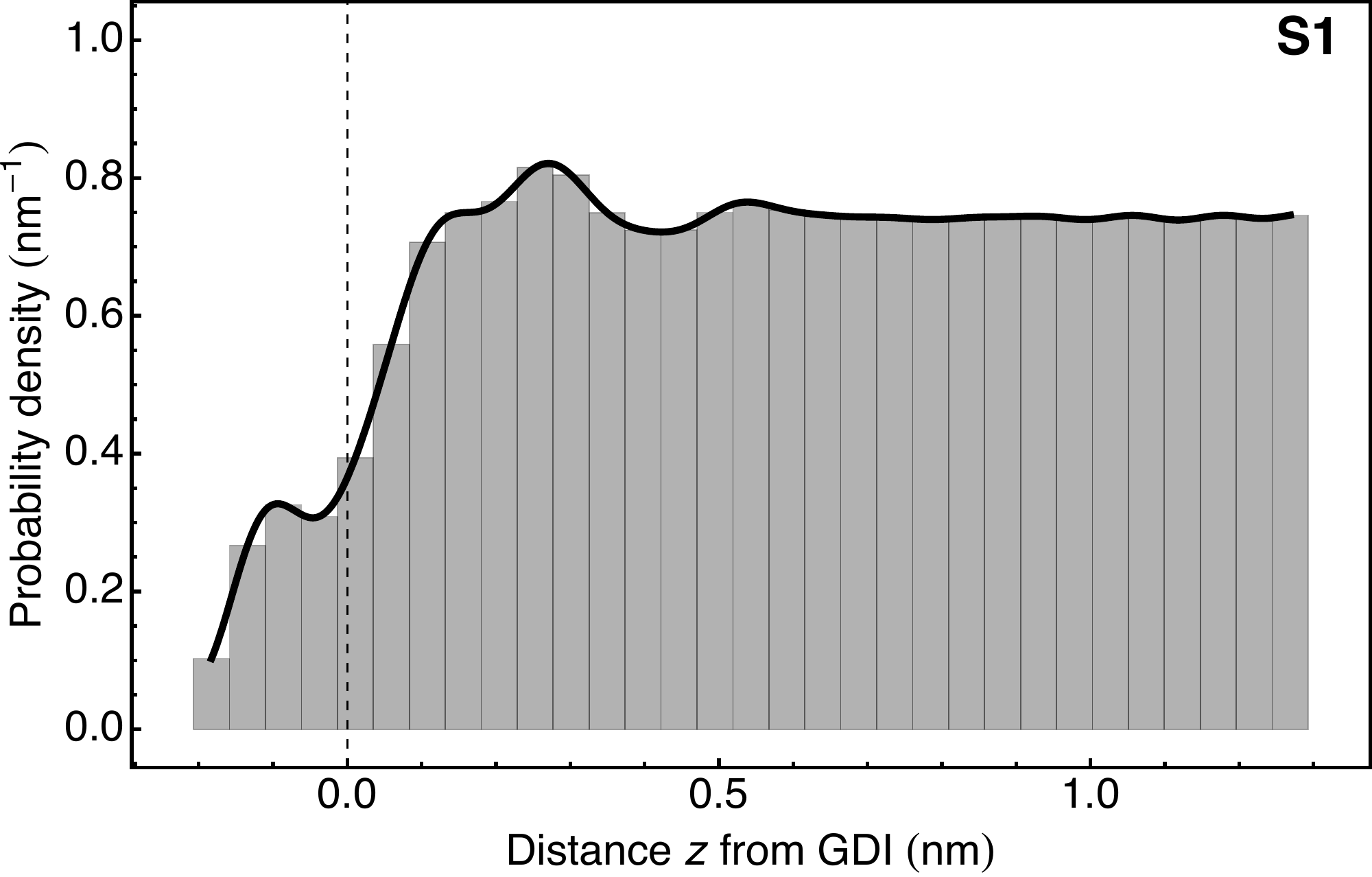} \hspace{1cm}
 \includegraphics[width=7cm]{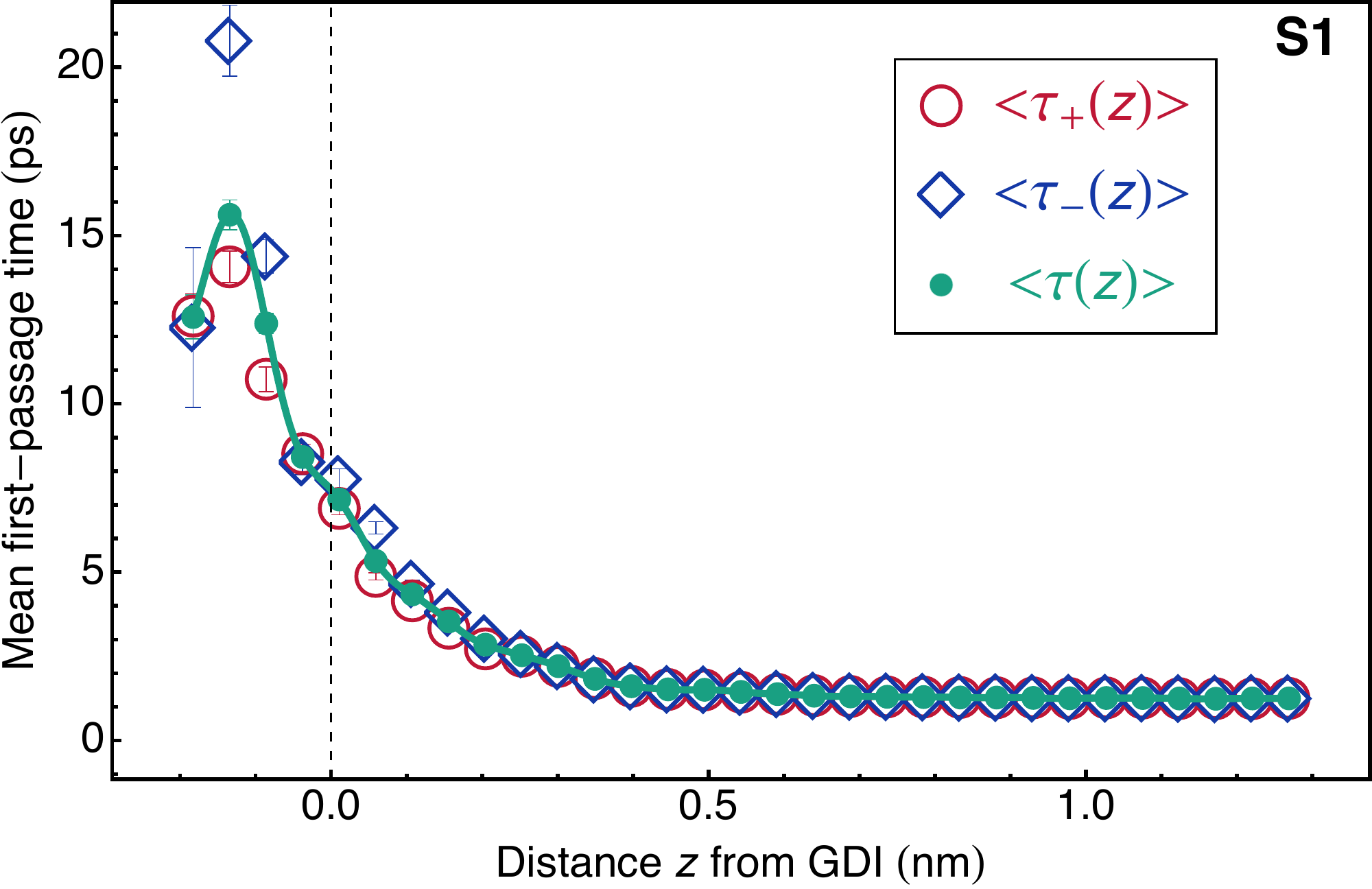}\\
  \includegraphics[width=7cm]{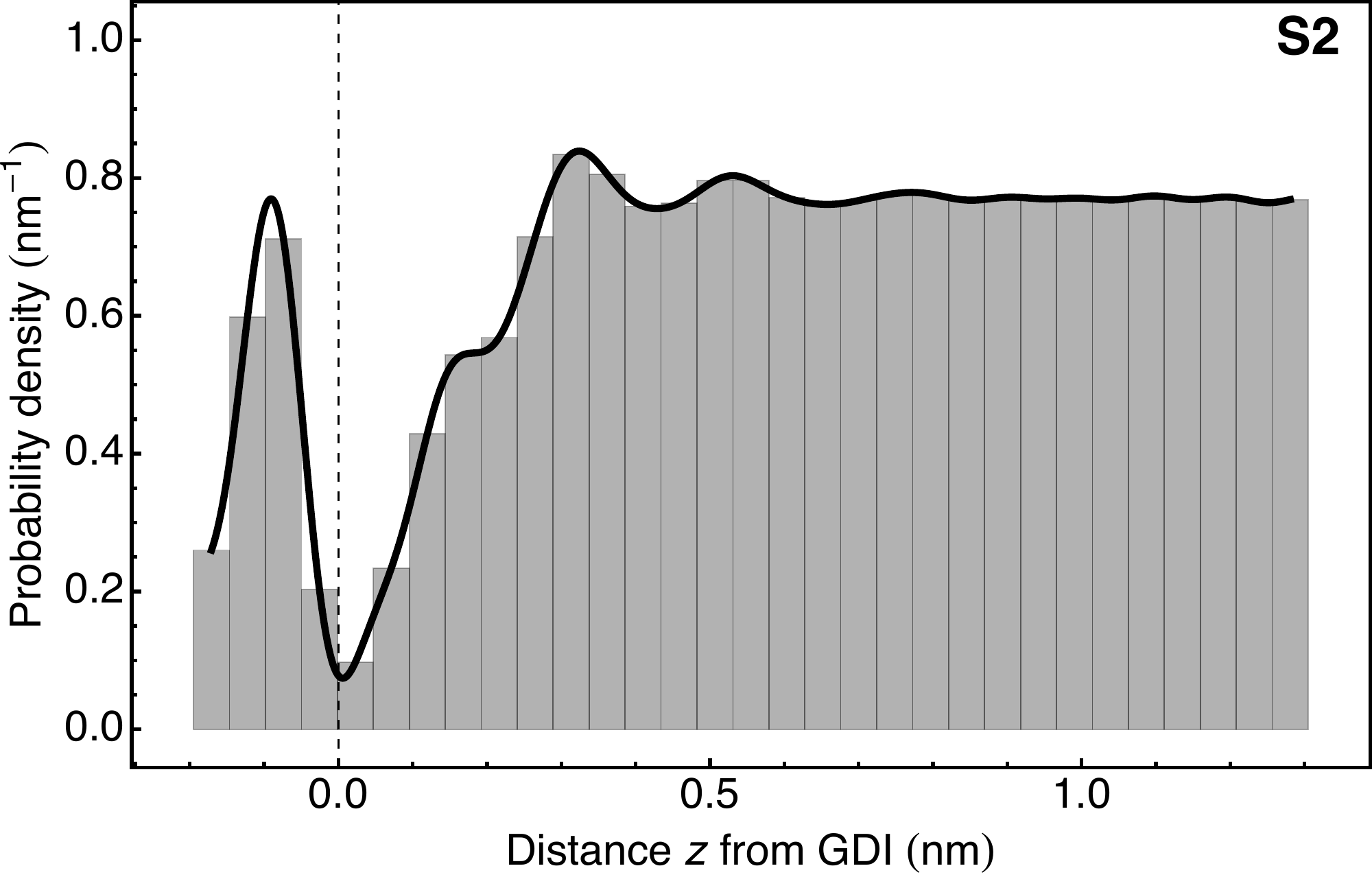} \hspace{1cm}
 \includegraphics[width=7cm]{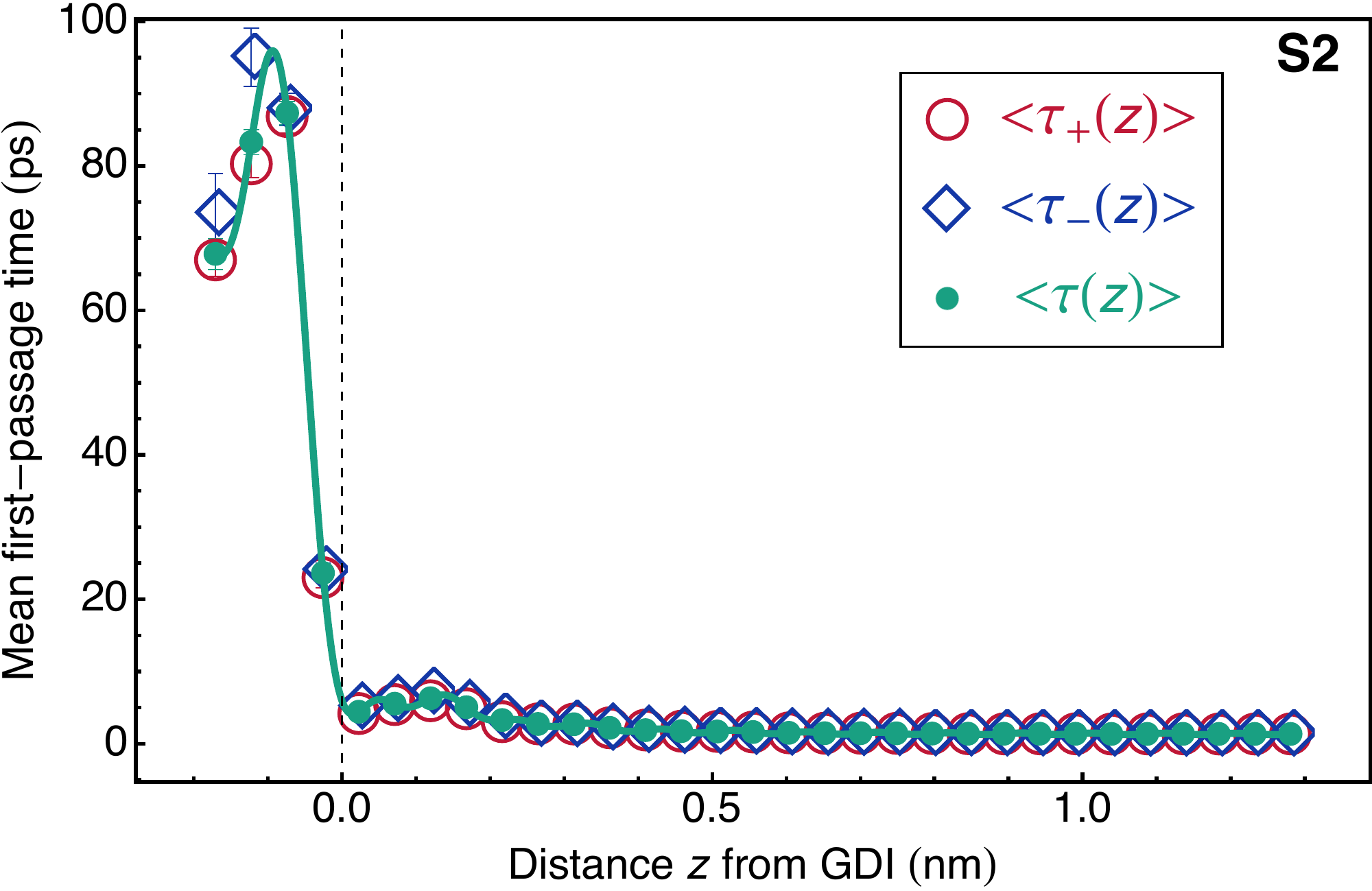}\\
  \includegraphics[width=7cm]{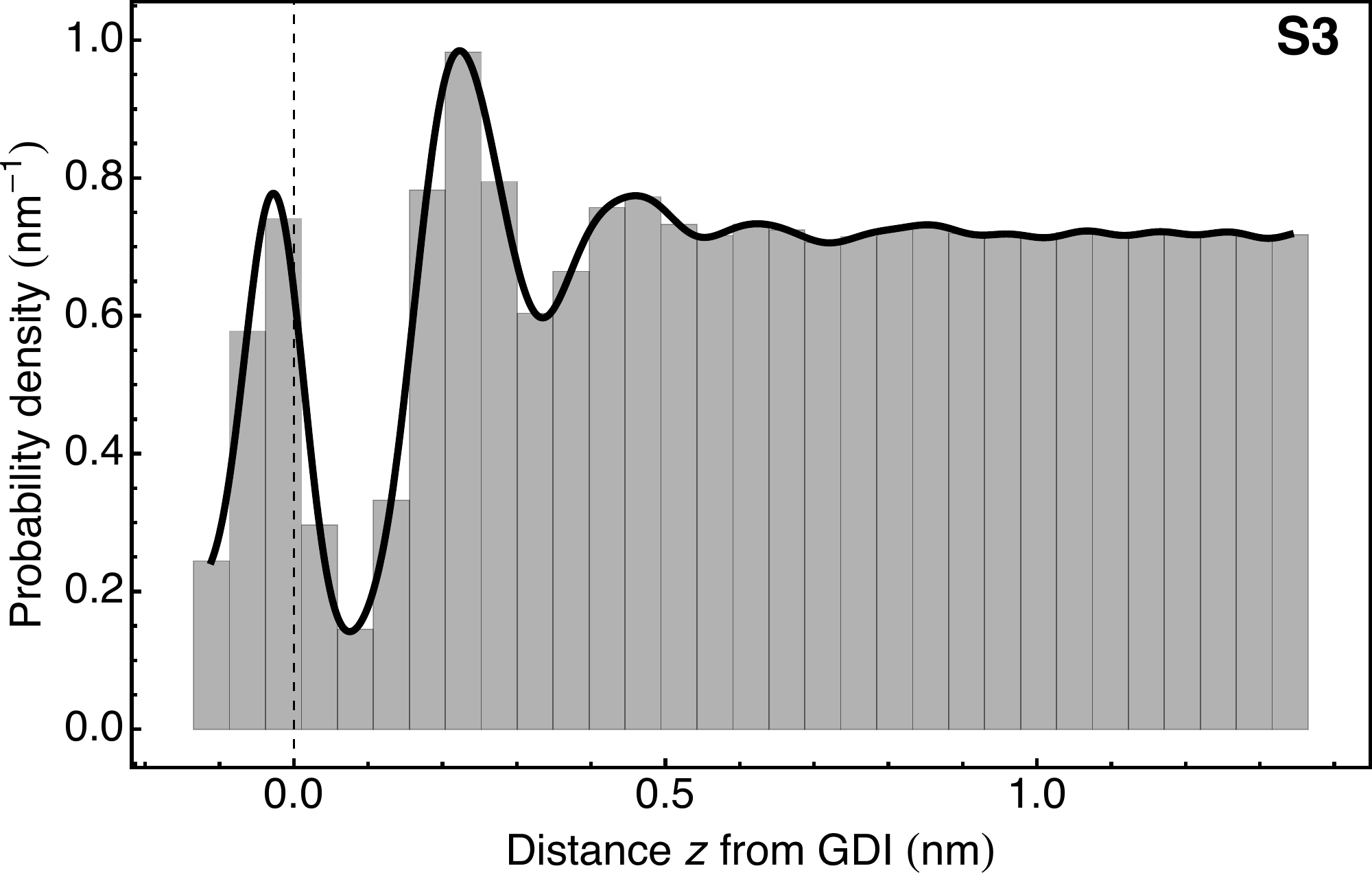} \hspace{1cm}
 \includegraphics[width=7cm]{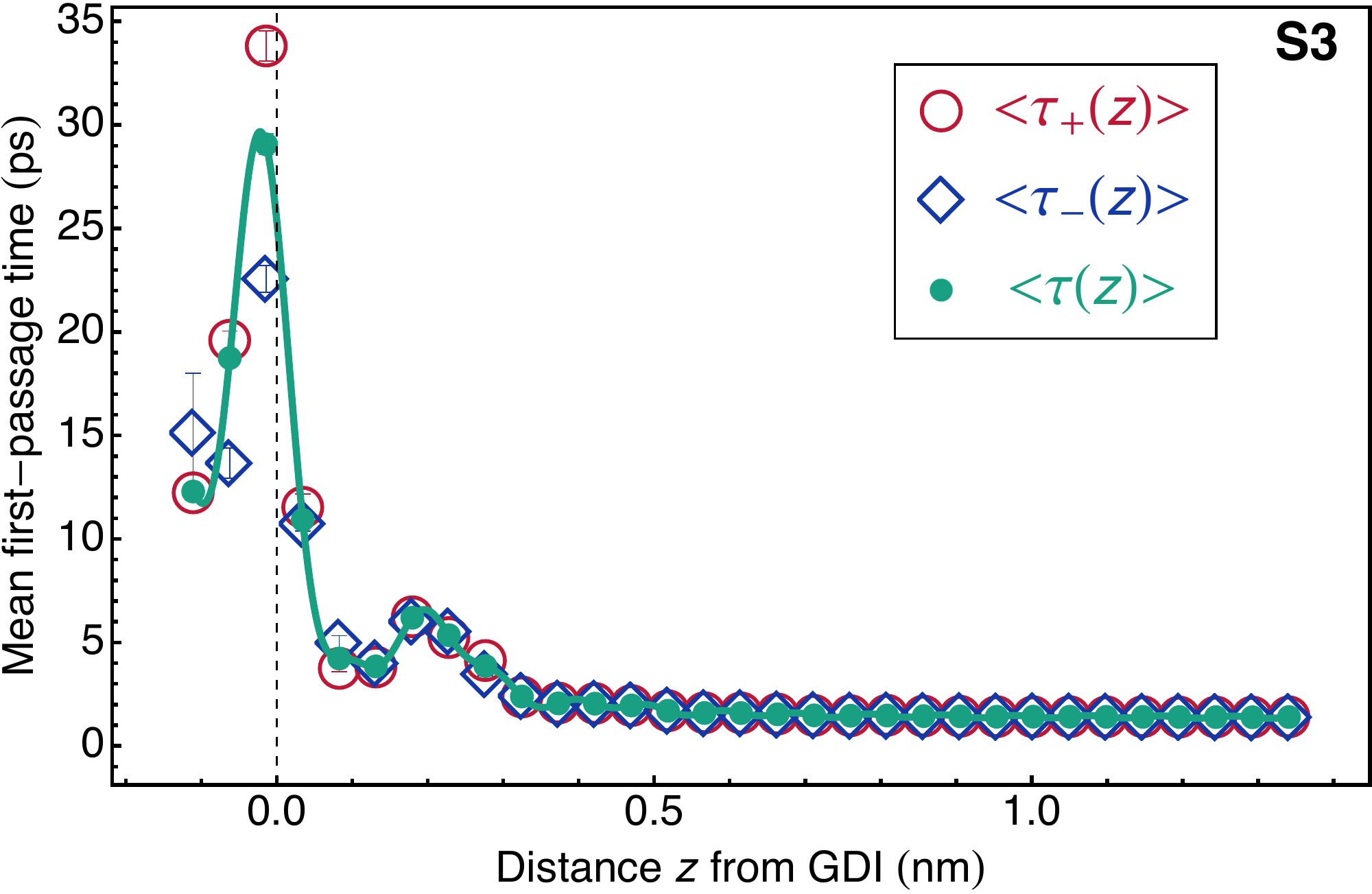}
 \caption{Density histograms (left panels) and 1D mean first-passage times  to escape the interval $[z-L,z+L]$ with  $L=\SI{1}{\angstrom}$ (right panels) for water molecules residing initially at a distance $z$ from the Gibbs dividing interface (GDI) of the
 glutamine crystals S1, S2 and S3. The vertical dashed lines mark the Gibbs dividing interface
 ($z=0$), whereas the solid lines represent trigonometric interpolations.}
 \label{fig:input}
\end{figure*}

In this section we study the dynamics of water molecules projected onto the $Z$
coordinate axis---the direction perpendicular to the surface of contact with the glutamine
crystals.  
Using our molecular-dynamics simulations, we measure the stationary spatial density of
water molecules $P(z)$ that varies with the distance $z$ from the Gibbs dividing
interface (Fig.~\ref{fig:input}, first column).~\cite{Qaisrani2019} For the three surfaces
we find that $P(z)$  increases  with
the separation distance from the interface, saturating to the bulk value
at distances greater than 5--6 \si{\angstrom} from the Gibbs dividing interfaces.
In the case of S2 and S3, the water density near these glutamine crystals
has a pronounced minimum close to the Gibbs dividing interface, and a maximum below it. This result reveals presence of hydrophilic \emph{pockets} coupled with the underlying
geometric roughness of the surfaces S2 and S3, which admit a deep penetration of
water.
For the glutamine crystal S1, we do not observe these phenomena but a smooth
increase of the water density as a function of the distance perpendicular to the surface.


To further understand the kinetics of water molecules in the direction perpendicular to the surfaces,
we measure the mean first-passage time for water molecules along the
$Z$ axis. As we will see below, the mean  first-passage time $\langle \tau(z) \rangle\equiv \langle \tau_Z(z) \rangle$
with  $L=\SI{1}{\angstrom}$ along the $Z$ axis (Fig.~\ref{fig:method})
provides insights into the transverse mobility of water molecules that reside at a distance
$z$ from the protein surface. Figure~\ref{fig:new} (bottom row) shows the values of {\em conditional} mean first-passage times
$\langle \tau_{+}(z) \rangle$ and $\langle \tau_{-}(z) \rangle$ for water molecules
to first escape the first-passage corridor $[z-L,z+L]$ through its positive or negative end, respectively.
In the wetting layer of all the three glutamine surfaces---below the Gibbs dividing
interfaces---we observe a strong anisotropy of the conditional mean first-passage
times $\avg{\tau_+(z)} \ne \avg{\tau_-(z)}$. As $z$ grows, the magnitude of these
two first-passage times decrease and eventually converge to the bulk value away from the interface,
$\avg{\tau(z)} = \avg{\tau_+(z)} = \avg{\tau_-(z)}$ for $z\gg 5$\Ang.
The decay of the mean first-passage times towards bulk value as a function of distance
from the surface, is not monotonous for the surfaces S2 and S3. In
particular, we observe a fast dynamics of water
molecules with a local minimum of the conditional mean first-passage times in a
region of low local density where the chemical potential is large.



Motivated by recent results describing fluctuations of water density at the nanoscale
with effective stochastic models,~\cite{Hummer2005,Sedlmeier2011,OlivaresRivas2013}
we investigate next whether our measurements of molecular density and mean first-passage
times can be explained by such a mathematical model. We consider a paradigmatic model of a one-dimensional inhomogeneous diffusion along the $Z$ axis with a space-dependent diffusion coefficient $D(z)$ in presence of a potential $U(z)$. In this model, the evolution of the density of the fluid $P(z,t)$, which depends on the spatial coordinate $z$ and time $t$,
obeys the Smoluchowski equation
\begin{equation}\label{eq:SE}
    \frac{\partial P(z,t)}{\partial t} = \frac{\partial}{\partial z} \left(
        \beta D(z)  P(z,t)\frac{\partial U(z)}{\partial z}
        + D(z) \frac{\partial P(z,t)}{\partial z}
    \right),
\end{equation}
where $\beta\equiv (k_{\rm B}T)^{-1}$,  with $k_\mathrm{B}$ Boltzmann's constant, $T$ the
temperature of the fluid. In Eq.~\eqref{eq:SE} we have assumed Einstein's relation for the
mobility $\mu(z)= \beta D(z)$. The right-hand side of the Smoluchowski equation represents
the negative divergence of the drift current $-\mu(z) P(z) U'(z)$ and Fick's diffusion
flow $-D(z) P'(z)$. 

For the Smoluchowski diffusion model described by Eq.~\eqref{eq:SE} quadrature formulas  for  mean first-passage times with two absorbing boundaries have been derived  (see Appendix~\ref{sec:FPTS}).~\cite{hanggi1990reaction,GoelRichterDyn2016}
These formulas are amenable to a complete analytical
treatment only for special cases  of $U(z)$ and $D(z)$. Here we consider the  limit  of a small passage  corridor width $L$ by taking a linear-order approximation  for $U(z)$: $U(z+\epsilon)\simeq U(z)+U'(z)\epsilon$ with $\epsilon \in [z-L,z+L]$. Furthermore,
as we show in Appendix~\ref{sec:LE2}, the diffusion coefficient of water near the
three glutamine surfaces S1, S2, and S3 varies much slower with $z$. As a result, it can be approximated by a constant value  over the entire first-passage corridor, i.e. $D(z+\epsilon)=D(z)$ for $\epsilon \in [z-L,z+L]$.  This approximation is consistent with previous work showing that the
effective diffusion coefficient in water-peptide systems varies slower than the relevant
interaction potential.~\cite{Hummer2005}
Taking the Smoluchowski model described by Eq.~\eqref{eq:SE} and  the aforementioned approximations   $U(z+\epsilon)\simeq U(z)+U'(z)\epsilon$ and $D(z+\epsilon)=D(z)$ for any $\epsilon \in [z-L,z+L]$, we derive the following formula  for the mean first-passage time
to escape the corridor $[z-L,z+L]$ (see Appendix~\ref{sec:FPTS})
\begin{equation}\label{eq:tau}
    \left\langle\tau(z)\right\rangle \simeq \frac{L\, \tanh\Big(
       L \beta U'(z)/2\Big)}{ D(z)\beta U'(z)}.
\end{equation}
Note that in the limit of $L\beta U'(z)$ being small, Eq.~\eqref{eq:tau} yields $\langle\tau(z)\rangle\simeq L^2/2D(z)$
and, thus, recovers the quadratic dependency of the mean first-passage time on $L$
obtained in the bulk, given by Eq.~\eqref{eq:mfptbulk}.

From our measurements of the local stationary density of water molecules $P(z)$ (Fig.~\ref{fig:input} left column) and their mean first-passage times $\left\langle\tau(z)\right\rangle$ (Fig.~\ref{fig:input} right column), we determine
the effective water-glutamine chemical potential $U(z)$ and then infer the space-dependent diffusivity $D(z)$ for the three glutamine crystal structures. The inference method
proceeds as follows. First, we obtain  the effective potential  as $\beta U(z)=-\ln P(z)$. Second, plugging this result into Eq.~\eqref{eq:tau} we obtain an estimate for the diffusion coefficient  ${D}(z)\equiv L\, \tanh( L  \beta U'(z)/2)/  U'(z)\langle{\tau}(z)\rangle$ in terms of $L$, the mean first-passage time $\langle\hat{\tau}(z)\rangle$ and the local density $P(z)$.\footnote{The spatial derivatives were  evaluated numerically,
 using a pseudospectral approach (see Appendix~\ref{sec:data}).} Im our estimates
 we used the trajectories from
molecular-dynamics simulations previously described in Sec.~\ref{sec:2} and a narrow passage corridor of
half width $L=\SI{1}{\angstrom}$.

\begin{figure}
 \centering
 \includegraphics[width=8cm]{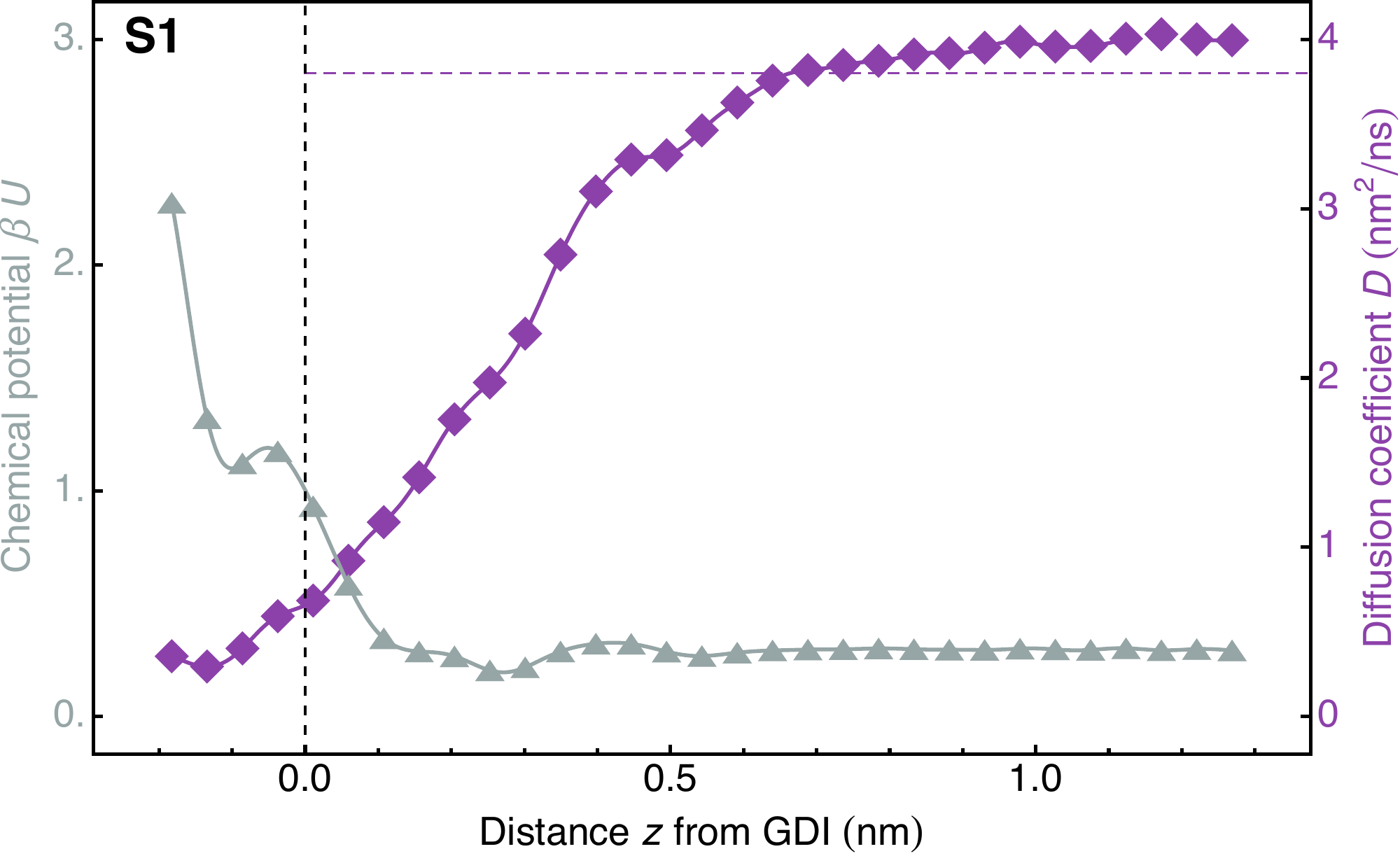}\\ \vspace{0.5cm}
  \includegraphics[width=8cm]{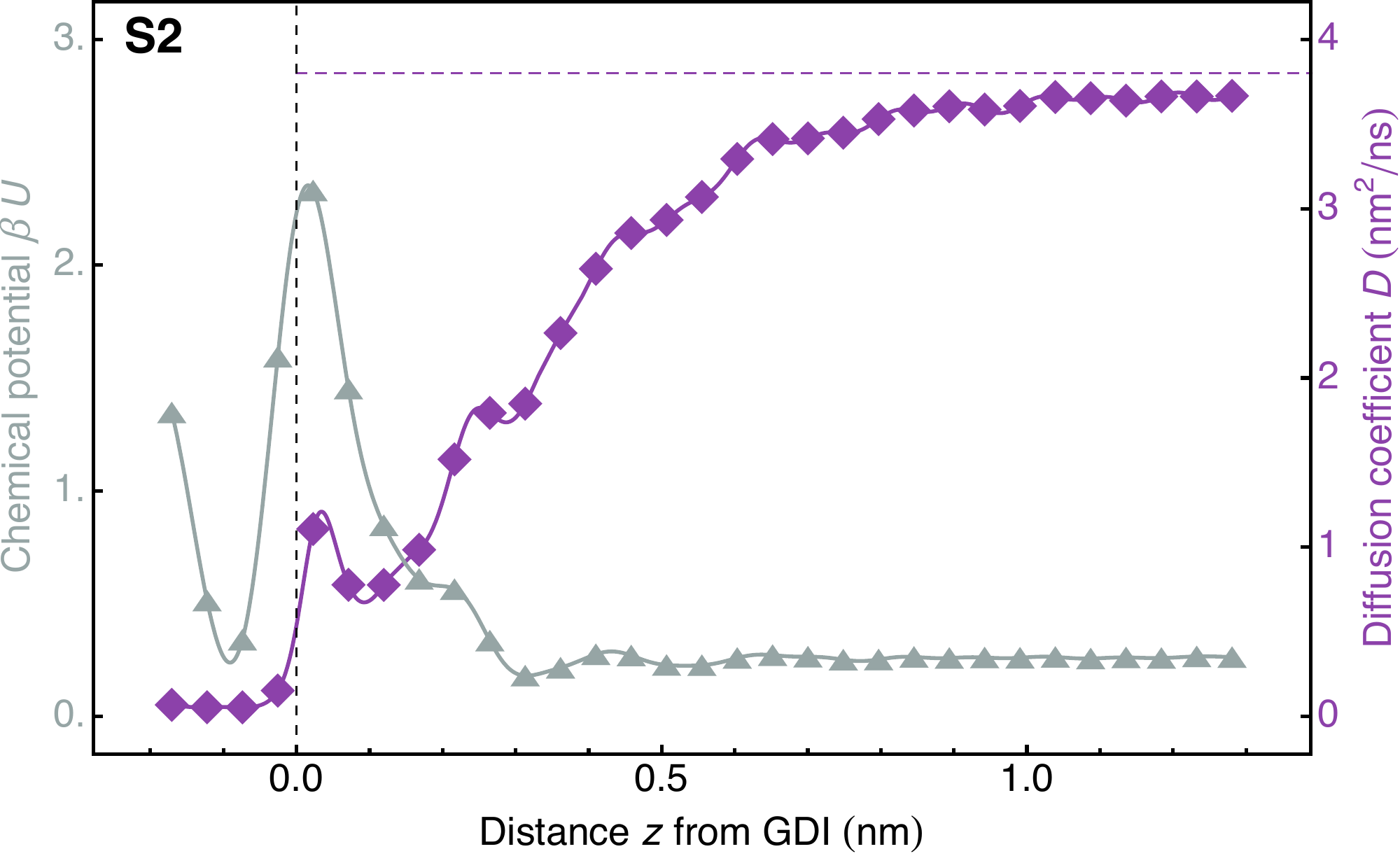}\\ \vspace{0.5cm}
 \includegraphics[width=8cm]{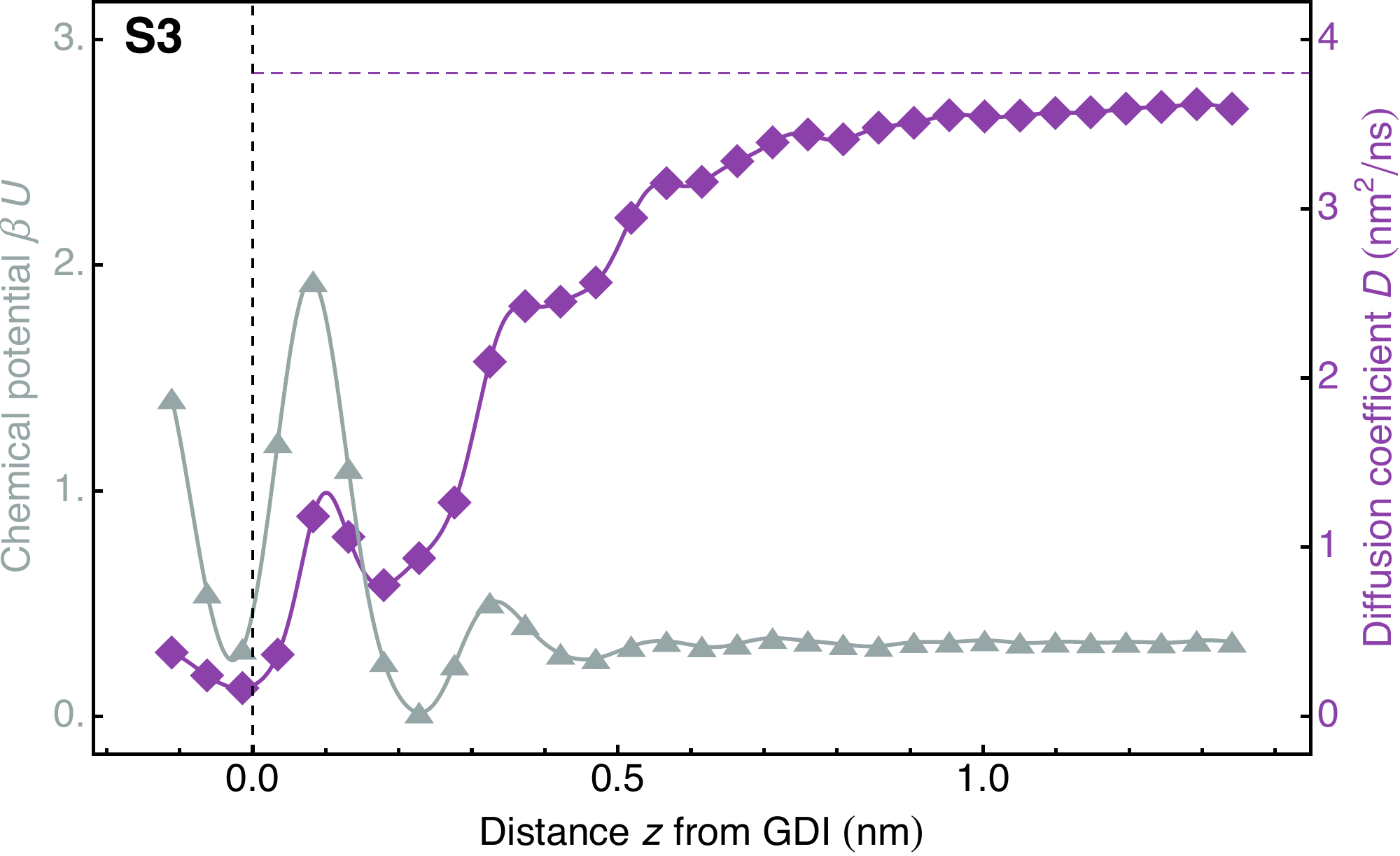}\\
 \caption{Inference of space-dependent effective diffusion coefficient and potential from first-passage and density measurements of water molecules near glutamine.
 Estimates of the space-dependent transverse diffusion coefficient $D(z)$ (purple diamonds, right axis)
 and the effective chemical potential $U(z)$ (gray triangles, left  axis) for water
 molecules as functions of the distance $z$ to the Gibbs dividing interface of  crystalline-glutamine surfaces S1 (top), S2 (middle), and S3 (bottom). Solid lines are
 trigonometric interpolations of the data. The purple dashed horizontal lines are set at the diffusion constant
 value of the bulk water $D_\mathrm{bulk}=\SI{3.81}{nm^2/ns}$.
 }
 \label{fig:main}
\end{figure}

Figure~\ref{fig:main} shows the values  of $U(z)$ and $D(z)$, inferred from our molecular-dynamics
simulations for water molecules near the three surfaces of the glutamine crystal.
Consistent with the water density profiles, the effective potentials associated with
the surfaces S2 and S3 are characterised by large energy barriers that separate
the protein-wetting layer from the liquid phase above the Gibbs dividing interface. The
space-dependent diffusion coefficient also features some oscillations near
the surface. Specifically, the peaks in the local diffusion constant close to the interface
appear to be located close to the positions of the maxima associated with the underlying potential.
More interesting is the difference of the
length scales over which the potential and the diffusion coefficient approach their
bulk values. For all the three water-glutamine contacts simulated, the potential $U(z)$
flattens out at distances of about $\SI{5}{\angstrom}$ from the respective Gibbs
dividing interface. In contrast, the space-dependent diffusivity extends over a longer
range: the diffusion coefficient approaches its bulk value at about
$\SI{10}{\angstrom}$ from the surfaces.  For large distances, the relative error of our
estimates of $D(z)$ with respect to the bulk value \SIrange{5}{6}{\%}, a result that
stems from  the uncertainty of the inferred slope of the  potential $U'(z)$.

The specific origin of why the diffusion coefficient
saturates over a longer length scale in comparison with the range of the
associated chemical potential is not so obvious and we cannot pinpoint any particular
reasons at this moment. We remark however, that the density $P(z)$ and the mean first-passage time $\langle\tau(z)\rangle$,
that are used to reconstruct $U(z)$ and $D(z)$, integrate out other degrees of freedom that
may be playing a role in tuning the dynamics. In our previous work, we also examined
how the orientational correlations and the charge density of the water molecules
change as functions of the distance from the Gibbs dividing interface (see Figure~4
in Ref.~\cite{Qaisrani2019}). We observed therein subtle spatial patterns of the
water charge density, suggesting that the glutamine surface induces perturbations of the
electrostatic potential created by the solvent, which are not reflected in the mass
density.

\section{\label{sec:5}Exploring the validity of Langevin models}

%
%

\begin{figure*}
 \centering
  \includegraphics[height=5cm]{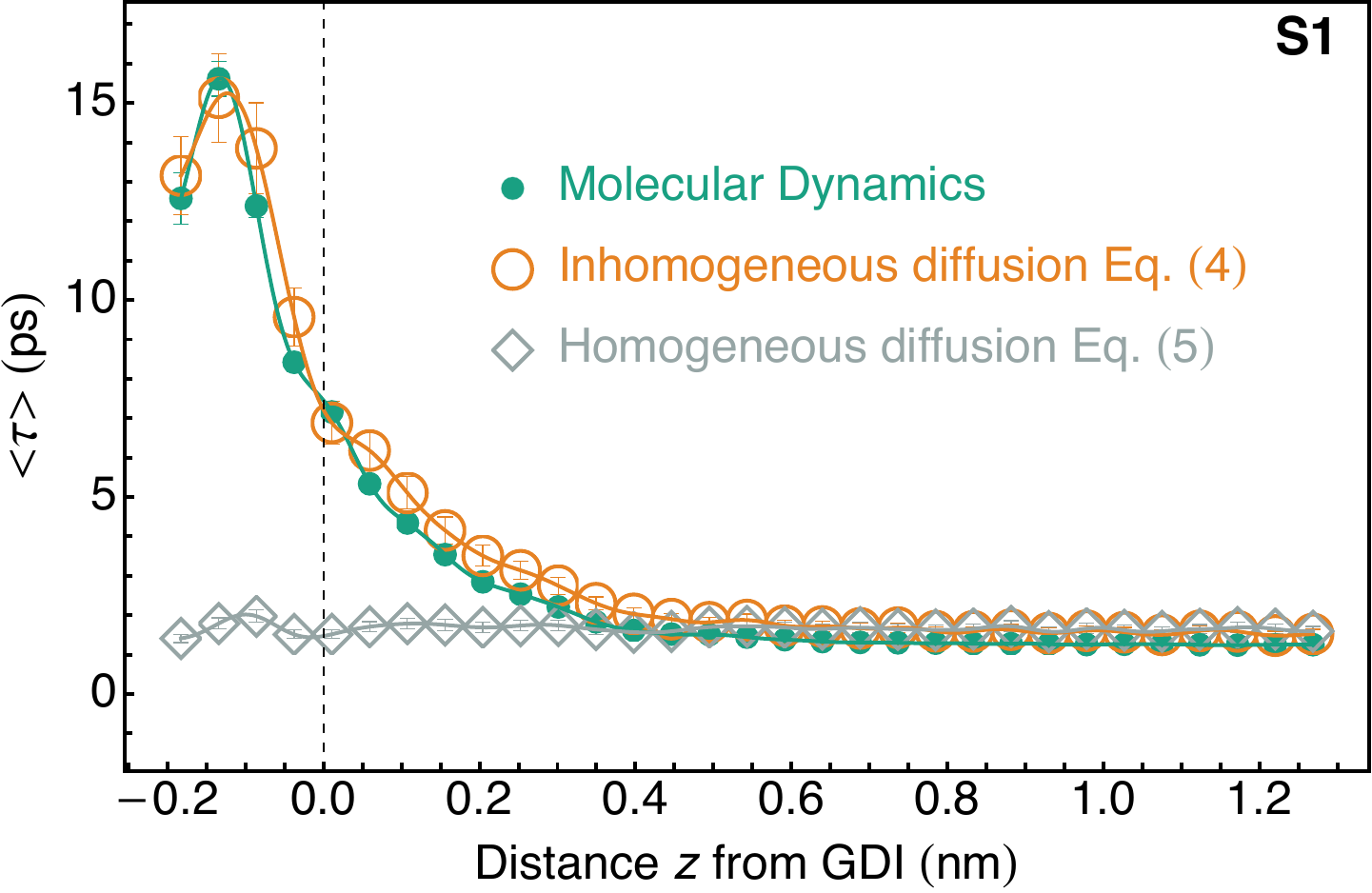} \hspace{1cm}
   \includegraphics[height=5cm]{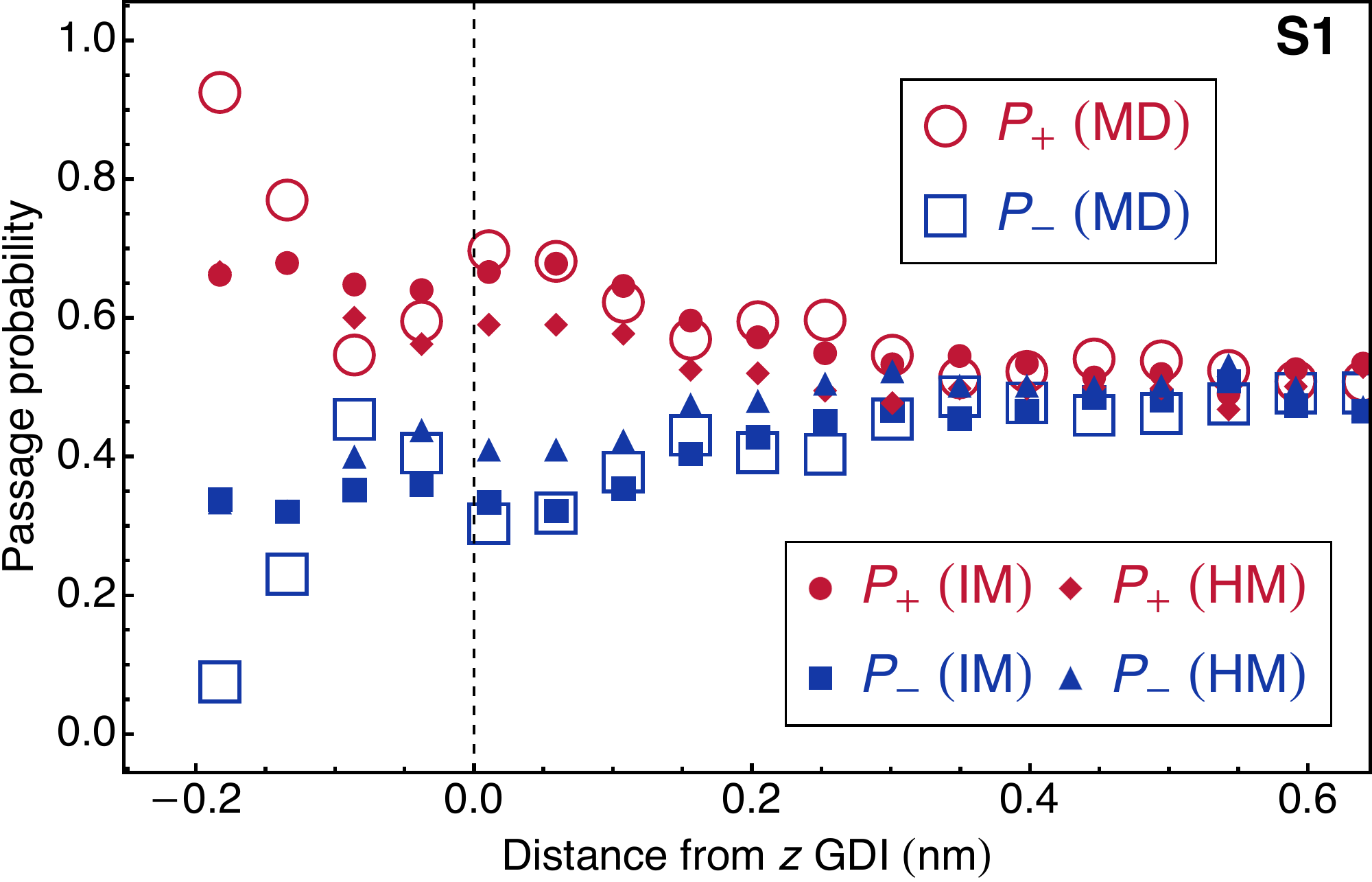}\\
   \includegraphics[height=5cm]{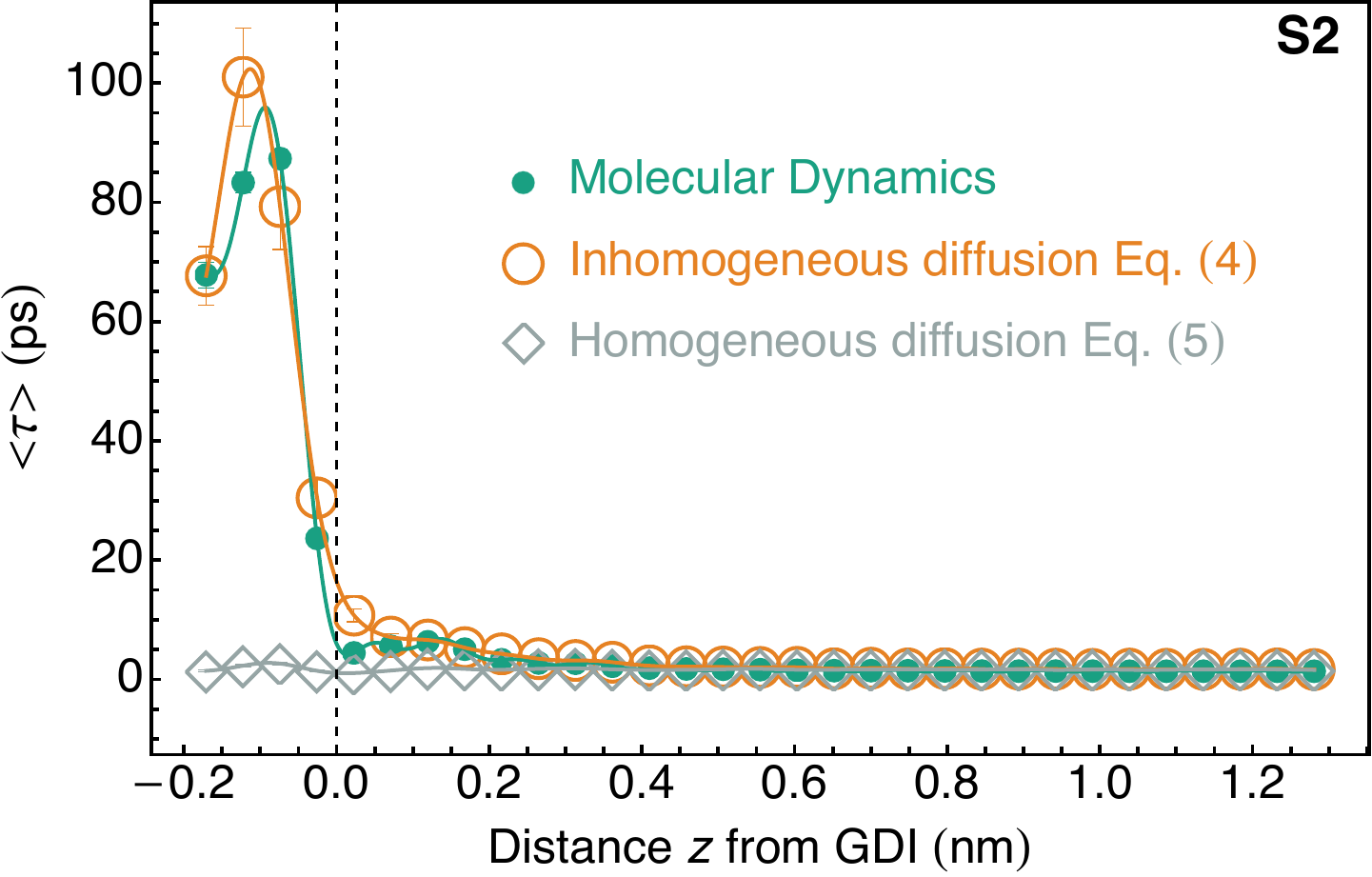} \hspace{1cm}
    \includegraphics[height=5cm]{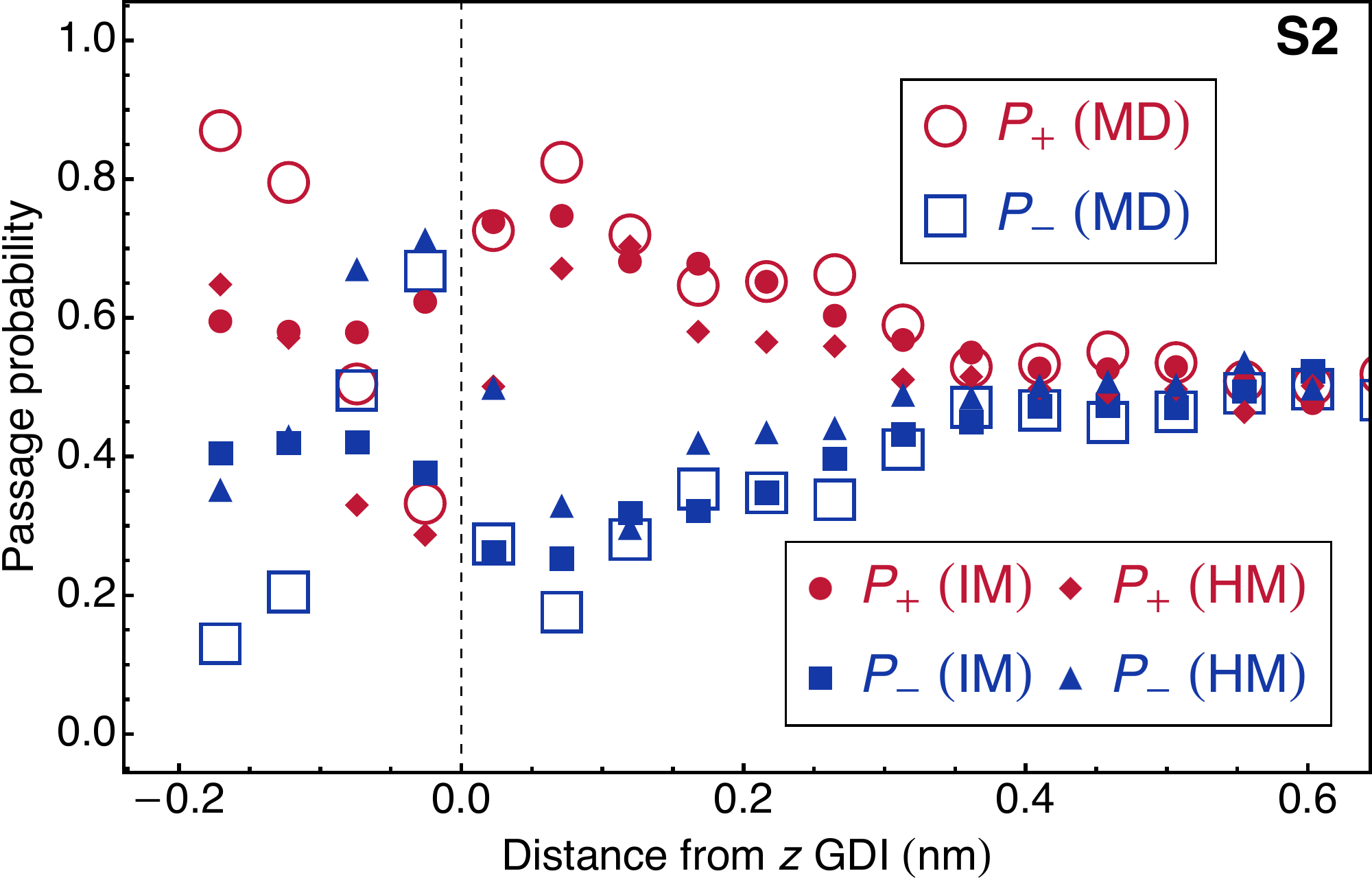} \\
    \includegraphics[height=5cm]{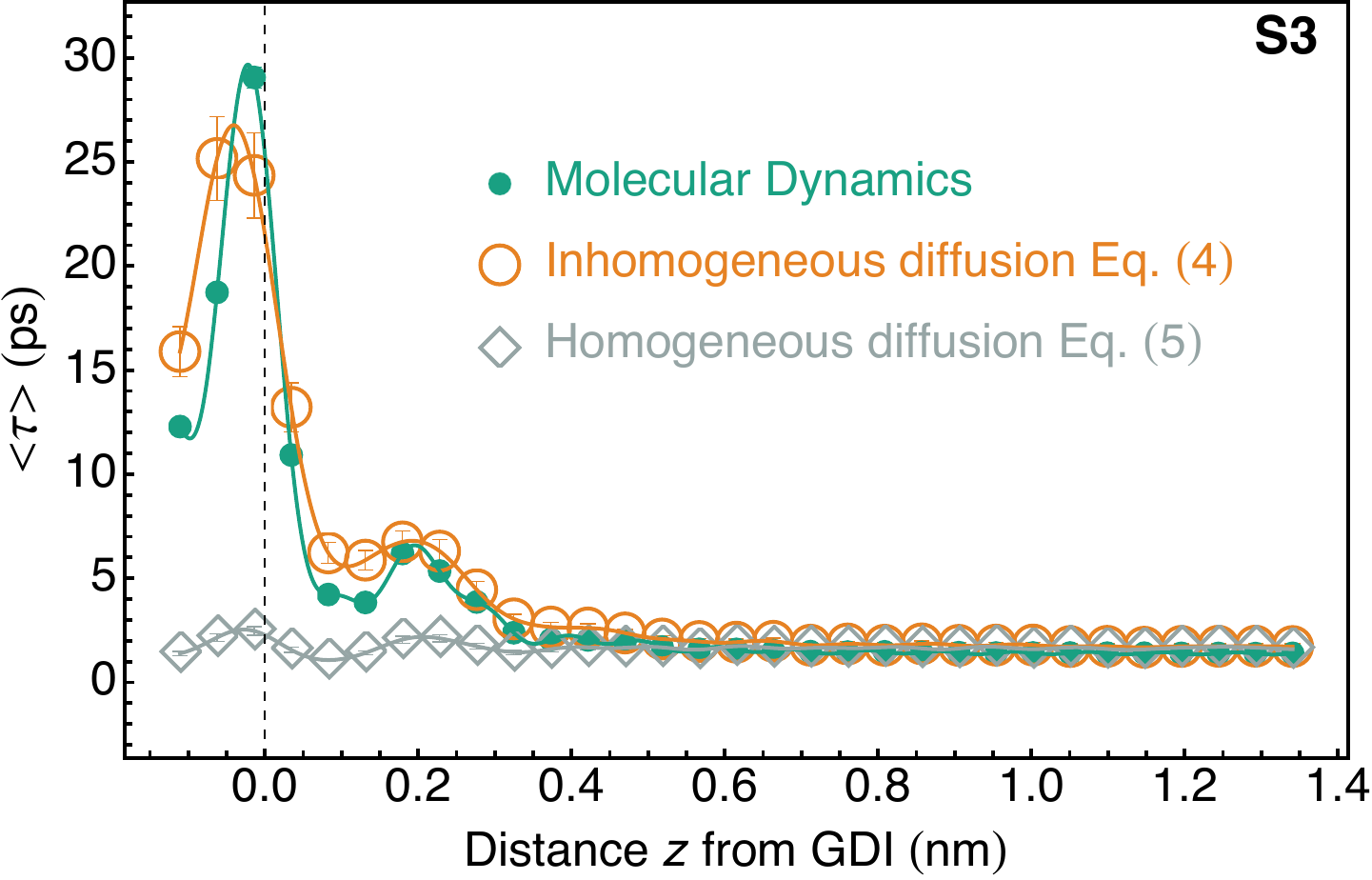} \hspace{1cm}
     \includegraphics[height=5cm]{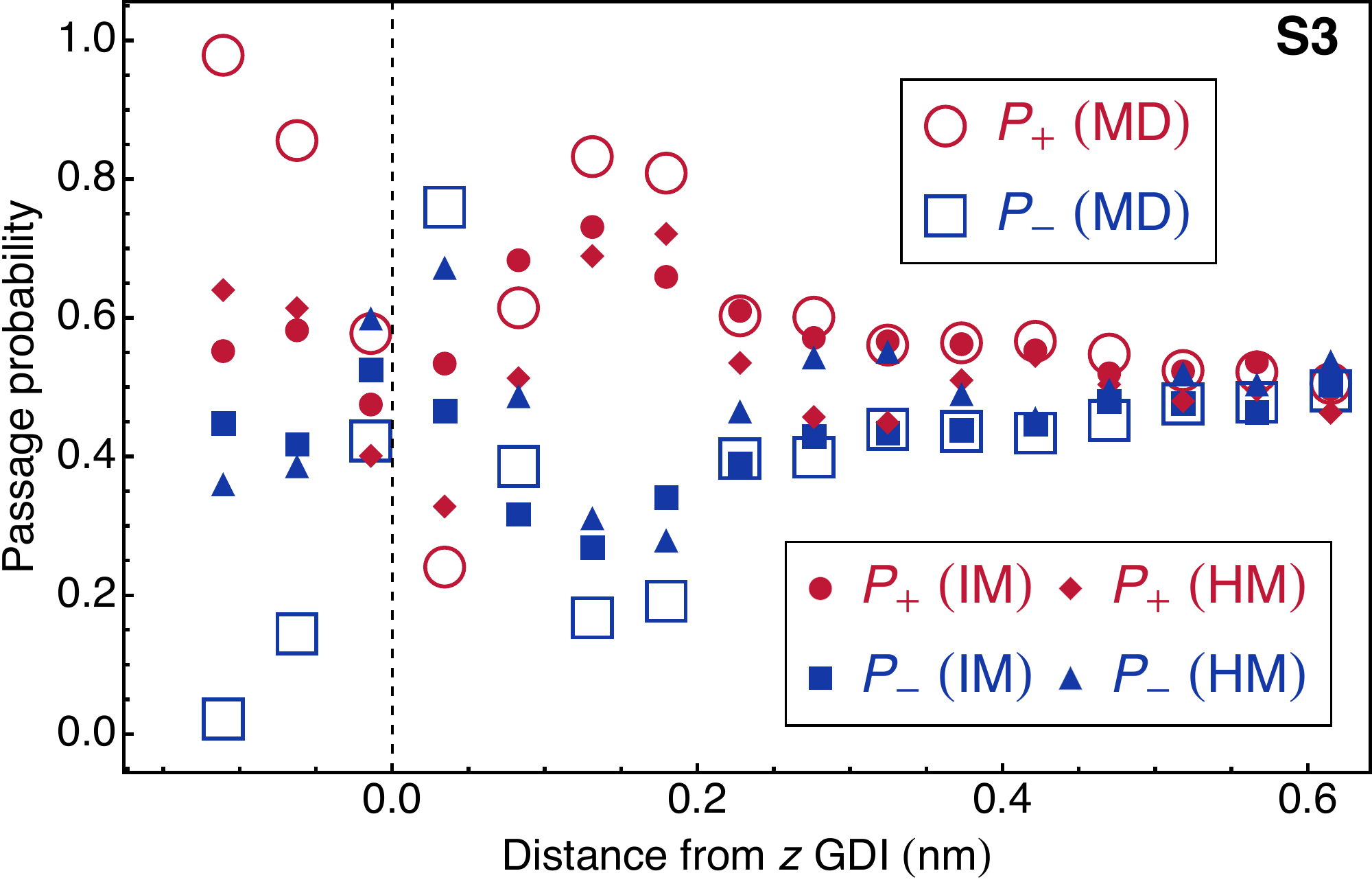}
 \caption{
 Comparison between the first-passage statistics obtained in our molecular-dynamics simulations
 of the water-glutamine contacts S1--3 (MD) with the same statistics measured in numerical simulations of
 two stochastic equations---the inhomogeneous stochastic diffusion model [IM, Eq.~\eqref{eq:LD}]
 and the homogeneous stochastic diffusion model [HM, Eq.~\eqref{eq:fake}]. Left panels:
 mean first-passage time $\avg{\tau(z)}$ as a function of the distance from the
 Gibbs-dividing interface. Right panels: passage probabilities $\mathsf{P}_+(z)$ and $\mathsf{P}_-(z)$
 for water molecules to first reach the positive and negative end of the first-passage corridor
 $[z-L, z+L]$ ($L=\SI{1}{\angstrom}$). The solid lines
 are trigonometric interpolations of the data points. The total number of first-passage events   analyzed  for  each value of $z$ was
  $10^4$--$10^5$ in our molecular-dynamics simulations  and  $10^3$  in our stochastic simulations.}
 \label{fig:main2}
\end{figure*}

The  preceding analysis has revealed a slowdown of the water dynamics due to the
the strong electrostatic potential arising from the charged chemical groups
of the glutamine molecule.
A key assumption in our first-passage analysis is that the
 translational dynamics of water is well described by a Markovian diffusion model,
which at the ensemble level, is
characterised by the Smoluchowski equation~\eqref{eq:SE}.
To verify this assumption and to lend more credence to this theoretical approach,
  we compare in Fig.~\ref{fig:main2} first-passage time statistics obtained in our water-glutamine molecular dynamics simulations (Sec.~\ref{sec:4}) with first-passage times obtained from numerical stochastic simulations of space-dependent diffusions.
For the latter, we performed {\em a posteriori} numerical simulations of the overdamped Ito-Langevin equation
\begin{equation}\label{eq:LD}
    \frac{\text{d}z}{\text{d}t}= - \beta D(z) U'(z) + D'(z) + \sqrt{2 D(z)} \cdot \xi,
\end{equation}
where $\xi(t)$ is Gaussian white noise, whereas $U(z)$ and $D(z)$ are, respectively,
the effective potential and the space-dependent diffusion coefficient of water inferred from
our molecular-dynamics simulations (see Sec.~\ref{sec:4} and Appendix~\ref{sec:data}).
Note that the so-called spurious drift term $D'(z)$ in Eq.~\eqref{eq:LD} ensures that the Fokker-Planck equation describing ensembles of trajectories generated by~\eqref{eq:LD},  coincides with the Smoluchowski
equation~\eqref{eq:SE}.~\cite{LauLubensky2007}
In order to examine the sensitivity of our results to the presence of a space dependent
diffusion constant, we  also perform numerical simulations of an overdamped Langevin
equation
\begin{equation}\label{eq:fake}
  \frac{\text{d}z}{\text{d}t}= - \beta D_{\rm bulk} U'(z)  + \sqrt{2  D_{\rm bulk}} \cdot \xi.
\end{equation}
with  $D_\mathrm{bulk} = \SI{3.81}{nm^2/ns}$ bulk water's (space-independent) diffusion coefficient
 $U(z)$ given by  water's effective chemical potential  as in  Eq.~\eqref{eq:LD}.

  From  numerical simulations  of Eqs.~\eqref{eq:LD}
  and~\eqref{eq:fake}, we evaluate  mean first-passage times $\langle
  \tau (z)\rangle$ and passage probabilities $\mathsf{P}_+(z)$ and
  $\mathsf{P}_-(z)$  (see Fig.~\ref{fig:method}) to escape the
  corridor $[z-L, z+L]$ with $L=\SI{1}{\angstrom}$. We compare these
  statistics  in Fig.~\ref{fig:main2} with the data  obtained from our
  molecular-dynamics simulations reported in Sec.~\ref{sec:4}.  The
  mean first-passage times of the inhomogeneous Langevin model
  Eq.~\eqref{eq:LD} are in good agreement with  those obtained in the
  molecular-dynamics simulations, both capturing the slowdown
  associated with the mobility of water near all the three glutamine
  surfaces. Above the Gibbs dividing interface, the Langevin model
  yields   slightly larger first-passage times than molecular dynamics
  in the bulk liquid, as reported previously in Ref.~\cite{OlivaresRivas2013}. On the
  other hand, the homogeneous
  diffusion model~\eqref{eq:fake} fails completely to capture the
  complexity as well as the magnitude of the mean first-passage  time of
  water molecules (left column of Fig.~\ref{fig:main2}) at distances
  $z<\SI{0.5}{nm}$ from any of the three glutamine surfaces.

The passage probabilities
$\mathsf{P}_+(z)$ and  $\mathsf{P}_-(z)$ (Fig.~\ref{fig:main2}, right
column) provide further insights into the behavior of water near the glutamine
surface.  Interestingly, the passage probabilities obtained from
simulations of the Langevin model~\eqref{eq:LD} are in excellent
agreement with those obtained from the molecular dynamics,
reproducing local maxima and minima above the Gibbs dividing interface
for the three glutamine crystals S1, S2 and S3.  On the other hand,
the Langevin model given by Eq.~\eqref{eq:LD} is not able to capture
accurately the non-trivial behaviour of the passage probabilities
$\mathsf{P}_+(z)$ and $\mathsf{P}_-(z)$ for water molecules that are closely
pinned to the glutamine surface. This
phenomenon is also observed for the mean first-passage time
$\langle\tau (z)\rangle$ (Fig.~\ref{fig:main2}, left column).

%
%
%
%


To obtain a more quantitative perspective of the validity of the Langevin
model, we employ some recent results invoking the extent of {\em
  first-passage time symmetry}  derived within the context of
stochastic thermodynamics and random
walks.~\cite{Roldan2015,neri2017statistics,krapivsky2018first}   For
an homogeneous Langevin equation   $\text{d}z/\text{d}t= v  + \sqrt{2
  D} \cdot \xi$ with $v$ and $D>0$ as two given parameters, the
probability densities  of the conditional first-passage times
$\tau_+(z)$ and $\tau_-(z)$ are
identical.~\cite{Roldan2015,neri2017statistics,krapivsky2018first}
This result implies a symmetry between the conditional mean
first-passage times
\begin{equation}\label{eq:fpts}
\langle\tau_+(z)\rangle=\langle\tau_-(z)\rangle,
\end{equation}
which holds for any  values of the drift $v$ and diffusion coefficient
$D$.  We put Eq.~\eqref{eq:fpts} to test using the data from our
molecular dynamics simulations.  For narrow first-passage corridors,
$U(z)$ and $D(z)$ in Section~\ref{sec:4}, Eq.~\eqref{eq:LD} yields an
effective local homogeneous Langevin equation
$\frac{\text{d}z}{\text{d}t}\simeq  v(z) +  \sqrt{2  D(z)} \cdot \xi$
with $v_Z(z)=\beta U'(z)$ and $D(z)$ referring to the local slope of the potential
and diffusion coefficient, respectively. Therefore, if our approximations are viable,
we expect the symmetry Eq.~\eqref{eq:fpts}, \textit{i.e.}
$\langle\tau_+(z)\rangle=\langle\tau_-(z)\rangle = \avg{\tau(z)}$,
to hold within narrow passage corridors at any distance from the glutamine
surface. To quantify the degree of  violation of the first-passage time symmetry~\eqref{eq:fpts}
we introduce a measure of accuracy
\begin{equation}
\delta_{\pm}(z) = \left(
  1 - \frac{|\avg{\tau_+(z)} - \avg{\tau_-(z)}|}{\avg{\tau(z)}}
\right)\times 100,
\label{eq:op}
\end{equation}
which characterises the relative difference (in \%) between the two
conditional mean first-passage times with respect to the total mean
first-passage time along the $Z$ axis. An accuracy $\delta_\pm(z)=100$ implies a
perfect agreement, whereas $\delta_\pm(z)=0$ a strong violation of the first-passage symmetry~\eqref{eq:fpts}.

\begin{figure}
\centering\hspace{-0.5cm}
\includegraphics[width=0.95\columnwidth]{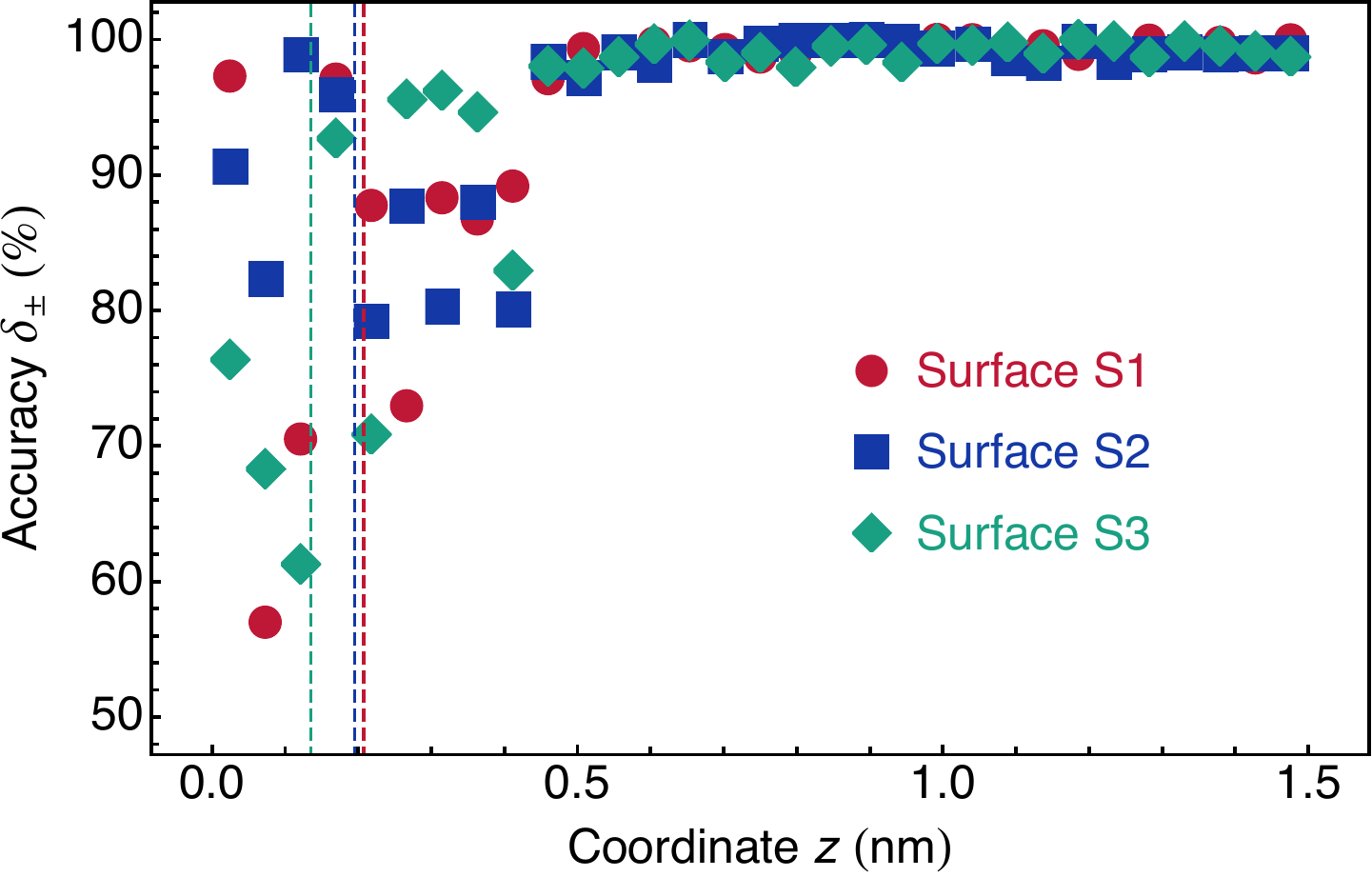}
\caption{Accuracy  measure $\delta_\pm(z)$, given by Eq.~\eqref{eq:op}, of the first-passage time symmetry~\eqref{eq:fpts}  obtained for our water-glutamine molecular-dynamics simulations as a function of the distance $z$ to the contact with the three glutamine surfaces (see legend). The vertical dashed lines indicate the location of  the Gibbs dividing interfaces between the liquid and glutamine crystal.
\label{fig:dpm}}
\end{figure}

We report values of
$\delta_{\pm}(z)$ given by Eq.~\eqref{eq:op} in Fig.~\ref{fig:dpm},
which shows that the first-passage symmetry  holds within an accuracy
of at least~\SI{75}{\%} at distances $z\gtrsim\SI{1}{\angstrom}$ from
the Gibbs dividing interfaces of water with all the three glutamine
surfaces S1, S2, and S3 (Fig.~\ref{fig:dpm}).  For water molecules
located closer to the glutamine surface, we find
statistically significant deviations from the symmetry Eq.~\eqref{eq:fpts}---a
result that challenges the validity of the Langevin model~\eqref{eq:LD} and the approximations used in our estimates below the protein-wetting layer. In order to verify that this
discrepancy is not caused by linear-order effects of the
inhomogeneous diffusion, we have estimated the mobility of water
molecules by using a more complex diffusion model
(Appendix~\ref{sec:LE2}) and  found no appreciable  changes in our
results (Fig.~\ref{fig:extra}).~\footnote{We remark that the magnitude of the first-passage corridor $L$
  cannot be reduced below \SI{1}{\angstrom}---the length scale of a covalent bond.}

In conclusion, the preceding analysis reveals the limits of the ability of
the Smoluchowski diffusion model to explain the dynamics of water molecules
that remain for a long time closely pinned to the glutamine surface.
Earlier, we showed by examining the 3D first-passage statistics that
the diffusive dynamics of molecules with the same coordinate $z$ and
various coordinates $x$ and $y$ can differ substantially. Hence, the
projection of this dynamics on a single coordinate axis may mix
populations of water molecules with different dynamical properties. In
such a complex environment alternative approaches that take into
account inhomogeneous distributions of microscopic displacements
and/or nonlocal and memory
effects~\cite{Metzler1999,Barkai2000,Sheu2010,Sharma2018,loos2019non}
may provide further insights into water dynamics below the Gibbs
dividing interface.

\section{Conclusions}
\label{sec:6}
In this work, we have shown that first-passage times of water molecules at atomic scales carry information about their dynamical interaction with  glutamine crystals. We have used a combination of atomistic molecular-dynamics simulations
and stochastic methods of stochastic theory to characterise the mobility
of water molecules near crystalline glutamine.
 This system is a useful
model for understanding how the diffusive dynamics of water is perturbed by protein
aggregates, a problem that is highly relevant in studies of neurodegenerative diseases
and crystal nucleation.

First-passage statistics of water molecules in the molecular-dynamics simulations
were analysed as a function of distance from three crystallographic planes of the glutamine
crystal. We showed that the mean time and a plethora of related first-passage statistics
provide fresh insights for modelling the diffusive dynamics of water molecules in
contact with a crystal surface and more generally biological systems.
Interestingly, the passage probabilities of water molecules appear to be more sensitive to anisotropies in different directions as a function of distance from the surface compared to the mean first passage times.

Fitting our data to an inhomogeneous diffusion model, we have reconstructed the transverse space-dependent
diffusion coefficient of water molecules as a function of distance to each of the
three glutamine surfaces.  Our measurements reveal a slowdown of the solvent's diffusive dynamics near the protein
almost by a factor of 4 relative to the bulk. Curiously, the chemical potential of
the water molecules does not extend far beyond \SI{0.5}{nm} from the Gibbs dividing
interface, whereas the diffusion constant saturates slowly to the
bulk value approximately at $\SI{1}{nm}$. This may reflect the longer range of orientational
and charge correlations in the water as well as the fact that the diffusion is more
sensitive to the excluded volume induced by the presence of the crystal surface.

The combination of both the molecular-dynamics simulations and stochastic modelling
provides us with a robust method to estimate space-dependent diffusion coefficients from mean first-passage times to escape narrow corridors, valid not only in atomic but also at mesoscopic scales.
This approach can applied to a plethora of different systems and opens up the possibility
for interesting applications including understanding the diffusion tensor of water
molecules around more complex and disordered interfaces.~\cite{chemrev2016,gibertihassanali2017,Poli2020} Besides this, the first-passage distributions open up some interesting directions on trying to use them to identify chemical fingerprints for hydrophobic and hydrophilic
environments. The application of these methods will be the subject of forthcoming
studies in our group. This will also help in the interpretation of the experimental
measurements of water structure and dynamics near biological surfaces.~\cite{Klass2017,Tros2017}

\section*{Conflicts of interest}
There are no conflicts to declare.

\section*{Acknowledgements}
We thank Francesco Sciortino,   Jacopo Grilli, the Grill lab, the Pavin group, the Tolic lab, and Peter H\"anggi  for stimulating discussions.

\appendix
\section*{\huge Appendix}
\section{\label{sec:MD}Molecular dynamics.}
All the molecular dynamics simulations for the three surfaces were performed with the GROMACS
package.~\cite{abraham2015gromacs} In all these simulations, we used the
OPLS-AA\cite{jorgensen1996development} force field together with the TIP4P
water model.~\cite{jorgensen1983comparison,jorgensen1985temperature}
A non-bonded pair list was created with a cut-off radius of 1.4 \si{nm} and updated after
every 10 time integration steps. For the shifted Lennard-Jones potential,
the cut-off was set at 1.2 \si{nm}. The long-range electrostatic forces were
taken into account by using the Particle Mesh Ewald-Switch\cite{darden1993particle} method with
a Coulomb switching cut-off 1.2 \si{nm}.
A long-range dispersion correction for the pressure and energy
was applied to the truncated van der Waals interactions. All bonds were constrained using the
LINCS algorithm.~\cite{hess1997lincs} A timestep of 2fs was used for the Verlet integrator.
All simulations were performed in the canonical ensemble
(NVT) at 300K using the velocity-rescale thermostat\cite{bussi2007canonical} with a
time-constant of 0.1 ps.

\section{\label{sec:data}Data analysis.~~} Statistical data were collected from 15-\si{ns}-long trajectories for each of the three water-glutamine interfaces, and from a 2-ns-long trajectory
for the bulk water. Positions of water molecules were identified with the coordinates
of the oxygen atoms in the \ce{H2O} groups.

The effective potential $U(z)$ associated with the glutamine surfaces was extracted
from histograms of the water density (Sec.~\ref{sec:4}). We inspected an interval
of length $\ell = \SI{1.5}{nm}$ along the $Z$ coordinate axis in the immediate contact
with the glutamine crystal. This interval was partitioned into $\nu=31$ bins of
equal size. Values of the histogram density were assigned to the bins' centres $z_{i=1,2...}$,
which thus form an equidistant grid of points. In addition we assume zero-gradient
boundary conditions: thermodynamic properties have no spatial variation for large
$z$ in contact with the bulk liquid, whereas at the protein end of the inspected
interval we impose a fictitious reflective boundary condition.

A pseudospectral basis set,\cite{Trefethen2000} that is appropriate for the above
boundary conditions and nodes, is given by trigonometric functions
\begin{equation}\label{eq:phi}
    \phi_n(z) = \begin{cases}
        \sqrt{2/\ell}\qquad\qquad\qquad\text{ if } n=0,\\
    \cos(\pi n z/\ell)/\sqrt{\ell}\qquad\text{ if } 0 < n < 30,
    \end{cases}
\end{equation}
which are orthonormal with respect to the weight $\delta{z}=\ell/\nu \approx 0.03\,\si{nm}$
and the scalar product
$$
    \avg{\phi_n(z), \phi_m(z)} = \delta{z} \sum_{i=0}^{\nu-1}
    \phi_n(z_i) \phi_m(z_i).
$$
Any function $f(z)$ that satisfies the same boundary conditions, such as $U(z)$ and
$\ln D(z)$, can therefore be represented by
\begin{equation}\label{eq:trig}
    f(z) = \sum_{n=0}^{\nu-1} f_n \phi_n(z),\quad
    f_n = \avg{\phi_n(z) f(z)}.
\end{equation}
This trigonometric interpolation ensures spectral accuracy of numerical calculations
with empirical functions.~\cite{Trefethen2000} We have verified that the pseudospectral
representation of the steady-state density with the chosen number of the bins $\nu$
renders a density estimate that is as efficient as a smooth histogram with the Epanechnikov
kernel.

The first-passage events were detected with a time resolution of $\delta{t}=\SI{50}{fs}$
for molecules' displacements of length $L=\SI{1}{\angstrom}$ in any direction from the
initial positions $z(0)=z$. To mitigate the discretization effect of the data acquisition
we subtracted from each time measurement $t$ of the first-passage event a linear-order
correction
$$\tau(z) = t-\delta{t}\frac{L - |z(t-\delta{t}) - z|}{|z(t)-z(t-\delta{t})|}.$$
We have also checked that our choice of $\delta{t}$ does not have a strong influence
on the results, by using a higher resolution $\delta{t}=\SI{6}{fs}$ with trajectories
of a shorter duration ($\SI{5}{ns}$).

The first-passage times were averaged over the water molecules residing initially
in the same density histogram bin. Hence the resulting value $\avg{\tau(z)}$ was
attributed to the bins' centers $z = z_i\pm\delta{z}/2$.

\section{\label{sec:FPTS} Mean first-passage time: proof of Eq.~\eqref{eq:tau}}

We first review exact results for the first-passage time statistics of one-dimensional
diffusion processes. Let $P(x,t|x_0,0)$ be the probability density for the process
to be in a state $x(t) = x$ at a time $t$ given that its initial state was $x(0)=x_0$.
We assume that this probability density evolves according to a general Fokker-Planck
equation
\begin{equation}\label{eq:fpeg}
\frac{\partial P}{\partial t} = - \frac{\partial}{\partial x}[a(x)P]
  +\frac{1}{2}\frac{\partial^2}{\partial x^2}[b(x)P]
\end{equation}
with two real-valued functions $a(x)$ and $b(x)>0$. Exact formal expressions for
several first-passage time statistics have been previously reported for this class
of systems.~\cite{hanggi1990reaction,GoelRichterDyn2016} We consider here escape
of a molecule with the initial position $x(0)=z$ through one of the two boundaries
of the passage corridor $[z-L,z+L]$ with $L>0$. The probability $\mathsf{P}_-(z)$
and $\mathsf{P}_+(z)$ for the system to first reach the boundary at $z-L$ and $z+L$,
respectively, are given by
\begin{equation}\label{eq:ppluspminus}
  \mathsf{P}_-(z) = \frac{\int_{z}^{z+L} \text{d}y\,\pi(y)}{\int_{z-L}^{z+L} \text{d}y\,\pi(y)}, \quad
\mathsf{P}_+(z) = \frac{\int_{z-L}^z \text{d}y\,\pi(y)}{\int_{z-L}^{z+L} \text{d}y\,\pi(y)},
\end{equation}
with the auxiliary function $\pi(y)$ defined as
\begin{equation}
\pi(y) \equiv \exp \left(-\int^y \text{d}x\frac{2a(x)}{b(x)} \right).
\label{eq:ibf}
\end{equation}
The mean first-passage time for the system to first reach either the boundary at
$z-L$ or the boundary at $z+L$ is then given by
\begin{equation}\label{eq:MFPTzth}
\langle\tau(z)\rangle = \int_z^{z+L} \text{d}y\, \pi(y) I(y) -\mathsf{P}_-(z)\int_{z-L}^{z+L} \text{d}y\, \pi(y) I(y) ,
\end{equation}
in which the functions $\mathsf{P}_-(z)$ and $\pi(y)$ are given by Eqs.~\eqref{eq:ppluspminus} and~\eqref{eq:ibf}, respectively, whereas
\begin{equation}\label{eq:Izeta}
  I(y) \equiv \int_{z-L}^y  \frac{2\text{d}x}{b(x) \pi(x)}.
    \end{equation}

The diffusion model that we use to describe fluctuations of water density is governed
by the Fokker-Planck equation [Eq.(1)], copied here for convenience
\begin{equation*}
  \frac{\partial P(z,t)}{\partial t} = \frac{\partial}{\partial z} \left(
      \beta D(z)  P(z,t)\frac{\partial U(z)}{\partial z}
      + D(z) \frac{\partial P(z,t)}{\partial z}
  \right),
\end{equation*}
which corresponds to Eq.~\eqref{eq:fpeg} with
\begin{equation}\label{eq:ab}
a(x) = -\beta D(x) U'(x) + D'(x),\quad b(x) = 2D(x)\,.
\end{equation}
Specializing Eqs.~\eqref{eq:ibf} and~\eqref{eq:Izeta} to functions given by Eq.~\eqref{eq:ab}
we evaluate the auxiliary functions
\begin{equation}\label{eq:aux}
  \pi(y) =  \frac{\me^{\beta U(y)}}{D(y)},\quad I(y) \equiv \int_{z-L}^y \text{d}x\,\me^{-\beta U(x)},
\end{equation}
which resemble the inverse Boltzmann factor and the partition function, respectively.
By substituting Eq.~\eqref{eq:aux} into \eqref{eq:ppluspminus} and \eqref{eq:MFPTzth},
we obtain the following explicit formulas for the passage probabilities
\begin{equation}
\label{eq:ppluspminus2}
  \mathsf{P}_-(z) = \frac{\int_{z}^{z+L} \text{d}y\,\me^{\beta U(y)}/D(y)}{\int_{z-L}^{z+L} \text{d}y\,\me^{\beta U(y)}/D(y)}, \quad
\mathsf{P}_+(z) = \frac{\int_{z-L}^z \text{d}y\,\me^{\beta U(y)}/D(y)}{\int_{z-L}^{z+L} \text{d}y\,\me^{\beta U(y)}/D(y)},
\end{equation}
and the mean first-passage time
\begin{eqnarray}\label{eq:MFPTzth2}
\langle\tau(z)\rangle &=& \int_z^{z+L}  \frac{\text{d}y}{D(y)} \int_{z-L}^y \text{d}x\,\me^{-\beta [U(x)-U(y)]}\\
&-&\mathsf{P}_-(z)\int_{z-L}^{z+L}   \frac{\text{d}y}{D(y)} \int_{z-L}^y \text{d}x\,\me^{-\beta [U(x)-U(y)]}\;.\nonumber
\end{eqnarray}
Equations~\eqref{eq:ppluspminus2} and~\eqref{eq:MFPTzth2} are amenable to a complete analytical
treatment only in special cases. For a sufficiently small $L$ we may approximate
the effective potential by a truncated power-series expansion
\begin{equation}\label{eq:Uz}
  U(y) \simeq U(z) +  (y-z)h + O(y^2),
\end{equation}
which holds for $-L < y - z < L$ with $h \equiv U'(z)$. Because the diffusion coefficient
is a strictly positive quantity, to preserve this property we use the approximation
\begin{equation}\label{eq:Dz}
D(y)\simeq D(z)\me^{\kappa(y-z) + O(y^2)},
\end{equation}
in which $\kappa=D'(z)/D(z)$ is the logarithmic derivative of $D(z)$. The above equation
implies a linear expansion of the logarithm of the diffusion coefficient
$\ln D(y) \simeq \ln D(z) + \kappa (y-z) + O(y^2)$. For the approximate Eqs.~\eqref{eq:Uz}
and~\eqref{eq:Dz} the formulas of the passage probabilities [Eq.~\eqref{eq:ppluspminus2}]
read
\begin{equation} \label{eq:P+P-app}
\mathsf{P}_-(z) \simeq \frac{1}{1+\me^{-L (\beta h - \kappa)}},\quad \mathsf{P}_+(z) \simeq \frac{1}{1+\me^{L (\beta h -\kappa)}}.
\end{equation}
Eqs.~\eqref{eq:Uz}--~\eqref{eq:P+P-app} plugged into \eqref{eq:MFPTzth2} lead to
the following approximate expression for the mean first-passage time
\begin{equation}\label{eq:MFPTzth3}
\langle\tau(z)\rangle \simeq \frac{\cosh[(\beta h + \kappa) L/2]
  /\cosh[(\beta h - \kappa) L/2] - 1}{\kappa \beta h D(z)} .
\end{equation}
The limit $\kappa\to0$ of Eq.~\eqref{eq:MFPTzth3} yields Eq.~\eqref{eq:tau}, \textit{viz.}
\begin{equation}
\langle\tau(z)\rangle \simeq \frac{L\, \tanh(L \beta k/2)}{ D(z)\beta k}.
\end{equation}
The limit $h\to0$ of Eq.~\eqref{eq:tau} gives in its turn the formula of the mean
first-passage time in bulk water
\begin{equation}
  \avg{\tau_\mathrm{bulk}} = \frac{L^2}{2 D_\mathrm{bulk}},
\end{equation}
in which we have identified $\avg{\tau(z)}\to\avg{\tau_\mathrm{bulk}}$ and
$D(z)\to D_\mathrm{bulk}$.

\section{\label{sec:LE2} Linear-order effect of inhomogeneous diffusion}
In this Appendix we show that the linear-order effect of inhomogeneous diffusion
is negligible in our estimations of the coefficient $D(z)$ (Sec.~\ref{sec:4}). From
Eq.~\eqref{eq:MFPTzth3} we can express the diffusion coefficient as
\begin{equation}\label{eq:Dzfull}
  D(z) = \theta(h,\kappa) /  \avg{\tau(z)},
\end{equation}
in which
$$
  \theta(h,\kappa) = \frac{\cosh[(\beta h + \kappa) L/2]
  /\cosh[(\beta h - \kappa) L/2] - 1}{\kappa \beta h}
$$
depends only on the slope of the effective potential $h$ and the unknown linear-order
diffusion parameter $\kappa$.

The effective potential of water molecules and the mean first-passage times $\avg{\tau(z)}$
have been extracted from our molecular-dynamics simulations (Sec.~\ref{sec:4}).
Therefore we need to find only the parameter $\kappa$ in order to evaluate Eq.~\eqref{eq:Dzfull}
and, in particular, the coefficient $\theta(h,\kappa)$ to the linear-order in $z$.
The parameter $\kappa$ can be inferred from the first-passage probabilities that
we reported in Sec.~\ref{sec:5} (right panels in Fig.~\eqref{fig:main2}) by using
Eq.~\eqref{eq:P+P-app}:
$$
  \kappa = \beta h + \log[P_+(z)/P_-(z)]/L.
$$

\begin{figure}
 \centering
 \includegraphics[height=6cm]{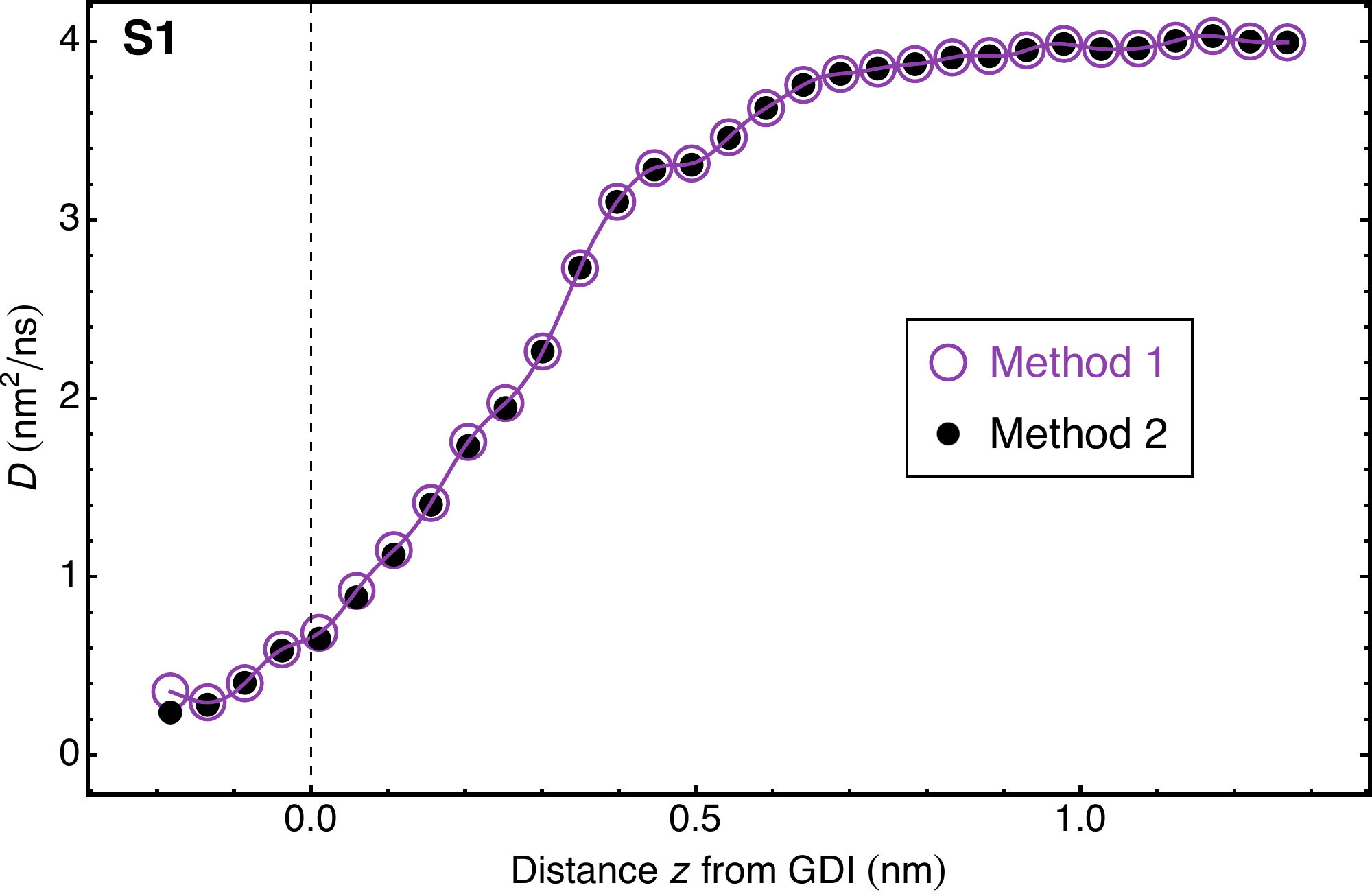}\\
 \includegraphics[height=6cm]{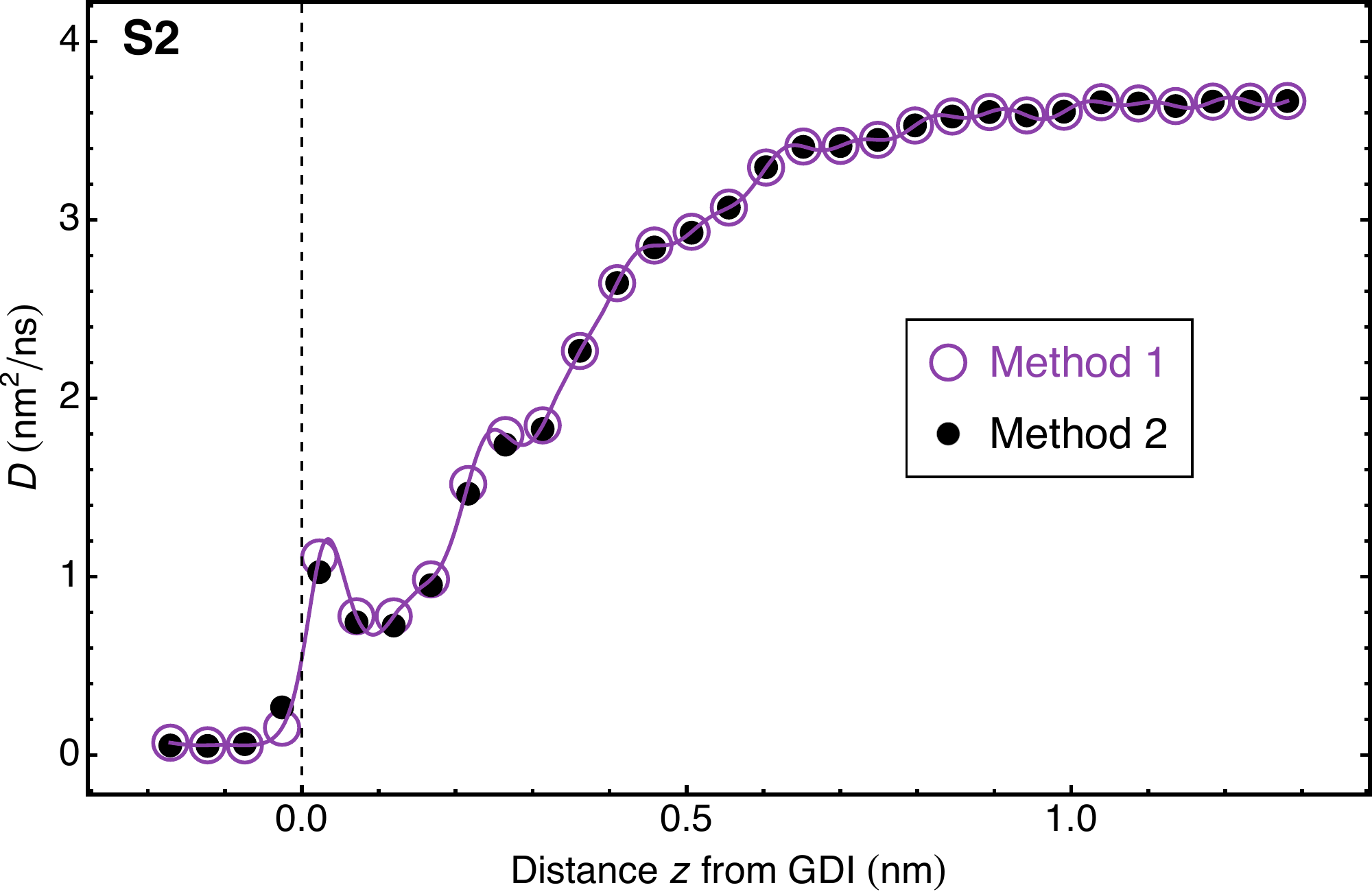}\\
 \includegraphics[height=6cm]{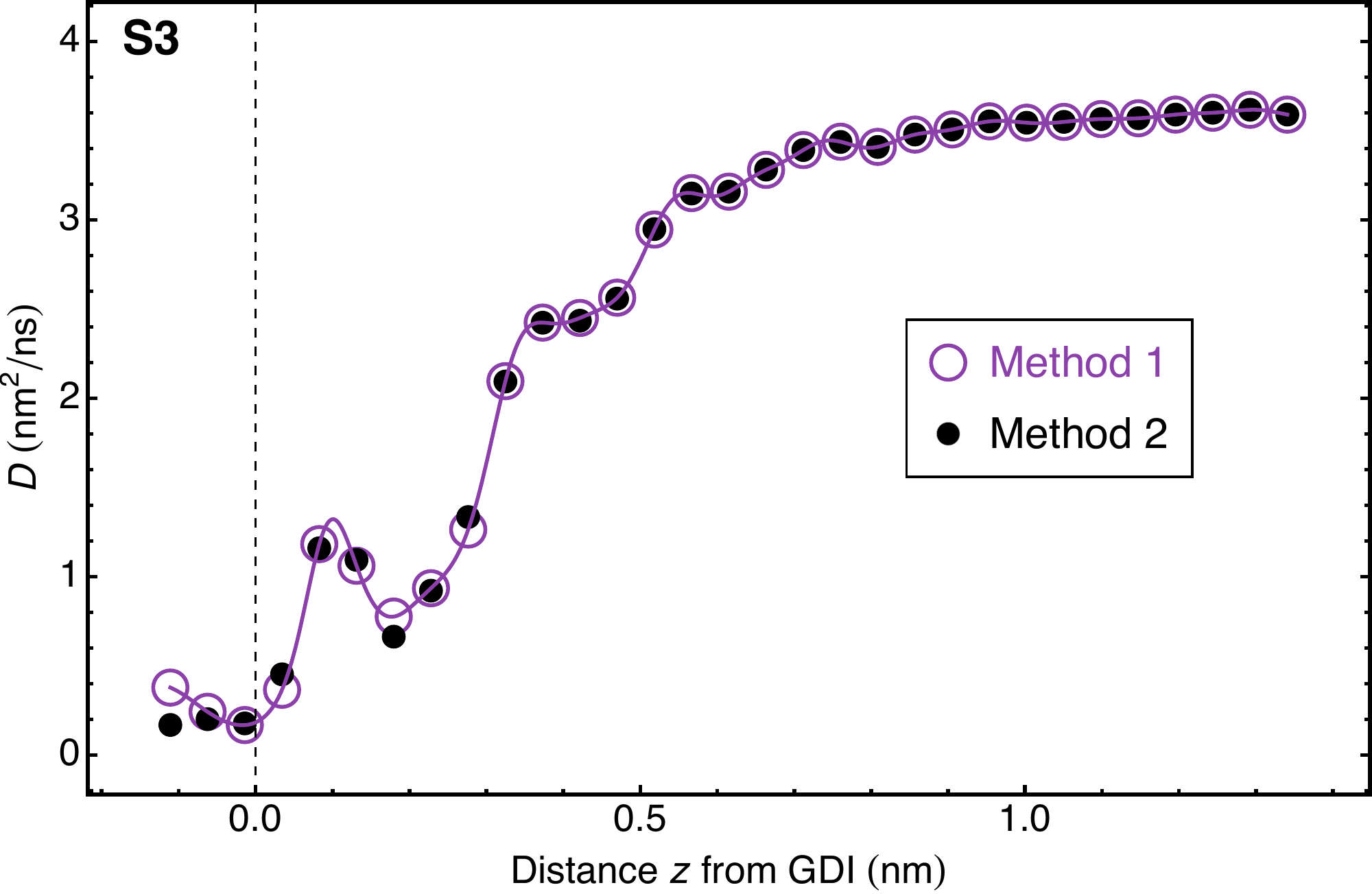}
 \caption{
   Comparison of the diffusion coefficient values estimated by the method of Sec.~\ref{sec:5}
   (Method 1) and the method of Appendix~\ref{sec:LE2} (Method 2) that takes into
   account the linear-order variation of $D(z)$ as a function of the distance to
   the Gibbs dividing interface with one of the three glutamine surfaces.
 }
 \label{fig:extra}
\end{figure}

As shown in Fig.~\ref{fig:extra}, Equation~\eqref{eq:Dzfull} that incorporates the
linear-order correction due to the variation of $D(z)$ gives estimates almost identical
to those of Sec.~\ref{sec:5}. This fact confirms that the mobility of water molecules
changes much slower than their effective potential.

\clearpage
\newpage

\balance

\providecommand*{\mcitethebibliography}{\thebibliography}
\csname @ifundefined\endcsname{endmcitethebibliography}
{\let\endmcitethebibliography\endthebibliography}{}

\bibliographystyle{rsc}

\begin{mcitethebibliography}{107}
\providecommand*{\natexlab}[1]{#1}
\providecommand*{\mciteSetBstSublistMode}[1]{}
\providecommand*{\mciteSetBstMaxWidthForm}[2]{}
\providecommand*{\mciteBstWouldAddEndPuncttrue}
  {\def\EndOfBibitem{\unskip.}}
\providecommand*{\mciteBstWouldAddEndPunctfalse}
  {\let\EndOfBibitem\relax}
\providecommand*{\mciteSetBstMidEndSepPunct}[3]{}
\providecommand*{\mciteSetBstSublistLabelBeginEnd}[3]{}
\providecommand*{\EndOfBibitem}{}
\mciteSetBstSublistMode{f}
\mciteSetBstMaxWidthForm{subitem}
{(\emph{\alph{mcitesubitemcount}})}
\mciteSetBstSublistLabelBeginEnd{\mcitemaxwidthsubitemform\space}
{\relax}{\relax}

\bibitem[Fenimore \emph{et~al.}(2002)Fenimore, Frauenfelder, McMahon, and
  Parak]{Fenimore2002}
P.~W. Fenimore, H.~Frauenfelder, B.~H. McMahon and F.~G. Parak,
  \emph{Proceedings of the National Academy of Sciences of the United States of
  America}, 2002, \textbf{99}, 16047--16051\relax
\mciteBstWouldAddEndPuncttrue
\mciteSetBstMidEndSepPunct{\mcitedefaultmidpunct}
{\mcitedefaultendpunct}{\mcitedefaultseppunct}\relax
\EndOfBibitem
\bibitem[Yang \emph{et~al.}(2007)Yang, Onuchic, Garc{\'{i}}a, and
  Levine]{Yang2007}
S.~Yang, J.~N. Onuchic, A.~E. Garc{\'{i}}a and H.~Levine, \emph{Journal of
  Molecular Biology}, 2007, \textbf{372}, 756--763\relax
\mciteBstWouldAddEndPuncttrue
\mciteSetBstMidEndSepPunct{\mcitedefaultmidpunct}
{\mcitedefaultendpunct}{\mcitedefaultseppunct}\relax
\EndOfBibitem
\bibitem[{Van Hijkoop} \emph{et~al.}(2007){Van Hijkoop}, Dammers, Malek, and
  Coppens]{VanHijkoop2007}
V.~J. {Van Hijkoop}, A.~J. Dammers, K.~Malek and M.~O. Coppens, \emph{Journal
  of Chemical Physics}, 2007, \textbf{127}, year\relax
\mciteBstWouldAddEndPuncttrue
\mciteSetBstMidEndSepPunct{\mcitedefaultmidpunct}
{\mcitedefaultendpunct}{\mcitedefaultseppunct}\relax
\EndOfBibitem
\bibitem[Best and Hummer(2010)]{Best2010}
R.~B. Best and G.~Hummer, \emph{Proceedings of the National Academy of Sciences
  of the United States of America}, 2010, \textbf{107}, 1088--1093\relax
\mciteBstWouldAddEndPuncttrue
\mciteSetBstMidEndSepPunct{\mcitedefaultmidpunct}
{\mcitedefaultendpunct}{\mcitedefaultseppunct}\relax
\EndOfBibitem
\bibitem[Best \emph{et~al.}(2013)Best, Hummer, and Eaton]{Best2013}
R.~B. Best, G.~Hummer and W.~A. Eaton, \emph{Proceedings of the National
  Academy of Sciences of the United States of America}, 2013, \textbf{110},
  17874--17879\relax
\mciteBstWouldAddEndPuncttrue
\mciteSetBstMidEndSepPunct{\mcitedefaultmidpunct}
{\mcitedefaultendpunct}{\mcitedefaultseppunct}\relax
\EndOfBibitem
\bibitem[Hinczewski \emph{et~al.}(2010)Hinczewski, {Von Hansen}, Dzubiella, and
  Netz]{Hinczewski2010}
M.~Hinczewski, Y.~{Von Hansen}, J.~Dzubiella and R.~R. Netz, \emph{Journal of
  Chemical Physics}, 2010, \textbf{132}, year\relax
\mciteBstWouldAddEndPuncttrue
\mciteSetBstMidEndSepPunct{\mcitedefaultmidpunct}
{\mcitedefaultendpunct}{\mcitedefaultseppunct}\relax
\EndOfBibitem
\bibitem[von Hansen \emph{et~al.}(2010)von Hansen, Kalcher, and
  Dzubiella]{VonHansen2010}
Y.~von Hansen, I.~Kalcher and J.~Dzubiella, \emph{The Journal of Physical
  Chemistry B}, 2010, \textbf{114}, 13815--13822\relax
\mciteBstWouldAddEndPuncttrue
\mciteSetBstMidEndSepPunct{\mcitedefaultmidpunct}
{\mcitedefaultendpunct}{\mcitedefaultseppunct}\relax
\EndOfBibitem
\bibitem[Wong-ekkabut and Karttunen(2016)]{Wongekkabut2016}
J.~Wong-ekkabut and M.~Karttunen, \emph{Journal of Biological Physics}, 2016,
  \textbf{42}, 133--146\relax
\mciteBstWouldAddEndPuncttrue
\mciteSetBstMidEndSepPunct{\mcitedefaultmidpunct}
{\mcitedefaultendpunct}{\mcitedefaultseppunct}\relax
\EndOfBibitem
\bibitem[Sharma and Biswas(2018)]{Sharma2018}
S.~Sharma and P.~Biswas, \emph{Journal of Physics Condensed Matter}, 2018,
  \textbf{30}, year\relax
\mciteBstWouldAddEndPuncttrue
\mciteSetBstMidEndSepPunct{\mcitedefaultmidpunct}
{\mcitedefaultendpunct}{\mcitedefaultseppunct}\relax
\EndOfBibitem
\bibitem[Knapp and Muegge(1993)]{Knapp1993}
E.~W. Knapp and I.~Muegge, \emph{The Journal of Physical Chemistry}, 1993,
  \textbf{97}, 11339--11343\relax
\mciteBstWouldAddEndPuncttrue
\mciteSetBstMidEndSepPunct{\mcitedefaultmidpunct}
{\mcitedefaultendpunct}{\mcitedefaultseppunct}\relax
\EndOfBibitem
\bibitem[Muegge and Knapp(1995)]{Muegge1995}
I.~Muegge and E.~W. Knapp, \emph{Physical Chemistry The Journal}, 1995,
  \textbf{0}, year\relax
\mciteBstWouldAddEndPuncttrue
\mciteSetBstMidEndSepPunct{\mcitedefaultmidpunct}
{\mcitedefaultendpunct}{\mcitedefaultseppunct}\relax
\EndOfBibitem
\bibitem[Sedlmeier \emph{et~al.}(2011)Sedlmeier, von Hansen, Mengyu, Horinek,
  and Netz]{Sedlmeier2011}
F.~Sedlmeier, Y.~von Hansen, L.~Mengyu, D.~Horinek and R.~R. Netz,
  \emph{Journal of Statistical Physics}, 2011, \textbf{145}, 240--252\relax
\mciteBstWouldAddEndPuncttrue
\mciteSetBstMidEndSepPunct{\mcitedefaultmidpunct}
{\mcitedefaultendpunct}{\mcitedefaultseppunct}\relax
\EndOfBibitem
\bibitem[Amann-Winkel \emph{et~al.}(2016)Amann-Winkel, Bellissent-Funel, Bove,
  Loerting, Nilsson, Paciaroni, Schlesinger, and Skinner]{amann2016x}
K.~Amann-Winkel, M.-C. Bellissent-Funel, L.~E. Bove, T.~Loerting, A.~Nilsson,
  A.~Paciaroni, D.~Schlesinger and L.~Skinner, \emph{Chemical reviews}, 2016,
  \textbf{116}, 7570--7589\relax
\mciteBstWouldAddEndPuncttrue
\mciteSetBstMidEndSepPunct{\mcitedefaultmidpunct}
{\mcitedefaultendpunct}{\mcitedefaultseppunct}\relax
\EndOfBibitem
\bibitem[Pearson and Pascher(1979)]{pearson1979molecular}
R.~H. Pearson and I.~Pascher, \emph{Nature}, 1979, \textbf{281}, 499\relax
\mciteBstWouldAddEndPuncttrue
\mciteSetBstMidEndSepPunct{\mcitedefaultmidpunct}
{\mcitedefaultendpunct}{\mcitedefaultseppunct}\relax
\EndOfBibitem
\bibitem[Finer-Moore \emph{et~al.}(1992)Finer-Moore, Kossiakoff, Hurley,
  Earnest, and Stroud]{finer1992solvent}
J.~S. Finer-Moore, A.~A. Kossiakoff, J.~H. Hurley, T.~Earnest and R.~M. Stroud,
  \emph{Proteins: Structure, Function, and Bioinformatics}, 1992, \textbf{12},
  203--222\relax
\mciteBstWouldAddEndPuncttrue
\mciteSetBstMidEndSepPunct{\mcitedefaultmidpunct}
{\mcitedefaultendpunct}{\mcitedefaultseppunct}\relax
\EndOfBibitem
\bibitem[Svergun \emph{et~al.}(1998)Svergun, Richard, Koch, Sayers, Kuprin, and
  Zaccai]{svergun1998protein}
D.~Svergun, S.~Richard, M.~Koch, Z.~Sayers, S.~Kuprin and G.~Zaccai,
  \emph{Proceedings of the National Academy of Sciences}, 1998, \textbf{95},
  2267--2272\relax
\mciteBstWouldAddEndPuncttrue
\mciteSetBstMidEndSepPunct{\mcitedefaultmidpunct}
{\mcitedefaultendpunct}{\mcitedefaultseppunct}\relax
\EndOfBibitem
\bibitem[Merzel and Smith(2002)]{merzel2002first}
F.~Merzel and J.~C. Smith, \emph{Proceedings of the National Academy of
  Sciences}, 2002, \textbf{99}, 5378--5383\relax
\mciteBstWouldAddEndPuncttrue
\mciteSetBstMidEndSepPunct{\mcitedefaultmidpunct}
{\mcitedefaultendpunct}{\mcitedefaultseppunct}\relax
\EndOfBibitem
\bibitem[Otting \emph{et~al.}(1991)Otting, Liepinsh, and
  Wuthrich]{otting1991protein}
G.~Otting, E.~Liepinsh and K.~Wuthrich, \emph{Science}, 1991, \textbf{254},
  974--980\relax
\mciteBstWouldAddEndPuncttrue
\mciteSetBstMidEndSepPunct{\mcitedefaultmidpunct}
{\mcitedefaultendpunct}{\mcitedefaultseppunct}\relax
\EndOfBibitem
\bibitem[Liepinsh \emph{et~al.}(1992)Liepinsh, Otting, and
  W{\"u}thrich]{liepinsh1992nmr}
E.~Liepinsh, G.~Otting and K.~W{\"u}thrich, \emph{Nucleic acids research},
  1992, \textbf{20}, 6549--6553\relax
\mciteBstWouldAddEndPuncttrue
\mciteSetBstMidEndSepPunct{\mcitedefaultmidpunct}
{\mcitedefaultendpunct}{\mcitedefaultseppunct}\relax
\EndOfBibitem
\bibitem[Nucci \emph{et~al.}(2011)Nucci, Pometun, and Wand]{nucci2011mapping}
N.~V. Nucci, M.~S. Pometun and A.~J. Wand, \emph{Journal of the American
  Chemical Society}, 2011, \textbf{133}, 12326--12329\relax
\mciteBstWouldAddEndPuncttrue
\mciteSetBstMidEndSepPunct{\mcitedefaultmidpunct}
{\mcitedefaultendpunct}{\mcitedefaultseppunct}\relax
\EndOfBibitem
\bibitem[Nucci \emph{et~al.}(2011)Nucci, Pometun, and Wand]{nucci2011site}
N.~V. Nucci, M.~S. Pometun and A.~J. Wand, \emph{Nature structural \& molecular
  biology}, 2011, \textbf{18}, 245\relax
\mciteBstWouldAddEndPuncttrue
\mciteSetBstMidEndSepPunct{\mcitedefaultmidpunct}
{\mcitedefaultendpunct}{\mcitedefaultseppunct}\relax
\EndOfBibitem
\bibitem[Brotzakis \emph{et~al.}(2016)Brotzakis, Groot, Brandeburgo, Bakker,
  and Bolhuis]{brotzakis2016dynamics}
Z.~Brotzakis, C.~Groot, W.~H. Brandeburgo, H.~Bakker and P.~Bolhuis, \emph{The
  Journal of Physical Chemistry B}, 2016, \textbf{120}, 4756--4766\relax
\mciteBstWouldAddEndPuncttrue
\mciteSetBstMidEndSepPunct{\mcitedefaultmidpunct}
{\mcitedefaultendpunct}{\mcitedefaultseppunct}\relax
\EndOfBibitem
\bibitem[Russo \emph{et~al.}(2004)Russo, Hura, and
  Head-Gordon]{russo2004hydration}
D.~Russo, G.~Hura and T.~Head-Gordon, \emph{Biophysical journal}, 2004,
  \textbf{86}, 1852--1862\relax
\mciteBstWouldAddEndPuncttrue
\mciteSetBstMidEndSepPunct{\mcitedefaultmidpunct}
{\mcitedefaultendpunct}{\mcitedefaultseppunct}\relax
\EndOfBibitem
\bibitem[Wood \emph{et~al.}(2007)Wood, Plazanet, Gabel, Kessler, Oesterhelt,
  Tobias, Zaccai, and Weik]{wood2007coupling}
K.~Wood, M.~Plazanet, F.~Gabel, B.~Kessler, D.~Oesterhelt, D.~Tobias, G.~Zaccai
  and M.~Weik, \emph{Proceedings of the National Academy of Sciences}, 2007,
  \textbf{104}, 18049--18054\relax
\mciteBstWouldAddEndPuncttrue
\mciteSetBstMidEndSepPunct{\mcitedefaultmidpunct}
{\mcitedefaultendpunct}{\mcitedefaultseppunct}\relax
\EndOfBibitem
\bibitem[Schir{\`o} \emph{et~al.}(2015)Schir{\`o}, Fichou, Gallat, Wood, Gabel,
  Moulin, H{\"a}rtlein, Heyden, Colletier,
  Orecchini,\emph{et~al.}]{schiro2015translational}
G.~Schir{\`o}, Y.~Fichou, F.-X. Gallat, K.~Wood, F.~Gabel, M.~Moulin,
  M.~H{\"a}rtlein, M.~Heyden, J.-P. Colletier, A.~Orecchini \emph{et~al.},
  \emph{Nature communications}, 2015, \textbf{6}, 6490\relax
\mciteBstWouldAddEndPuncttrue
\mciteSetBstMidEndSepPunct{\mcitedefaultmidpunct}
{\mcitedefaultendpunct}{\mcitedefaultseppunct}\relax
\EndOfBibitem
\bibitem[Russo \emph{et~al.}(2005)Russo, Murarka, Copley, and
  Head-Gordon]{russo2005molecular}
D.~Russo, R.~K. Murarka, J.~R. Copley and T.~Head-Gordon, \emph{The Journal of
  Physical Chemistry B}, 2005, \textbf{109}, 12966--12975\relax
\mciteBstWouldAddEndPuncttrue
\mciteSetBstMidEndSepPunct{\mcitedefaultmidpunct}
{\mcitedefaultendpunct}{\mcitedefaultseppunct}\relax
\EndOfBibitem
\bibitem[Bizzarri and Cannistraro(2002)]{bizzarri2002molecular}
A.~R. Bizzarri and S.~Cannistraro, \emph{Molecular dynamics of water at the
  protein- solvent interface}, 2002\relax
\mciteBstWouldAddEndPuncttrue
\mciteSetBstMidEndSepPunct{\mcitedefaultmidpunct}
{\mcitedefaultendpunct}{\mcitedefaultseppunct}\relax
\EndOfBibitem
\bibitem[Pizzitutti \emph{et~al.}(2007)Pizzitutti, Marchi, Sterpone, and
  Rossky]{pizzitutti2007protein}
F.~Pizzitutti, M.~Marchi, F.~Sterpone and P.~J. Rossky, \emph{The Journal of
  Physical Chemistry B}, 2007, \textbf{111}, 7584--7590\relax
\mciteBstWouldAddEndPuncttrue
\mciteSetBstMidEndSepPunct{\mcitedefaultmidpunct}
{\mcitedefaultendpunct}{\mcitedefaultseppunct}\relax
\EndOfBibitem
\bibitem[Makarov \emph{et~al.}(2000)Makarov, Andrews, Smith, and
  Pettitt]{makarov2000residence}
V.~A. Makarov, B.~K. Andrews, P.~E. Smith and B.~M. Pettitt, \emph{Biophysical
  journal}, 2000, \textbf{79}, 2966--2974\relax
\mciteBstWouldAddEndPuncttrue
\mciteSetBstMidEndSepPunct{\mcitedefaultmidpunct}
{\mcitedefaultendpunct}{\mcitedefaultseppunct}\relax
\EndOfBibitem
\bibitem[Marchi \emph{et~al.}(2002)Marchi, Sterpone, and
  Ceccarelli]{marchi2002water}
M.~Marchi, F.~Sterpone and M.~Ceccarelli, \emph{Journal of the American
  Chemical Society}, 2002, \textbf{124}, 6787--6791\relax
\mciteBstWouldAddEndPuncttrue
\mciteSetBstMidEndSepPunct{\mcitedefaultmidpunct}
{\mcitedefaultendpunct}{\mcitedefaultseppunct}\relax
\EndOfBibitem
\bibitem[Henchman and McCammon(2002)]{henchman2002structural}
R.~H. Henchman and J.~A. McCammon, \emph{Protein Science}, 2002, \textbf{11},
  2080--2090\relax
\mciteBstWouldAddEndPuncttrue
\mciteSetBstMidEndSepPunct{\mcitedefaultmidpunct}
{\mcitedefaultendpunct}{\mcitedefaultseppunct}\relax
\EndOfBibitem
\bibitem[Li \emph{et~al.}(2007)Li, Hassanali, Kao, Zhong, and
  Singer]{li2007hydration}
T.~Li, A.~A. Hassanali, Y.-T. Kao, D.~Zhong and S.~J. Singer, \emph{Journal of
  the American Chemical Society}, 2007, \textbf{129}, 3376--3382\relax
\mciteBstWouldAddEndPuncttrue
\mciteSetBstMidEndSepPunct{\mcitedefaultmidpunct}
{\mcitedefaultendpunct}{\mcitedefaultseppunct}\relax
\EndOfBibitem
\bibitem[Fogarty and Laage(2014)]{fogarty2014water}
A.~C. Fogarty and D.~Laage, \emph{The Journal of Physical Chemistry B}, 2014,
  \textbf{118}, 7715--7729\relax
\mciteBstWouldAddEndPuncttrue
\mciteSetBstMidEndSepPunct{\mcitedefaultmidpunct}
{\mcitedefaultendpunct}{\mcitedefaultseppunct}\relax
\EndOfBibitem
\bibitem[Luise \emph{et~al.}(2000)Luise, Falconi, and
  Desideri]{luise2000molecular}
A.~Luise, M.~Falconi and A.~Desideri, \emph{Proteins: Structure, Function, and
  Bioinformatics}, 2000, \textbf{39}, 56--67\relax
\mciteBstWouldAddEndPuncttrue
\mciteSetBstMidEndSepPunct{\mcitedefaultmidpunct}
{\mcitedefaultendpunct}{\mcitedefaultseppunct}\relax
\EndOfBibitem
\bibitem[Rossky and Karplus(1979)]{rossky1979solvation}
P.~J. Rossky and M.~Karplus, \emph{Journal of the American Chemical Society},
  1979, \textbf{101}, 1913--1937\relax
\mciteBstWouldAddEndPuncttrue
\mciteSetBstMidEndSepPunct{\mcitedefaultmidpunct}
{\mcitedefaultendpunct}{\mcitedefaultseppunct}\relax
\EndOfBibitem
\bibitem[Heyden and Tobias(2013)]{heyden2013spatial}
M.~Heyden and D.~J. Tobias, \emph{Physical review letters}, 2013, \textbf{111},
  218101\relax
\mciteBstWouldAddEndPuncttrue
\mciteSetBstMidEndSepPunct{\mcitedefaultmidpunct}
{\mcitedefaultendpunct}{\mcitedefaultseppunct}\relax
\EndOfBibitem
\bibitem[Garc{\'\i}a and Hummer(2000)]{garcia2000water}
A.~E. Garc{\'\i}a and G.~Hummer, \emph{Proteins: Structure, Function, and
  Bioinformatics}, 2000, \textbf{38}, 261--272\relax
\mciteBstWouldAddEndPuncttrue
\mciteSetBstMidEndSepPunct{\mcitedefaultmidpunct}
{\mcitedefaultendpunct}{\mcitedefaultseppunct}\relax
\EndOfBibitem
\bibitem[Laage \emph{et~al.}(2009)Laage, Stirnemann, and Hynes]{laage2009water}
D.~Laage, G.~Stirnemann and J.~T. Hynes, \emph{The Journal of Physical
  Chemistry B}, 2009, \textbf{113}, 2428--2435\relax
\mciteBstWouldAddEndPuncttrue
\mciteSetBstMidEndSepPunct{\mcitedefaultmidpunct}
{\mcitedefaultendpunct}{\mcitedefaultseppunct}\relax
\EndOfBibitem
\bibitem[Bellissent-Funel \emph{et~al.}(2016)Bellissent-Funel, Hassanali,
  Havenith, Henchman, Pohl, Sterpone, van~der Spoel, Xu, and
  Garcia]{bellissent2016water}
M.-C. Bellissent-Funel, A.~Hassanali, M.~Havenith, R.~Henchman, P.~Pohl,
  F.~Sterpone, D.~van~der Spoel, Y.~Xu and A.~E. Garcia, \emph{Chemical
  Reviews}, 2016, \textbf{116}, 7673--7697\relax
\mciteBstWouldAddEndPuncttrue
\mciteSetBstMidEndSepPunct{\mcitedefaultmidpunct}
{\mcitedefaultendpunct}{\mcitedefaultseppunct}\relax
\EndOfBibitem
\bibitem[Rani and Biswas(2015)]{rani2015diffusion}
P.~Rani and P.~Biswas, \emph{The Journal of Physical Chemistry B}, 2015,
  \textbf{119}, 13262--13270\relax
\mciteBstWouldAddEndPuncttrue
\mciteSetBstMidEndSepPunct{\mcitedefaultmidpunct}
{\mcitedefaultendpunct}{\mcitedefaultseppunct}\relax
\EndOfBibitem
\bibitem[Pronk \emph{et~al.}(2014)Pronk, Lindahl, and Kasson]{pronk2014dynamic}
S.~Pronk, E.~Lindahl and P.~M. Kasson, \emph{Nature communications}, 2014,
  \textbf{5}, 3034\relax
\mciteBstWouldAddEndPuncttrue
\mciteSetBstMidEndSepPunct{\mcitedefaultmidpunct}
{\mcitedefaultendpunct}{\mcitedefaultseppunct}\relax
\EndOfBibitem
\bibitem[Dellerue and Bellissent-Funel(2000)]{dellerue2000relaxational}
S.~Dellerue and M.-C. Bellissent-Funel, \emph{Chemical Physics}, 2000,
  \textbf{258}, 315--325\relax
\mciteBstWouldAddEndPuncttrue
\mciteSetBstMidEndSepPunct{\mcitedefaultmidpunct}
{\mcitedefaultendpunct}{\mcitedefaultseppunct}\relax
\EndOfBibitem
\bibitem[Laage and Hynes(2007)]{laage2007reorientional}
D.~Laage and J.~T. Hynes, \emph{Proceedings of the National Academy of Sciences
  of the United States of America}, 2007, \textbf{104}, 11167--11172\relax
\mciteBstWouldAddEndPuncttrue
\mciteSetBstMidEndSepPunct{\mcitedefaultmidpunct}
{\mcitedefaultendpunct}{\mcitedefaultseppunct}\relax
\EndOfBibitem
\bibitem[Laage \emph{et~al.}(2017)Laage, Elsaesser, and Hynes]{laagehynes2017}
D.~Laage, T.~Elsaesser and J.~T. Hynes, \emph{Chemical Reviews}, 2017,
  \textbf{117}, 10694--10725\relax
\mciteBstWouldAddEndPuncttrue
\mciteSetBstMidEndSepPunct{\mcitedefaultmidpunct}
{\mcitedefaultendpunct}{\mcitedefaultseppunct}\relax
\EndOfBibitem
\bibitem[Ball(2008)]{PhilipBall2008}
P.~Ball, \emph{Chemical Reviews}, 2008, \textbf{108}, 74--108\relax
\mciteBstWouldAddEndPuncttrue
\mciteSetBstMidEndSepPunct{\mcitedefaultmidpunct}
{\mcitedefaultendpunct}{\mcitedefaultseppunct}\relax
\EndOfBibitem
\bibitem[Levy and Onuchic(2006)]{Levy2006}
Y.~Levy and J.~N. Onuchic, \emph{Annual Review of Biophysics and Biomolecular
  Structure}, 2006, \textbf{35}, 389--415\relax
\mciteBstWouldAddEndPuncttrue
\mciteSetBstMidEndSepPunct{\mcitedefaultmidpunct}
{\mcitedefaultendpunct}{\mcitedefaultseppunct}\relax
\EndOfBibitem
\bibitem[Che(2002)]{Chen2002}
\emph{Proceedings of the National Academy of Sciences of the United States of
  America}, 2002, \textbf{99}, 11884--11889\relax
\mciteBstWouldAddEndPuncttrue
\mciteSetBstMidEndSepPunct{\mcitedefaultmidpunct}
{\mcitedefaultendpunct}{\mcitedefaultseppunct}\relax
\EndOfBibitem
\bibitem[Nelson \emph{et~al.}(2005)Nelson, Sawaya, Balbirnie, Madsen, Riekel,
  Grothe, and Eisenberg]{eisenberg2005}
R.~Nelson, M.~R. Sawaya, M.~Balbirnie, A.~{\O}. Madsen, C.~Riekel, R.~Grothe
  and D.~Eisenberg, \emph{Nature}, 2005, \textbf{435}, 773--778\relax
\mciteBstWouldAddEndPuncttrue
\mciteSetBstMidEndSepPunct{\mcitedefaultmidpunct}
{\mcitedefaultendpunct}{\mcitedefaultseppunct}\relax
\EndOfBibitem
\bibitem[Fitzpatrick \emph{et~al.}(2013)Fitzpatrick, Debelouchina, Bayro,
  Clare, Caporini, Bajaj, Jaroniec, Wang, Ladizhansky, M{\"u}ller, MacPhee,
  Waudby, Mott, De~Simone, Knowles, Saibil, Vendruscolo, Orlova, Griffin, and
  Dobson]{Fitzpatrick5468}
A.~W.~P. Fitzpatrick, G.~T. Debelouchina, M.~J. Bayro, D.~K. Clare, M.~A.
  Caporini, V.~S. Bajaj, C.~P. Jaroniec, L.~Wang, V.~Ladizhansky, S.~A.
  M{\"u}ller, C.~E. MacPhee, C.~A. Waudby, H.~R. Mott, A.~De~Simone, T.~P.~J.
  Knowles, H.~R. Saibil, M.~Vendruscolo, E.~V. Orlova, R.~G. Griffin and C.~M.
  Dobson, \emph{Proceedings of the National Academy of Sciences}, 2013,
  \textbf{110}, 5468--5473\relax
\mciteBstWouldAddEndPuncttrue
\mciteSetBstMidEndSepPunct{\mcitedefaultmidpunct}
{\mcitedefaultendpunct}{\mcitedefaultseppunct}\relax
\EndOfBibitem
\bibitem[W{\"a}lti \emph{et~al.}(2016)W{\"a}lti, Ravotti, Arai, Glabe, Wall,
  B{\"o}ckmann, G{\"u}ntert, Meier, and Riek]{roland2016}
M.~A. W{\"a}lti, F.~Ravotti, H.~Arai, C.~G. Glabe, J.~S. Wall, A.~B{\"o}ckmann,
  P.~G{\"u}ntert, B.~H. Meier and R.~Riek, \emph{Proceedings of the National
  Academy of Sciences}, 2016, \textbf{113}, E4976--E4984\relax
\mciteBstWouldAddEndPuncttrue
\mciteSetBstMidEndSepPunct{\mcitedefaultmidpunct}
{\mcitedefaultendpunct}{\mcitedefaultseppunct}\relax
\EndOfBibitem
\bibitem[Shimanovich \emph{et~al.}(2018)Shimanovich, Pinotsi, Shimanovich, Yu,
  Bolisetty, Adamcik, Mezzenga, Charmet, Vollrath, Gazit, Dobson, Schierle,
  Holland, Kaminski, and Knowles]{gazitsilk}
U.~Shimanovich, D.~Pinotsi, K.~Shimanovich, N.~Yu, S.~Bolisetty, J.~Adamcik,
  R.~Mezzenga, J.~Charmet, F.~Vollrath, E.~Gazit, C.~M. Dobson, G.~K. Schierle,
  C.~Holland, C.~F. Kaminski and T.~P.~J. Knowles, \emph{Macromolecular
  Bioscience}, 2018, \textbf{18}, 1700295\relax
\mciteBstWouldAddEndPuncttrue
\mciteSetBstMidEndSepPunct{\mcitedefaultmidpunct}
{\mcitedefaultendpunct}{\mcitedefaultseppunct}\relax
\EndOfBibitem
\bibitem[Pinotsi \emph{et~al.}(2016)Pinotsi, Grisanti, Mahou, Gebauer,
  Kaminski, Hassanali, and Kaminski~Schierle]{hassanali1}
D.~Pinotsi, L.~Grisanti, P.~Mahou, R.~Gebauer, C.~F. Kaminski, A.~Hassanali and
  G.~S. Kaminski~Schierle, \emph{Journal of the American Chemical Society},
  2016, \textbf{138}, 3046--3057\relax
\mciteBstWouldAddEndPuncttrue
\mciteSetBstMidEndSepPunct{\mcitedefaultmidpunct}
{\mcitedefaultendpunct}{\mcitedefaultseppunct}\relax
\EndOfBibitem
\bibitem[Jong \emph{et~al.}(2019)Jong, Azar, Grisanti, Stephens, Jones,
  Credgington, Kaminski~Schierle, and Hassanali]{hassanali2}
K.~H. Jong, Y.~T. Azar, L.~Grisanti, A.~D. Stephens, S.~T.~E. Jones,
  D.~Credgington, G.~S. Kaminski~Schierle and A.~Hassanali, \emph{Phys. Chem.
  Chem. Phys.}, 2019, \textbf{21}, 23931--23942\relax
\mciteBstWouldAddEndPuncttrue
\mciteSetBstMidEndSepPunct{\mcitedefaultmidpunct}
{\mcitedefaultendpunct}{\mcitedefaultseppunct}\relax
\EndOfBibitem
\bibitem[Stephens \emph{et~al.}(2020)Stephens, Qaisrani, Ruggiero, Jones, Poli,
  Bond, Woodhams, Kleist, Grisanti, Gebauer, Zeitler, Credgington, Hassanali,
  and Kaminski~Schierle]{hassanali3}
A.~D. Stephens, M.~N. Qaisrani, M.~T. Ruggiero, S.~T. Jones, E.~Poli, A.~D.
  Bond, P.~J. Woodhams, E.~M. Kleist, L.~Grisanti, R.~Gebauer, J.~A. Zeitler,
  D.~Credgington, A.~Hassanali and G.~S. Kaminski~Schierle, \emph{bioRxiv},
  2020\relax
\mciteBstWouldAddEndPuncttrue
\mciteSetBstMidEndSepPunct{\mcitedefaultmidpunct}
{\mcitedefaultendpunct}{\mcitedefaultseppunct}\relax
\EndOfBibitem
\bibitem[Grisanti \emph{et~al.}(2017)Grisanti, Pinotsi, Gebauer,
  Kaminski~Schierle, and Hassanali]{hassanali4}
L.~Grisanti, D.~Pinotsi, R.~Gebauer, G.~S. Kaminski~Schierle and A.~A.
  Hassanali, \emph{Phys. Chem. Chem. Phys.}, 2017, \textbf{19},
  4030--4040\relax
\mciteBstWouldAddEndPuncttrue
\mciteSetBstMidEndSepPunct{\mcitedefaultmidpunct}
{\mcitedefaultendpunct}{\mcitedefaultseppunct}\relax
\EndOfBibitem
\bibitem[Wang \emph{et~al.}(2017)Wang, Jo, DeGrado, and Hong]{wang2017}
T.~Wang, H.~Jo, W.~F. DeGrado and M.~Hong, \emph{Journal of the American
  Chemical Society}, 2017, \textbf{139}, 6242--6252\relax
\mciteBstWouldAddEndPuncttrue
\mciteSetBstMidEndSepPunct{\mcitedefaultmidpunct}
{\mcitedefaultendpunct}{\mcitedefaultseppunct}\relax
\EndOfBibitem
\bibitem[Liu \emph{et~al.}(2004)Liu, Harder, and Berne]{Liu2004}
P.~Liu, E.~Harder and B.~J. Berne, \emph{Journal of Physical Chemistry B},
  2004, \textbf{108}, 6595--6602\relax
\mciteBstWouldAddEndPuncttrue
\mciteSetBstMidEndSepPunct{\mcitedefaultmidpunct}
{\mcitedefaultendpunct}{\mcitedefaultseppunct}\relax
\EndOfBibitem
\bibitem[Hummer(2005)]{Hummer2005}
G.~Hummer, \emph{New Journal of Physics}, 2005, \textbf{7}, year\relax
\mciteBstWouldAddEndPuncttrue
\mciteSetBstMidEndSepPunct{\mcitedefaultmidpunct}
{\mcitedefaultendpunct}{\mcitedefaultseppunct}\relax
\EndOfBibitem
\bibitem[Olivares-Rivas \emph{et~al.}(2013)Olivares-Rivas, Colmenares, and
  L{\'{o}}pez]{OlivaresRivas2013}
W.~Olivares-Rivas, P.~J. Colmenares and F.~L{\'{o}}pez, \emph{The Journal of
  Chemical Physics}, 2013, \textbf{139}, 074103\relax
\mciteBstWouldAddEndPuncttrue
\mciteSetBstMidEndSepPunct{\mcitedefaultmidpunct}
{\mcitedefaultendpunct}{\mcitedefaultseppunct}\relax
\EndOfBibitem
\bibitem[Qaisrani \emph{et~al.}(2019)Qaisrani, Grisanti, Gebauer, and
  Hassanali]{Qaisrani2019}
M.~N. Qaisrani, L.~Grisanti, R.~Gebauer and A.~Hassanali, \emph{Physical
  Chemistry Chemical Physics}, 2019, \textbf{21}, 16083--16094\relax
\mciteBstWouldAddEndPuncttrue
\mciteSetBstMidEndSepPunct{\mcitedefaultmidpunct}
{\mcitedefaultendpunct}{\mcitedefaultseppunct}\relax
\EndOfBibitem
\bibitem[Redner(2001)]{redner2001guide}
S.~Redner, \emph{A guide to first-passage processes}, Cambridge University
  Press, 2001\relax
\mciteBstWouldAddEndPuncttrue
\mciteSetBstMidEndSepPunct{\mcitedefaultmidpunct}
{\mcitedefaultendpunct}{\mcitedefaultseppunct}\relax
\EndOfBibitem
\bibitem[Metzler \emph{et~al.}(2014)Metzler, Oshanin, and
  Redner]{metzler2014first}
R.~Metzler, G.~Oshanin and S.~Redner, \emph{First-passage phenomena and their
  applications}, World Scientific, 2014\relax
\mciteBstWouldAddEndPuncttrue
\mciteSetBstMidEndSepPunct{\mcitedefaultmidpunct}
{\mcitedefaultendpunct}{\mcitedefaultseppunct}\relax
\EndOfBibitem
\bibitem[Masoliver \emph{et~al.}(1987)Masoliver, West, and
  Lindenberg]{masoliver1987bistability}
J.~Masoliver, B.~J. West and K.~Lindenberg, \emph{Physical Review A}, 1987,
  \textbf{35}, 3086\relax
\mciteBstWouldAddEndPuncttrue
\mciteSetBstMidEndSepPunct{\mcitedefaultmidpunct}
{\mcitedefaultendpunct}{\mcitedefaultseppunct}\relax
\EndOfBibitem
\bibitem[H{\"a}nggi \emph{et~al.}(1990)H{\"a}nggi, Talkner, and
  Borkovec]{hanggi1990reaction}
P.~H{\"a}nggi, P.~Talkner and M.~Borkovec, \emph{Reviews of modern physics},
  1990, \textbf{62}, 251\relax
\mciteBstWouldAddEndPuncttrue
\mciteSetBstMidEndSepPunct{\mcitedefaultmidpunct}
{\mcitedefaultendpunct}{\mcitedefaultseppunct}\relax
\EndOfBibitem
\bibitem[Sokolov(2003)]{sokolov2003cyclization}
I.~Sokolov, \emph{Physical review letters}, 2003, \textbf{90}, 080601\relax
\mciteBstWouldAddEndPuncttrue
\mciteSetBstMidEndSepPunct{\mcitedefaultmidpunct}
{\mcitedefaultendpunct}{\mcitedefaultseppunct}\relax
\EndOfBibitem
\bibitem[Condamin \emph{et~al.}(2007)Condamin, B{\'e}nichou, Tejedor,
  Voituriez, and Klafter]{condamin2007first}
S.~Condamin, O.~B{\'e}nichou, V.~Tejedor, R.~Voituriez and J.~Klafter,
  \emph{Nature}, 2007, \textbf{450}, 77--80\relax
\mciteBstWouldAddEndPuncttrue
\mciteSetBstMidEndSepPunct{\mcitedefaultmidpunct}
{\mcitedefaultendpunct}{\mcitedefaultseppunct}\relax
\EndOfBibitem
\bibitem[Koren \emph{et~al.}(2007)Koren, Lomholt, Chechkin, Klafter, and
  Metzler]{koren2007leapover}
T.~Koren, M.~A. Lomholt, A.~V. Chechkin, J.~Klafter and R.~Metzler,
  \emph{Physical review letters}, 2007, \textbf{99}, 160602\relax
\mciteBstWouldAddEndPuncttrue
\mciteSetBstMidEndSepPunct{\mcitedefaultmidpunct}
{\mcitedefaultendpunct}{\mcitedefaultseppunct}\relax
\EndOfBibitem
\bibitem[Mattos \emph{et~al.}(2012)Mattos, Mej{\'\i}a-Monasterio, Metzler, and
  Oshanin]{mattos2012first}
T.~G. Mattos, C.~Mej{\'\i}a-Monasterio, R.~Metzler and G.~Oshanin,
  \emph{Physical Review E}, 2012, \textbf{86}, 031143\relax
\mciteBstWouldAddEndPuncttrue
\mciteSetBstMidEndSepPunct{\mcitedefaultmidpunct}
{\mcitedefaultendpunct}{\mcitedefaultseppunct}\relax
\EndOfBibitem
\bibitem[Bray \emph{et~al.}(2013)Bray, Majumdar, and
  Schehr]{bray2013persistence}
A.~J. Bray, S.~N. Majumdar and G.~Schehr, \emph{Advances in Physics}, 2013,
  \textbf{62}, 225--361\relax
\mciteBstWouldAddEndPuncttrue
\mciteSetBstMidEndSepPunct{\mcitedefaultmidpunct}
{\mcitedefaultendpunct}{\mcitedefaultseppunct}\relax
\EndOfBibitem
\bibitem[Pal and Reuveni(2017)]{pal2017first}
A.~Pal and S.~Reuveni, \emph{Physical review letters}, 2017, \textbf{118},
  030603\relax
\mciteBstWouldAddEndPuncttrue
\mciteSetBstMidEndSepPunct{\mcitedefaultmidpunct}
{\mcitedefaultendpunct}{\mcitedefaultseppunct}\relax
\EndOfBibitem
\bibitem[Hartich and Godec(2019)]{hartich2019interlacing}
D.~Hartich and A.~Godec, \emph{Journal of Statistical Mechanics: Theory and
  Experiment}, 2019, \textbf{2019}, 024002\relax
\mciteBstWouldAddEndPuncttrue
\mciteSetBstMidEndSepPunct{\mcitedefaultmidpunct}
{\mcitedefaultendpunct}{\mcitedefaultseppunct}\relax
\EndOfBibitem
\bibitem[Szabo \emph{et~al.}(1980)Szabo, Schulten, and
  Schulten]{szabo1980first}
A.~Szabo, K.~Schulten and Z.~Schulten, \emph{The Journal of chemical physics},
  1980, \textbf{72}, 4350--4357\relax
\mciteBstWouldAddEndPuncttrue
\mciteSetBstMidEndSepPunct{\mcitedefaultmidpunct}
{\mcitedefaultendpunct}{\mcitedefaultseppunct}\relax
\EndOfBibitem
\bibitem[Galburt \emph{et~al.}(2007)Galburt, Grill, Wiedmann, Lubkowska, Choy,
  Nogales, Kashlev, and Bustamante]{galburt2007backtracking}
E.~A. Galburt, S.~W. Grill, A.~Wiedmann, L.~Lubkowska, J.~Choy, E.~Nogales,
  M.~Kashlev and C.~Bustamante, \emph{Nature}, 2007, \textbf{446},
  820--823\relax
\mciteBstWouldAddEndPuncttrue
\mciteSetBstMidEndSepPunct{\mcitedefaultmidpunct}
{\mcitedefaultendpunct}{\mcitedefaultseppunct}\relax
\EndOfBibitem
\bibitem[Rold{\'a}n \emph{et~al.}(2016)Rold{\'a}n, Lisica,
  S{\'a}nchez-Taltavull, and Grill]{roldan2016stochastic}
{\'E}.~Rold{\'a}n, A.~Lisica, D.~S{\'a}nchez-Taltavull and S.~W. Grill,
  \emph{Physical Review E}, 2016, \textbf{93}, 062411\relax
\mciteBstWouldAddEndPuncttrue
\mciteSetBstMidEndSepPunct{\mcitedefaultmidpunct}
{\mcitedefaultendpunct}{\mcitedefaultseppunct}\relax
\EndOfBibitem
\bibitem[Yang \emph{et~al.}(2017)Yang, Liu, Li, Marchesoni, H{\"a}nggi, and
  Zhang]{yang2017hydrodynamic}
X.~Yang, C.~Liu, Y.~Li, F.~Marchesoni, P.~H{\"a}nggi and H.~Zhang,
  \emph{Proceedings of the National Academy of Sciences}, 2017, \textbf{114},
  9564--9569\relax
\mciteBstWouldAddEndPuncttrue
\mciteSetBstMidEndSepPunct{\mcitedefaultmidpunct}
{\mcitedefaultendpunct}{\mcitedefaultseppunct}\relax
\EndOfBibitem
\bibitem[Gladrow \emph{et~al.}(2019)Gladrow, Ribezzi-Crivellari, Ritort, and
  Keyser]{gladrow2019experimental}
J.~Gladrow, M.~Ribezzi-Crivellari, F.~Ritort and U.~F. Keyser, \emph{Nature
  communications}, 2019, \textbf{10}, 1--9\relax
\mciteBstWouldAddEndPuncttrue
\mciteSetBstMidEndSepPunct{\mcitedefaultmidpunct}
{\mcitedefaultendpunct}{\mcitedefaultseppunct}\relax
\EndOfBibitem
\bibitem[Chandrasekhar(1943)]{chandrasekhar1943dynamical}
S.~Chandrasekhar, \emph{The Astrophysical Journal}, 1943, \textbf{97},
  263\relax
\mciteBstWouldAddEndPuncttrue
\mciteSetBstMidEndSepPunct{\mcitedefaultmidpunct}
{\mcitedefaultendpunct}{\mcitedefaultseppunct}\relax
\EndOfBibitem
\bibitem[Majumdar(2005)]{majumdar2005brownian}
S.~N. Majumdar, \emph{Curr. Sci.}, 2005, \textbf{88}, 2076--2092\relax
\mciteBstWouldAddEndPuncttrue
\mciteSetBstMidEndSepPunct{\mcitedefaultmidpunct}
{\mcitedefaultendpunct}{\mcitedefaultseppunct}\relax
\EndOfBibitem
\bibitem[Wergen \emph{et~al.}(2011)Wergen, Bogner, and Krug]{wergen2011record}
G.~Wergen, M.~Bogner and J.~Krug, \emph{Physical Review E}, 2011, \textbf{83},
  051109\relax
\mciteBstWouldAddEndPuncttrue
\mciteSetBstMidEndSepPunct{\mcitedefaultmidpunct}
{\mcitedefaultendpunct}{\mcitedefaultseppunct}\relax
\EndOfBibitem
\bibitem[Singh \emph{et~al.}(2019)Singh, Menczel, Golubev, Khaymovich,
  Peltonen, Flindt, Saito, Rold{\'a}n, and Pekola]{singh2019universal}
S.~Singh, P.~Menczel, D.~S. Golubev, I.~M. Khaymovich, J.~T. Peltonen,
  C.~Flindt, K.~Saito, {\'E}.~Rold{\'a}n and J.~P. Pekola, \emph{Physical
  review letters}, 2019, \textbf{122}, 230602\relax
\mciteBstWouldAddEndPuncttrue
\mciteSetBstMidEndSepPunct{\mcitedefaultmidpunct}
{\mcitedefaultendpunct}{\mcitedefaultseppunct}\relax
\EndOfBibitem
\bibitem[Berezhkovskii and Hummer(2002)]{berezhkovskii2002single}
A.~Berezhkovskii and G.~Hummer, \emph{Physical review letters}, 2002,
  \textbf{89}, 064503\relax
\mciteBstWouldAddEndPuncttrue
\mciteSetBstMidEndSepPunct{\mcitedefaultmidpunct}
{\mcitedefaultendpunct}{\mcitedefaultseppunct}\relax
\EndOfBibitem
\bibitem[van Hijkoop \emph{et~al.}(2007)van Hijkoop, Dammers, Malek, and
  Coppens]{van2007water}
V.~J. van Hijkoop, A.~J. Dammers, K.~Malek and M.-O. Coppens, \emph{The Journal
  of chemical physics}, 2007, \textbf{127}, 08B613\relax
\mciteBstWouldAddEndPuncttrue
\mciteSetBstMidEndSepPunct{\mcitedefaultmidpunct}
{\mcitedefaultendpunct}{\mcitedefaultseppunct}\relax
\EndOfBibitem
\bibitem[Sedlmeier \emph{et~al.}(2011)Sedlmeier, von Hansen, Mengyu, Horinek,
  and Netz]{sedlmeier2011water}
F.~Sedlmeier, Y.~von Hansen, L.~Mengyu, D.~Horinek and R.~R. Netz,
  \emph{Journal of Statistical Physics}, 2011, \textbf{145}, 240--252\relax
\mciteBstWouldAddEndPuncttrue
\mciteSetBstMidEndSepPunct{\mcitedefaultmidpunct}
{\mcitedefaultendpunct}{\mcitedefaultseppunct}\relax
\EndOfBibitem
\bibitem[Calero \emph{et~al.}(2011)Calero, Faraudo, and
  Aguilella-Arzo]{calero2011first}
C.~Calero, J.~Faraudo and M.~Aguilella-Arzo, \emph{Physical Review E}, 2011,
  \textbf{83}, 021908\relax
\mciteBstWouldAddEndPuncttrue
\mciteSetBstMidEndSepPunct{\mcitedefaultmidpunct}
{\mcitedefaultendpunct}{\mcitedefaultseppunct}\relax
\EndOfBibitem
\bibitem[Sharma and Biswas(2017)]{sharma2017hydration}
S.~Sharma and P.~Biswas, \emph{Journal of Physics: Condensed Matter}, 2017,
  \textbf{30}, 035101\relax
\mciteBstWouldAddEndPuncttrue
\mciteSetBstMidEndSepPunct{\mcitedefaultmidpunct}
{\mcitedefaultendpunct}{\mcitedefaultseppunct}\relax
\EndOfBibitem
\bibitem[Abraham \emph{et~al.}(2015)Abraham, Murtola, Schulz, P{\'a}ll, Smith,
  Hess, and Lindahl]{abraham2015gromacs}
M.~J. Abraham, T.~Murtola, R.~Schulz, S.~P{\'a}ll, J.~C. Smith, B.~Hess and
  E.~Lindahl, \emph{SoftwareX}, 2015, \textbf{1}, 19--25\relax
\mciteBstWouldAddEndPuncttrue
\mciteSetBstMidEndSepPunct{\mcitedefaultmidpunct}
{\mcitedefaultendpunct}{\mcitedefaultseppunct}\relax
\EndOfBibitem
\bibitem[Goel and Richter-Dyn(2016)]{GoelRichterDyn2016}
N.~S. Goel and N.~Richter-Dyn, \emph{{Stochastic Models in Biology}}, Elsevier,
  2016\relax
\mciteBstWouldAddEndPuncttrue
\mciteSetBstMidEndSepPunct{\mcitedefaultmidpunct}
{\mcitedefaultendpunct}{\mcitedefaultseppunct}\relax
\EndOfBibitem
\bibitem[Lau and Lubensky(2007)]{LauLubensky2007}
A.~W. Lau and T.~C. Lubensky, \emph{Physical Review E - Statistical, Nonlinear,
  and Soft Matter Physics}, 2007, \textbf{76}, year\relax
\mciteBstWouldAddEndPuncttrue
\mciteSetBstMidEndSepPunct{\mcitedefaultmidpunct}
{\mcitedefaultendpunct}{\mcitedefaultseppunct}\relax
\EndOfBibitem
\bibitem[Rold{\'{a}}n \emph{et~al.}(2015)Rold{\'{a}}n, Neri, D{\"{o}}rpinghaus,
  Meyr, and J{\"{u}}licher]{Roldan2015}
{\'{E}}.~Rold{\'{a}}n, I.~Neri, M.~D{\"{o}}rpinghaus, H.~Meyr and
  F.~J{\"{u}}licher, \emph{Physical Review Letters}, 2015, \textbf{115},
  year\relax
\mciteBstWouldAddEndPuncttrue
\mciteSetBstMidEndSepPunct{\mcitedefaultmidpunct}
{\mcitedefaultendpunct}{\mcitedefaultseppunct}\relax
\EndOfBibitem
\bibitem[Neri \emph{et~al.}(2017)Neri, Rold{\'a}n, and
  J{\"u}licher]{neri2017statistics}
I.~Neri, {\'E}.~Rold{\'a}n and F.~J{\"u}licher, \emph{Physical Review X}, 2017,
  \textbf{7}, 011019\relax
\mciteBstWouldAddEndPuncttrue
\mciteSetBstMidEndSepPunct{\mcitedefaultmidpunct}
{\mcitedefaultendpunct}{\mcitedefaultseppunct}\relax
\EndOfBibitem
\bibitem[Krapivsky and Redner(2018)]{krapivsky2018first}
P.~Krapivsky and S.~Redner, \emph{Journal of Statistical Mechanics: Theory and
  Experiment}, 2018, \textbf{2018}, 093208\relax
\mciteBstWouldAddEndPuncttrue
\mciteSetBstMidEndSepPunct{\mcitedefaultmidpunct}
{\mcitedefaultendpunct}{\mcitedefaultseppunct}\relax
\EndOfBibitem
\bibitem[Metzler \emph{et~al.}(1999)Metzler, Barkai, and Klafter]{Metzler1999}
R.~Metzler, E.~Barkai and J.~Klafter, \emph{{Deriving fractional Fokker-Planck
  equations from a generalised master equation}}, ~4, 1999\relax
\mciteBstWouldAddEndPuncttrue
\mciteSetBstMidEndSepPunct{\mcitedefaultmidpunct}
{\mcitedefaultendpunct}{\mcitedefaultseppunct}\relax
\EndOfBibitem
\bibitem[Barkai \emph{et~al.}()Barkai, Metzler, and Klafter]{Barkai2000}
E.~Barkai, R.~Metzler and J.~Klafter, \emph{{From continuous time random walks
  to the fractional Fokker-Planck equation}}\relax
\mciteBstWouldAddEndPuncttrue
\mciteSetBstMidEndSepPunct{\mcitedefaultmidpunct}
{\mcitedefaultendpunct}{\mcitedefaultseppunct}\relax
\EndOfBibitem
\bibitem[Sheu and Yang(2010)]{Sheu2010}
S.~Y. Sheu and D.~Y. Yang, \emph{Journal of Physical Chemistry B}, 2010,
  \textbf{114}, 16558--16566\relax
\mciteBstWouldAddEndPuncttrue
\mciteSetBstMidEndSepPunct{\mcitedefaultmidpunct}
{\mcitedefaultendpunct}{\mcitedefaultseppunct}\relax
\EndOfBibitem
\bibitem[Loos \emph{et~al.}(2019)Loos, Hermann, and Klapp]{loos2019non}
S.~A. Loos, S.~M. Hermann and S.~H. Klapp, \emph{arXiv preprint
  arXiv:1910.08372}, 2019\relax
\mciteBstWouldAddEndPuncttrue
\mciteSetBstMidEndSepPunct{\mcitedefaultmidpunct}
{\mcitedefaultendpunct}{\mcitedefaultseppunct}\relax
\EndOfBibitem
\bibitem[Björneholm \emph{et~al.}(2016)Björneholm, Hansen, Hodgson, Liu,
  Limmer, Michaelides, Pedevilla, Rossmeisl, Shen, Tocci, Tyrode, Walz, Werner,
  and Bluhm]{chemrev2016}
O.~Björneholm, M.~H. Hansen, A.~Hodgson, L.-M. Liu, D.~T. Limmer,
  A.~Michaelides, P.~Pedevilla, J.~Rossmeisl, H.~Shen, G.~Tocci, E.~Tyrode,
  M.-M. Walz, J.~Werner and H.~Bluhm, \emph{Chemical Reviews}, 2016,
  \textbf{116}, 7698--7726\relax
\mciteBstWouldAddEndPuncttrue
\mciteSetBstMidEndSepPunct{\mcitedefaultmidpunct}
{\mcitedefaultendpunct}{\mcitedefaultseppunct}\relax
\EndOfBibitem
\bibitem[Giberti and Hassanali(2017)]{gibertihassanali2017}
F.~Giberti and A.~A. Hassanali, \emph{The Journal of Chemical Physics}, 2017,
  \textbf{146}, 244703\relax
\mciteBstWouldAddEndPuncttrue
\mciteSetBstMidEndSepPunct{\mcitedefaultmidpunct}
{\mcitedefaultendpunct}{\mcitedefaultseppunct}\relax
\EndOfBibitem
\bibitem[Poli \emph{et~al.}(2020)Poli, Jong, and Hassanali]{Poli2020}
E.~Poli, K.~H. Jong and A.~Hassanali, \emph{Nature Communications}, 2020,
  \textbf{11}, 901\relax
\mciteBstWouldAddEndPuncttrue
\mciteSetBstMidEndSepPunct{\mcitedefaultmidpunct}
{\mcitedefaultendpunct}{\mcitedefaultseppunct}\relax
\EndOfBibitem
\bibitem[Ramakrishnan \emph{et~al.}(2017)Ramakrishnan, González-Jiménez,
  Lapthorn, and Wynne]{Klass2017}
G.~Ramakrishnan, M.~González-Jiménez, A.~J. Lapthorn and K.~Wynne, \emph{The
  Journal of Physical Chemistry Letters}, 2017, \textbf{8}, 2964--2970\relax
\mciteBstWouldAddEndPuncttrue
\mciteSetBstMidEndSepPunct{\mcitedefaultmidpunct}
{\mcitedefaultendpunct}{\mcitedefaultseppunct}\relax
\EndOfBibitem
\bibitem[Tros \emph{et~al.}(2017)Tros, Zheng, Hunger, Bonn, Bonn, Smits, and
  Woutersen]{Tros2017}
M.~Tros, L.~Zheng, J.~Hunger, M.~Bonn, D.~Bonn, G.~J. Smits and S.~Woutersen,
  \emph{Nature Communications}, 2017, \textbf{8}, 904\relax
\mciteBstWouldAddEndPuncttrue
\mciteSetBstMidEndSepPunct{\mcitedefaultmidpunct}
{\mcitedefaultendpunct}{\mcitedefaultseppunct}\relax
\EndOfBibitem
\bibitem[Jorgensen \emph{et~al.}(1996)Jorgensen, Maxwell, and
  Tirado-Rives]{jorgensen1996development}
W.~L. Jorgensen, D.~S. Maxwell and J.~Tirado-Rives, \emph{Journal of the
  American Chemical Society}, 1996, \textbf{118}, 11225--11236\relax
\mciteBstWouldAddEndPuncttrue
\mciteSetBstMidEndSepPunct{\mcitedefaultmidpunct}
{\mcitedefaultendpunct}{\mcitedefaultseppunct}\relax
\EndOfBibitem
\bibitem[Jorgensen \emph{et~al.}(1983)Jorgensen, Chandrasekhar, Madura, Impey,
  and Klein]{jorgensen1983comparison}
W.~L. Jorgensen, J.~Chandrasekhar, J.~D. Madura, R.~W. Impey and M.~L. Klein,
  \emph{The Journal of chemical physics}, 1983, \textbf{79}, 926--935\relax
\mciteBstWouldAddEndPuncttrue
\mciteSetBstMidEndSepPunct{\mcitedefaultmidpunct}
{\mcitedefaultendpunct}{\mcitedefaultseppunct}\relax
\EndOfBibitem
\bibitem[Jorgensen and Madura(1985)]{jorgensen1985temperature}
W.~L. Jorgensen and J.~D. Madura, \emph{Molecular Physics}, 1985, \textbf{56},
  1381--1392\relax
\mciteBstWouldAddEndPuncttrue
\mciteSetBstMidEndSepPunct{\mcitedefaultmidpunct}
{\mcitedefaultendpunct}{\mcitedefaultseppunct}\relax
\EndOfBibitem
\bibitem[Darden \emph{et~al.}(1993)Darden, York, and
  Pedersen]{darden1993particle}
T.~Darden, D.~York and L.~Pedersen, \emph{The Journal of chemical physics},
  1993, \textbf{98}, 10089--10092\relax
\mciteBstWouldAddEndPuncttrue
\mciteSetBstMidEndSepPunct{\mcitedefaultmidpunct}
{\mcitedefaultendpunct}{\mcitedefaultseppunct}\relax
\EndOfBibitem
\bibitem[Hess \emph{et~al.}(1997)Hess, Bekker, Berendsen, and
  Fraaije]{hess1997lincs}
B.~Hess, H.~Bekker, H.~J. Berendsen and J.~G. Fraaije, \emph{Journal of
  computational chemistry}, 1997, \textbf{18}, 1463--1472\relax
\mciteBstWouldAddEndPuncttrue
\mciteSetBstMidEndSepPunct{\mcitedefaultmidpunct}
{\mcitedefaultendpunct}{\mcitedefaultseppunct}\relax
\EndOfBibitem
\bibitem[Bussi \emph{et~al.}(2007)Bussi, Donadio, and
  Parrinello]{bussi2007canonical}
G.~Bussi, D.~Donadio and M.~Parrinello, \emph{The Journal of chemical physics},
  2007, \textbf{126}, 014101\relax
\mciteBstWouldAddEndPuncttrue
\mciteSetBstMidEndSepPunct{\mcitedefaultmidpunct}
{\mcitedefaultendpunct}{\mcitedefaultseppunct}\relax
\EndOfBibitem
\bibitem[Trefethen(2000)]{Trefethen2000}
L.~N. Trefethen, \emph{{Spectral Methods in MATLAB}}, SIAM, 2000\relax
\mciteBstWouldAddEndPuncttrue
\mciteSetBstMidEndSepPunct{\mcitedefaultmidpunct}
{\mcitedefaultendpunct}{\mcitedefaultseppunct}\relax
\EndOfBibitem
\end{mcitethebibliography}

\end{document}